%% file: Kleinert_Stephan_2015.tex
\documentclass[%
preprint,%
amsmath,amssymb]{elsarticle}

\usepackage{lineno,hyperref}

\journal{Physics Reports}

\input macro.inc                             %
\usepackage[utf8]{inputenc}
\usepackage{dsfont}
\usepackage{captcont}
\usepackage{subfigure}
\usepackage{booktabs}
\usepackage{amsmath}
\usepackage{amssymb}
\usepackage{natbib}
\usepackage[final]{showkeys}

\usepackage{graphicx}
\usepackage{dcolumn}
\usepackage{bm}

\biboptions{sort&compress}

\begin{document}

\begin{frontmatter}

\title{Representation-free description of light-pulse atom interferometry including non-inertial effects}

%


\author[Ulm]{Stephan Kleinert\corref{mycorrespondingauthor}}
\cortext[mycorrespondingauthor]{Corresponding author} 
\ead{kleinert.stephan@uni-ulm.de}

\author[Ulm]{Endre Kajari}%
\author[Ulm]{Albert Roura}%
\author[Ulm,Texas]{Wolfgang P. Schleich}

\address[Ulm]{Institut für Quantenphysik and Center for Integrated Quantum Science  and  Technology ($\text{I}Q^{ST}$), 
	  Universität Ulm, Albert-Einstein-Allee 11,  D-89081 Ulm, Germany.}

\address[Texas]{Texas A\&M University Institute for Advanced Study (TIAS), Institute for Quantum Science and Engineering (IQSE) and Department of Physics and Astronomy,  
	  Texas A\&M University College Station, Texas 77843-4242, USA.}
%
%
%
%
%
%
\begin{abstract}
Light-pulse atom interferometers rely on the wave nature of matter and
its manipulation with coherent laser pulses. They are used for precise
gravimetry and inertial sensing as well as for accurate measurements of
fundamental constants. Reaching higher precision requires longer
interferometer times which are naturally encountered in microgravity
environments such as drop-tower facilities, sounding rockets and
dedicated satellite missions aiming at fundamental quantum physics in
space. In all those cases, it is necessary to consider arbitrary
trajectories and varying orientations of the interferometer set-up in
non-inertial frames of reference.

Here we provide a versatile representation-free description of atom
interferometry entirely based on operator algebra to address this
general situation. We show how to analytically determine the phase shift
as well as the visibility of interferometers with an arbitrary number of
pulses including the effects of local gravitational accelerations,
gravity gradients, the rotation of the lasers and non-inertial frames of
reference. Our method conveniently unifies previous results and
facilitates the investigation of novel interferometer geometries.
\end{abstract}
\end{frontmatter}

\vspace*{1.5cm}

%
%

\pagebreak[4]

\tableofcontents
\vspace*{0.5cm}

%
%
%
%
%
%
\section{Introduction}
\label{sec_introduction}
The new era of matter-wave interferometry was initiated in 1924 by Louis de Broglie. 
The particle-wave complementarity combines the property of matter to behave as particles as well as waves. 
In analogy to light interferometer experiments~\cite{Michelson_1887,Perot_1899} the wave nature of matter has led to early 
proposals for matter-wave interferometry~\cite{Moellenstedt_1957,Lichte_1972,Rauch_1974,Colella_1975}. 
Significant progress in quantum optics, for instance the 
ability to manipulate internal 
atomic states by radio-frequency resonance demonstrated by Rabi et al.~\cite{Rabi_1938} 
and long-time coherent experiments by Ramsey~\cite{Ramsey_1949}, made accessible new standards in precise frequency measurement, 
nuclear magnetic resonance spectroscopy 
and actually provided quantum information gates. In the late 20th century techniques for a coherent manipulation of atoms 
pioneered atom interferometry~\cite{Clauser_1988,Borde_1989,Kasevich_1991}. 

In this article we present a compact and versatile description 
of light-pulse atom interferometry which provides a straightforward method for obtaining the phase shift and the visibility 
for arbitrary interferometer geometries taking 
into account local accelerations, gradients and rotations of the device. We pursue a description 
entirely based on operator algebra mathematically acting at the very heart of quantum mechanics.\\
%
%
%
%

We start in Section~\ref{sec_overview} with a review of the development and the state of the art of light-pulse atom interferometers.
In Section~\ref{sec_MWI_quadratic_potentials} we model matter-wave interferometry in the presence of a general quadratic potential. 
In particular, our representation-free approach incorporates the internal and the external dynamics in terms of unitary beam-splitter matrices and time evolution operators, respectively.
Section~\ref{sec:Compact_description_of_interferometry} introduces the generalized beam splitter in order to 
combine the internal and the external dynamics in a compact way. 
In Section~\ref{sec:Compact_description:MZ_geometry} we write the Mach-Zehnder interferometer as a sequence of 
generalized beam splitters. Moreover, it turns out in Section~\ref{Probability_of_ground_state_detection} that the 
total interferometer phase is a consequence of the non-commutativity of such generalized beam splitters. 
At the end of the section, we present the ground-state detection probability 
for the Mach-Zehnder and the multi-loop geometries. Thereby, we introduce the ``vertex rule'' as a useful graphical tool. 
Section~\ref{sec_characteristic_quantities} makes the link to the 
experiment while presenting explicit terms for the characteristic quantities of the Mach-Zehnder and the Butterfly geometries.
Finally, we apply our representation-free approach in Section~\ref{Non-inertial_frame} to interferometers in non-inertial reference 
frames and decompose the general coordinate transformation into elementary ones.

In order to keep the paper self-contained,
we summarize concepts and detailed calculations in several appendices.
In Appendix A we specify the external potential used throughout the paper.
In Appendix~\ref{app:math_excursus_Sympl_group} and \ref{app:tau_is_sympl} we study symplectic groups in order to derive a 
compact description of the dynamics in interferometers. In addition, it is 
essential for the generalized beam splitter to introduce the displacement operator in Appendix~\ref{app:displacement_operator} 
and do the transformation to the Heisenberg picture. 
Since the symplectic Fourier transform of the Wigner function significantly simplifies the calculation of the detection 
probability of the interferometer, we give a short introduction into the Wigner and the characteristic functions in 
Appendix~\ref{app:characteristic_function} and \ref{app:Wigner_func}.  
We calculate in Appendix~\ref{app:Mach-Zehnder_operator} and \ref{app:total_phase} the Mach-Zehnder operator 
and the total phase shift for the Mach-Zehnder as well as the Butterfly geometry.
We conclude in Appendix~\ref{app:Rotation_group_SO(3)} with a discussion of the rotation  group particularly relevant for the description of interferometry in non-inertial frames.
%
%
%
%
%
%
%
%
\section{Overview of light-pulse atom interferometry}
\label{sec_overview}

\subsection{Early developments}

The wave nature of matter particles, first proposed by Louis de Broglie, can be exploited to construct matter-wave interferometers; see refs.~\cite{Cronin_2009,Berman_1997} for some general reviews.
This was first realized with electrons propagating through a metal grating \cite{Joensson_1961,Joensson_1974} and later with neutrons diffracted off crystals~\cite{Rauch_1974}. Neutrons' larger mass and their vanishing electric charge, which imply a much shorter de Broglie wavelength and insensitivity to spurious electric fields, provided the required precision for performing a number of interesting experiments \cite{Greenberger_1983}. These included the realization of Wheeler's delayed-choice experiment, experiments on the sign change produced by a $360^\circ$\! rotation, and inertial sensing. Indeed, neutron interferometers were employed to measure Earth's rotation via the Sagnac effect for matter waves \cite{Werner_1979} as well as determining Earth's gravitational acceleration in the first experimental observation of the gravitational interaction directly affecting the quantum dynamics of a system \cite{Colella_1975}.

The next step was to perform interferometry with neutral atoms, which are easier to produce, using a double slit \cite{Carnal_1991} or matter gratings \cite{Keith_1991}. However, taking advantage of the ability to manipulate light in a well-controlled manner, higher-quality gratings (and which do not get clogged) can be achieved using light gratings generated with laser beams.
This was initially done with standing waves, both in the so-called Raman-Nath \cite{Gould_1986,Rasel_1995} and Bragg \cite{Martin_1988,Giltner_1995} regimes (corresponding respectively to thin and thick gratings).
On the other hand, laser beams consisting of running waves had been employed in schemes that generalized the Ramsey spectroscopy technique (based on the use of two $\pi/2$ pulses separated by a time $T$ \cite{Ramsey_1950} and being the key element in standard atomic clocks) to the case of optical transitions \cite{Keupp_2005,Poli_2014}. Such schemes involved a subsequent set of two additional $\pi/2$ pulses propagating in opposite direction to the first pair and could be naturally interpreted, as pointed out in Ref.~\cite{Borde_1989}, in terms of recoil-based matter-wave interferometers sensitive to inertial effects such as rotations and accelerations, which cause a frequency displacement in the Ramsey fringes (oscillations in the exit port population as a function of the frequency detuning). Furthermore, the internal-state labeling allows the read-out of the different exit ports without the need to spatially resolve them. The possibility of measuring rotation rates with this kind of interferometer was demonstrated experimentally for the first time in Ref.~\cite{Riehle_1991}.

The development of laser cooling techniques for neutral atoms was a crucial milestone on the road to precision atom interferometry which allowed longer interrogation times and narrower velocity distributions. Combining laser cooling with an atomic fountain configuration \cite{Kasevich_1989} enabled atomic clocks with longer times $T$ between the two microwave $\pi/2$ pulses (with the frequency resolution $\Delta \nu/\nu$ being inversely proportional to $T$) and a much larger number of Ramsey fringes (thanks to the narrower velocity spread), and the scheme has been employed in international standard time references based on microwave transitions ever since \cite{Sullivan_2001}.
A similar set-up based on an atomic fountain using laser-cooled atoms together with a pair of counterpropagating Raman beams along the vertical direction was shown to have a great potential for precision accelerometry and gravimetry measurements \cite{Kasevich_1991}. The atoms were exposed to a $\pi/2 - \pi - \pi/2$ pulse sequence (corresponding to a Mach-Zehnder configuration with a time $T$ between the pulses) which created a superposition of wave packets with central momenta differing by a double photon recoil (associated with the two-photon Raman transition) and eventually recombined them. The phase shift $\delta\phi$ between the two branches of the interferometer depends on the central position of the wave packets with respect to the laser wave fronts at the times of interaction with each pulse and is sensitive to the relative acceleration of the atoms: $\delta\phi = k_\text{eff}\, a\, T^2$, where $a$ is the acceleration of the atoms in the frame where the lasers are at rest and $\hbar k_\text{eff}$ is the momentum transfer associated with the two-photon recoil.
In contrast to previous schemes where the transverse motion of the atoms across a continuous beam determined the interaction time, here pulsed beams were employed to select the duration of the atom-light interaction and determine the amount of Rabi oscillation ($\pi/2$ other $\pi$).
This has become the standard set-up for light-pulse atom interferometers and can also be employed to measure rotation rates by choosing the direction of the Raman beams perpendicular to the initial velocity of the atoms \cite{Canuel_2006,Gauguet_2009,Stockton_2011,Berg_2015}.

Before turning to a more detailed discussion of various aspects of light-pulse atom interferometry, it is worth pointing out that in recent times matter-wave interferometry with much heavier objects (macromolecules and even nanoparticles) has been successfully performed employing matter gratings as well as light standing waves (acting as phase gratings or as absorption gratings) \cite{Haslinger_2013,Bateman_2014} and is a very active field \cite{Hornberger_2012,Kaltenbaek_2012,Kaltenbaek_2015} playing a crucial role in the investigation of quantum coherent phenomena with mesoscopic objects and the quest to explore the validity of quantum mechanics closer and closer to the macroscopic regime \cite{Nimmrichter_2013}.

\subsection{Diffraction mechanisms}
\label{sec:diffraction_mechanisms}

Two kinds of diffraction mechanisms are commonly employed in light-pulse atom interferometry: Raman and Bragg scattering. Both are based on laser pulses of finite duration in time and by adjusting their intensity and duration, one can generate so-called $\pi/2$ and $\pi$ pulses. The former, which gives rise to an equal-amplitude superposition of diffracted and undiffracted states, act as beam splitters, whereas $\pi$ pulses completely transfer the original state into the diffracted one and play a role analogous to mirrors in an optical interferometer.

When \emph{Raman} scattering is employed \cite{Kasevich_1991}, a pair of counterpropagating (sometimes copropagating) laser beams induce a two-photon transition between different hyperfine levels of the ground state. The process can be understood qualitatively as the absorption of a photon from one beam and the stimulated emission of another photon in the mode associated with the second beam. For counterpropagating beams this leads to an effective momentum transfer in the center-of-mass (COM) motion of the atom corresponding to twice the momentum of a single photon. Furthermore, in that case it is a velocity selective (also known as Doppler sensitive)
process and in order to obtain maximal diffraction efficiency, the frequency difference between the two beams needs to be resonantly tuned to account for the internal energy difference of the two hyperfine states (corresponding to several GHz) plus the recoil energy (tens of kHz) \cite{Moler_1992}. The finite duration of the pulse allows a certain deviation from this resonance condition (in agreement with Heisenberg's time-energy uncertainty relation) and for sufficiently short pulses (with accordingly higher intensity) one can typically achieve efficient diffraction of a band of momentum states around the resonant one with a velocity spread comparable to the value of the recoil velocity \cite{Moler_1992}.

The implementation of \emph{Bragg} scattering in light-pulse atom interferometry \cite{Kozuma_1999,Torii_2000} is similar to that described above for Raman scattering with counterpropagating laser beams. The main difference is that no change of internal state is involved, only a change of state for the COM motion. Therefore, the frequency difference needs to be tuned to a few tens of kHz (corresponding just to the recoil energy) and this can be easily accomplished with a single laser and acousto-optical modulators (AOMs) rather than the more complex set-up with a pair of phase-locked lasers typically required for Raman scattering.
On the other hand, there are in this case additional diffraction orders (associated with $2n$-photon transitions) whose resonance condition differs by a multiple, $n^2$, of the two-photon recoil energy. If one tried to used sufficiently short pulses to diffract a band of momentum states with a velocity spread comparable to the two-photon recoil velocity, the amplitude of exciting additional diffraction orders would be non-negligible. A rather narrow momentum distribution is, thus, necessary in order to select a single diffraction order (and also helps to spatially separate the exit ports, given the absence of internal state labeling in this case). This is a requirement which can be naturally met when working with Bose-Einstein condensates (BECs) \cite{Kozuma_1999} and Bragg diffraction is routinely employed in BEC-based atom interferometers \cite{Torii_2000,Debs_2011,Altin_2013,Muentinga_2013}.

Since the sensitivity of atom interferometers is typically proportional to the effective momentum transfer $k_\text{eff}$ (sometimes even quadratically), there have been notable efforts to attain higher values of $k_\text{eff}$. This can be done by including between the beam-splitter and mirror pulses a number of intermediate $\pi$ pulses which increase the relative velocity between the two interferometer branches \cite{McGuirk_2000,Leveque_2009a}. Alternatively, when Bragg scattering is employed, one can tune the frequency difference of the two beams so that the transition to a higher diffraction order becomes resonant and a $2n$-photon recoil is transferred \cite{Mueller_2008a,Chiow_2009}. Finally, one can try to find an optimal combination of both: in this way a total momentum transfer of 102 times a single-photon recoil was achieved in Ref.~\cite{Chiow_2011}.
Bloch oscillations have also been implemented, instead of additional $\pi$ pulses, as a way of increasing the relative velocity between the two interferometer branches \cite{Clade_2009,Mueller_2009,Kovachy_2010,Mcdonald_2013b,Mcdonald_2014b}, but very careful manipulation is required to avoid introducing small uncontrolled phase shifts in the process. This is, however, less problematic when used to accelerate both branches equally: there is then no increase of their relative velocity, but it can still lead to a significant sensitivity increase in certain recoil measurements \cite{Cadoret_2008,Bouchendira_2011}.

A retroreflection scheme where the two laser beams with different frequencies and properly chosen polarizations are injected together and reflected off a mirror covered with a quarter-lambda plate (which rotates the polarization plane) is commonly employed in high-precision measurements \cite{Peters_2001}. This results in two pairs of counterpropagating beams such that each pair corresponds to the configuration described above for Raman or Bragg scattering, and the propagation directions in one pair are opposite to those in the other pair. Such a scheme is employed to reduce the unwanted effects of wave-front distortions and part of the effects due to vibrations: because the two beams travel together up to the injection point, they are affected in the same way and the effects on the two-photon process essentially cancel out except for the vibrations of the mirror and any curvature imperfections of its surface. In atomic fountains the nonvanishing velocity of the atoms with respect to the mirror selects via Doppler effect only one of the counterpropagating pairs. This is not the case, however, in microgravity experiments, where the atoms are typically prepared with vanishing mean velocity. Both counterpropagating pairs induce then resonant transitions and give rise to \emph{double diffraction} processes with a richer phenomenology than the usual single diffraction \cite{Leveque_2009a,Giese_2013}. One can still have beam-splitter and mirror pulses, which generate then symmetric atom interferometer configurations where $k_\text{eff}$ is automatically doubled and a number of systematic effects and noise sources (including laser phase noise) cancel out. This was first implemented in Raman-based interferometers \cite{Leveque_2009a}, which become in addition insensitive to noise and systematics associated with the AC Stark shift because the internal state of the atoms is the same at any instant of time in both branches of the interferometer. Its extension to Bragg scattering was studied in detail in Ref.~\cite{Giese_2013} and has recently been implemented experimentally \cite{Kueber_2014,Ahlers_2015}.
Besides microgravity environments, double diffraction can be naturally employed in atom interferometers acting as gyroscopes where the Raman or Bragg beams are perpendicular to the velocity of the atoms (and to the gravitational acceleration)%
\footnote{If desired a single diffraction scheme can still be used in this case by tilting slightly the beams with respect to the direction of the atom velocity \cite{Canuel_2006}.}.
Furthermore, by considering an additional third laser beam in the retro-reflection scheme and a suitable choice of the three laser frequencies, a generalization of double diffraction which still retains many of its advantages can be employed in interferometers acting as gravimeters \cite{Malossi_2010}. Such a scheme has recently been exploited to perform precise tests of the equivalence principle with $^{85}\text{Rb}$ and $^{87}\text{Rb}$ \cite{Zhou_2015}.

In addition to Raman and Bragg scattering, diffraction by traveling laser waves based on single-photon optical transitions is sometimes used in atom interferometry. It was considered in the original proposal for Ramsey-Bordé interferometers \cite{Borde_1989} and has been employed in optical atomic clocks with free atoms \cite{Keupp_2005,Poli_2014}. More recently, it has received renewed attention as part of a novel scheme for differential phase-shift measurements with spatially separated atom interferometers sharing common laser beams \cite{Graham_2013,Hogan_2015}. In this new scheme the effects of laser phase noise are highly suppressed even when considering very long baselines (e.g.~millions of kilometers for certain space applications). It requires very long-lived metastable states (e.g.~the clock transition in $^{87}\text{Sr}$), so that spontaneous decay does not lead to significant loss of coherence even for long interferometer times.

\subsection{Applications to high-precision measurements}

In the last two decades light-pulse atom interferometers have demonstrated their great potential as highly accurate quantum sensors for both practical applications and fundamental measurements, which we summarize in this subsection.

\subsubsection{Inertial sensing}
\label{sec:inertial_sensing}

As already emphasized in Clauser's seminal paper of 1988 \cite{Clauser_1988}, matter-wave interferometers employing laser-cooled neutral atoms and diffraction gratings generated with laser beams can be exploited to construct very precise inertial sensors.
Clauser considered atomic beams crossing standing electromagnetic waves rather than light-pulse interferometers, but the essential idea is very similar and can be understood as follows. By considering neutral atoms in a magnetically insensitive state and using magnetic shielding to screen spurious magnetic fields and magnetic field gradients (and thus avoid forces due to the second order Zeeman effect) one can make sure that during their free evolution in the interferometer the motion of the atoms is only affected by gravitational and inertial forces to a very good approximation, so that they provide an excellent inertial reference. Accelerations and rotations of the interferometer device lead to changes in the location of the laser wave fronts relative to the atoms (acting as inertial references), which causes in turn a change of the phases acquired by the atoms in the diffraction process. And this finally gives rise to a net contribution to the phase shift between the interferometer branches, typically given by $\delta\phi \sim \vec k_\text{eff}^\text{T}\, \vec a\, T^2$ for accelerations and $\delta\phi \sim \vec k_\text{eff}^\text{T}\, (\vec v_0 \times \vec\Omega)\, T^2$ for rotations. Therefore, by monitoring appropriately their phase shift, one can use atom interferometers (or a combinations of them) as precise accelerometers and gyroscopes, which have found the following applications.
\begin{itemize}
\item \emph{Gravimetry and gradiometry:} For more than a decade atom interferometers have occupied a prominent place among the most accurate absolute gravimeters \cite{Peters_1999,Peters_2001,Merlet_2010,Farah_2014b,Hu_2013,Altin_2013}. Moreover, the development of mobile versions \cite{Hauth_2013} with comparable precision and accuracy, and even compact portable ones \cite{Bidel_2013}, make them particularly attractive for geophysics applications.
The main limitations are due to vibration noise of the retro-reflection mirror (as established by the equivalence principle, vibrations and accelerations are indistinguishable from gravitational forces). Using vibration isolation systems, sensitivities of $10^{-9} g$ can be achieved in about 10\,s. According to Ref.~\cite{Hu_2013}, there are prospects for reaching that sensitivity in 1\,s and improving the sensitivity by one order of magnitude after 100\,s of integration time.

On the other hand, by considering differential phase-shift measurements of spatially separated atom interferometers sharing common laser beams, one can perform gravity gradient measurements along the baseline with a precision which can exceed the limitations due to vibration noise for each single interferometer thanks to common-mode noise rejection in the differential measurement. This kind of devices are particularly sensitive to the properties and motion of local masses and have both geophysical and fundamental applications (some of them are discussed below).
Mainly set-ups with vertical baselines of the order of one meter have been considered so far \cite{Snadden_1998,McGuirk_2002,Sorrentino_2014}, but horizontal baselines of close to a meter have recently been reported as well \cite{Biedermann_2015}. A substantially enhanced version (with much longer interferometer times) aboard a dedicated satellite mission has been proposed for geodesy applications relying on precise measurements of gravity gradients at larger scales \cite{Carraz_2014}.

Furthermore, a gravitational antenna (MIGA) with a horizontal baseline of 200\,m is being built in a low-noise underground laboratory in France~\cite{Canuel_2014,Geiger_2015}. It will consist of several spatially separated atom interferometers distributed along the baseline and being interrogated by a common laser field inside a 200-meter-long optical cavity, and it will monitor changes in the local gravitational field with unprecedented sensitivity and with very interesting geophysical and hydrological applications.

Finally, it is also worth mentioning that the gravitational field curvature (corresponding to third-order derivatives of the potential) has recently been measured  employing a set-up similar to the vertical gradiometers described above \cite{Rosi_2015}.
\item \emph{Gyroscopes and navigation:} Atom interferometers employing thermal atomic beams \cite{Lenef_1997} have reached sensitivities better than $10^{-9}$\,rad/s \cite{Gustavson_1997,Gustavson_2000a} and have been further improved to meet the precision and stability requirements for navigation applications \cite{Durfee_2006}.
On the other hand, light-pulse interferometers with cold atoms can achieve high sensitivities with a significantly smaller size by using longer interferometer times (and smaller launch velocities $v_0$) \cite{Canuel_2006,Gauguet_2009,Stockton_2011,Barrett_2014,Berg_2015}, and have been shown to possess great potential for geodetic applications and inertial navigation \cite{Stockton_2011}. Moreover, a six-axis inertial sensor (for rotations and accelerations) has already been demonstrated \cite{Canuel_2006}.
\item \emph{Fundamental tests:} The equivalence principle is the cornerstone of general relativity and it encompasses three different but interrelated aspects: local Lorentz invariance (LLI), local position invariance (LPI) and the universality of free fall (UFF) \cite{Will_2014}.
There is, therefore, great interest in testing experimentally their validity with high precision, and it is very useful to have consistent theoretical frameworks describing such violations and allowing an unambiguous and well-defined parametrization.
The so-called Standard-Model Extension (SME) provides a rather general framework for describing violations of LLI \cite{Mattingly_2005}. Particle physics experiments together with cosmic ray observations \cite{Kostelecky_2011b} as well as atomic clock \cite{Wolf_2006} and atom-interferometric \cite{Mueller_2008c} measurements  have been exploited to put stringent bounds on many of its parameters. In addition to violations of LLI, the SME necessarily implies violations of LPI and UFF as well. Testing the latter two can be a good way of putting bounds on parameters within the SME that only affect gravitational phenomena \cite{Kostelecky_2011a}.
Furthermore, string-theory-inspired dilaton models \cite{Damour_2010} and related models giving rise to time-dependence of the fundamental constants \cite{Uzan_2011} give rise to violations of LPI and UFF while preserving LLI. LPI can be tested with gravitational redshift measurements comparing atomic clocks at different locations \cite{Altschul_2015} as well as with accurate measurements of atomic spectra from distant cosmological sources \cite{Uzan_2011}. UFF in turn has been tested at the $10^{-13}$ level with torsion balance experiments \cite{Schlamminger_2008,Adelberger_2009} and lunar laser ranging. Tests of UFF with atom interferometers \cite{Fray_2004,Bonnin_2013,Schlippert_2014,Zhou_2015,Bonnin_2015}  performing differential measurements with two different species (or related atomic experiments \cite{Tarallo_2014}) have reached at most the $10^{-8}$ level so far. However, they can still provide useful bounds on certain parameter combinations \cite{Hohensee_2011e,Hohensee_2013b,Hohensee_2011f,Schlippert_2014} because the atomic species employed are quite different from the materials used in the torsion balance experiments.
Moreover, there are plans for a dedicated satellite mission to test the UFF with atom interferometry at the $10^{-15}$ level~\cite{Aguilera_2014}.

In addition, atom-interferometry-based gradiometers like those described above have been employed to test Newton's inverse square law with unprecedented accuracy at 10-cm scales \cite{Biedermann_2015}. This was done by changing the position of well-characterized heavy masses near the center of the gradiometer baseline and measuring the corresponding changes in the gravity gradient.

\end{itemize}

\subsubsection{Measurement of fundamental constants}
\vspace*{1.0ex}

\begin{itemize}
\item \emph{Newton's gravitational constant $G$:} By changing the position of well characterized ring-shaped masses along the baseline separating the two atom interferometers in the vertical gradiometer configuration briefly described in Section~\ref{sec:inertial_sensing} and measuring the corresponding changes of the gravity gradient, one can measure the gravitational constant $G$ with an accuracy comparable to other state-of-the-art methods (mainly torsion balance experiments analogous to Cavendish's original experiment) \cite{Fixler_2007,Lamporesi_2008,Rosi_2014}. This is the less accurately determined fundamental constant of nature and there are conflicting results for different sets of measurements. The atom interferometric measurements provide a valuable addition with a very different kind of systematics from the other existing experiments.

It should also be mentioned that the possibility of using an alternative gradiometer configuration with a horizontal baseline has been investigated in Ref.~\cite{Biedermann_2015}.
\item \emph{Recoil measurements} and the determination of the \emph{fine-structure constant}~$\alpha$:
The phase shift in a Ramsey-Bordé interferometer with two-photon Raman pulses (or Bragg pulses) gets a contribution from the additional kinetic energy acquired by the atoms in one of the branches and directly related to the two-photon recoil. Since the wavelength of the photons can be specified with high accuracy, determining the recoil energy from the phase shift allows a precise measurement of $\hbar/m_X$, where $m_X$ is the mass of the atomic species employed \cite{Weiss_1993}. This precision can be increased by simultaneously increasing the velocity of the atoms in the two interferometer branches using Bloch oscillations \cite{Cadoret_2008,Bouchendira_2011,Bouchendira_2013}. Alternatively one can use higher-order Bragg pulses and take advantage of the fact that in this case the phase shift scales quadratically with $k_\text{eff}$ \cite{Lan_2013}.

The future redefinition of the kilogram which is under consideration implies fixing the value of the Planck constant. 
In that context the measurement of $\hbar/m_X$ provides a direct way of establishing an accurate mass scale in the atomic regime. The link with macroscopic scales could then be provided by the single-crystal $^{28}\text{Si}$ spheres of the Avogadro project, whose total number of atoms can be determined very precisely \cite{Bouchendira_2013,Lan_2013}.

In addition, an accurate measurement of $\hbar/m_X$ also provides an accurate way of determining the fine-structure constant~$\alpha$ \cite{Weiss_1993,Wicht_2002} thanks to the great accuracy with which the Rydberg constant $R_\infty$ is determined through spectroscopic measurements. In this way $\alpha$ has been determined \cite{Bouchendira_2011,Bouchendira_2013} with an accuracy comparable to the best existing results, obtained from the measurement of the anomalous magnetic moment of the electron. Comparing the two results can be regarded as a consistency check of QED calculations in particle physics from measurements in the atomic regime \cite{Bouchendira_2011,Bouchendira_2013}.
\end{itemize}

\subsubsection{General relativistic effects}
\vspace*{1.0ex}

\begin{itemize}
\item \emph{Lense-Thirring effect:}
A gyroscope following a certain trajectory in the gravitational field generated by a rotating mass distribution will experience a precession with respect to a distant fixed reference due to several general relativistic effects: the geodetic effect  (due to the relative velocity of the test particle with respect to the source), the Lense-Thirring effect (due to the frame dragging in the spacetime geometry surrounding a rotating source) and the Thomas precession (a special-relativistic effect due to the noncommutativity of Lorentz boosts along different directions). Exploiting the potential of atom interferometry for building compact and precise gyroscopes, a satellite mission for measuring the Lense-Thirring effect, which is of the order of $10^{-14}$\,rad/s for the case of the Earth, was proposed in Ref.~\cite{Jentsch_2004}.
\item \emph{PPN parameters:}
It has been suggested \cite{Dimopoulos_2008} that atom interferometers can be used to measure a linear combination of the PPN parameters $\beta$ and $\gamma$, which characterize the nonlinearity of the gravitational field and the spatial curvature of the spacetime geometry, respectively, to lowest order in the post-Newtonian expansion ($\beta = \gamma = 1$ for general relativity). As far as the dynamics of nonrelativistic atoms is concerned, the effect of such terms essentially amounts to replacing the gravitational potential $\phi$ with $\phi - (\beta + \gamma)\, \phi^2/c^2$. This extra contribution cannot be easily distinguished from the gravitational field itself, but it is in principle possible if one considers the gravity gradient along three perpendicular directions. This is because, in contrast with the Newtonian gravity gradient, its trace does not vanish in a region with vanishing density.
The measurement is, nevertheless, very challenging because the contribution to the gravity gradient is suppressed by a factor $10^{-9}$ compared to Earth's gravity gradient. Moreover, in principle one would need to guarantee the orthogonality of the three measured directions at the $10^{-9}$ level too. 
\item \emph{Gravitational wave detection:}
The use of single atom interferometers for gravitational wave detection was put forward in Ref.~\cite{Chiao_2004a}, but it was later shown to be based on a flawed analysis \cite{Roura_2006}. On the other hand, refs.~\cite{Dimopoulos_2008b,Hogan_2011} proposed a very different scheme where the gravitational waves directly affect the propagation of laser beams over long distances. These laser beams are shared by atom interferometers spatially separated by this long baseline and a differential phase-shift measurement is performed. This kind of gravitational antenna is, therefore, similar to those based on optical interferometers but with the atom interferometers replacing the freely suspended mirrors as inertial references. An important point to emphasize is that in order to minimize aliasing effects, a substantial number of atomic clouds need to be operated concurrently (with time separations among them much smaller than the total interferometer time for each one of them). 

More recently, a novel scheme based on single-photon optical transitions has been proposed \cite{Graham_2013}. It has the great advantage that the effects of laser phase noise are highly suppressed. Thanks to that, a single arm would be sufficient even for long baselines \cite{Hogan_2015,Chiow_2015} comparable to the $10^6$\,km of LISA. (Multiple arms would still be desirable because they provide additional information on the polarization of the gravitational waves, which can help to pinpoint the location of their source in the sky.)

As described in Section~\ref{sec:inertial_sensing}, MIGA is a prototype for a gravitational antenna with a 200-meter baseline which is being built in an underground laboratory in France \cite{Canuel_2014,Geiger_2015} and can serve as a testbed for some of these ideas.
\end{itemize}

\subsection{Long-time interferometry}

Since the sensitivity scales quadratically with the interferometer time in many cases, longer interferometer times (of the order of $10\, \text{s}$ or longer) are a key ingredient in order to achieve a substantial breakthrough in future high-precision measurements based on atom interferometry.
One possibility is to employ larger atomic fountains, and indeed total interferometer times over $2.5\, \text{s}$ have already been demonstrated in Stanford's 10-meter tower \cite{Dickerson_2013}. On the other hand, longer interferometer times can be naturally achieved with much more compact interferometer set-ups \cite{Rudolph_2015} in microgravity environments. There have recently been increasing efforts in this direction using parabolic flights \cite{Geiger_2011}, drop-tower facilities \cite{Zoest_2010,Muentinga_2013} and sounding rockets \cite{maius}. These can provide a very valuable testbed for the design and development of future experiments in the International Space Station \cite{cal}, or even dedicated satellite missions \cite{Aguilera_2014,ste-quest_general}.

Long-time interferometry offers the possibility of reaching unprecedented sensitivities, but it also poses serious challenges. A major one is preventing the size of the expanding atom cloud from becoming too large, which would lead to a number of drawbacks. First, if the atom density becomes too low, it is not possible to have a sufficiently high signal-to-noise ratio for the atom detection at the exit ports of the interferometer. Moreover, the cloud may even exceed the size of the set-up, so that a non-negligible fraction of the atoms is lost and does not contribute to the signal at all.
Second, a larger cloud is more sensitive to wave-front distortions. These can be mitigated by employing a retroreflection scheme as described in Section~\ref{sec:diffraction_mechanisms}, but for high-precision measurements and large atom clouds the requirements on the  regularity of the retroreflecting mirror's curvature may still become exceedingly high.
Third, rotations and gravity gradients lead to relative shifts in the central position and momentum of the two interfering wave packets at each exit port and this causes the appearance of a fringe pattern in the density profile as well as a contrast reduction in the oscillations (as a function of the phase shift) of the integrated particle number in each port. The effect is more important for larger atom clouds, when the size of the envelope (the size of the cloud) is larger than the fringe spacing. Fortunately, mitigation strategies based on the use of a tip-tilt mirror for retroreflection \cite{Hogan_2008,Lan_2012,Dickerson_2013,Hauth_2013} and a suitable adjustment of the pulse timing \cite{Roura_2014} have been proposed to overcome such loss of contrast, which become increasingly relevant for long interferometer times (e.g.\ the relative shift due to gravity gradients grows cubically with the interferometer time). However, the effectiveness of the mitigation strategy associated with gravity gradients (or the direct read-out of the phase shift from the location of the fringes in the fringe pattern of the density profile \cite{Sugarbaker_2013,Zeller_2015}) is eventually limited for sufficiently long interferometer times when considering thermal clouds \cite{Roura_2014}.

Minimizing the growth of the atom cloud at late times requires preparing an initial state with a very narrow momentum distribution, which will also have the added benefit of very high diffraction efficiencies and negligible dispersion (velocity-dependent) effects in the diffraction process. Promising techniques for achieving such narrow momentum distributions will be discussed in the next subsection.

\subsection{Bose-Einstein condensates and ``delta-kick cooling''}

There are several ways of generating the kind of narrow momentum distributions needed for long-time interferometry as discussed above. One possibility is to produce colder thermal ensembles (e.g.\ via evaporative cooling), but one should keep in mind that for sufficiently low temperatures the quantum degeneracy regime will eventually be reached (giving rise to a non-negligible condensate fraction for bosonic atomic species). The momentum spread can be further reduced by adiabatically opening the trap where the evaporatively cooled atoms are confined before release. However, achieving sufficiently narrow momentum distributions would require rather long times which would severely hamper the repetition rate and make this method unsuitable for high-precision experiments and for microgravity platforms with a limited time available per shot, such as the experiments in drop towers or parabolic flights.
There is fortunately an alternative method, known as \emph{``delta-kick cooling''} (DKC)\footnote{Strictly speaking one should not use the term ``cooling'' since the phase-space volume is preserved in the process. It would instead be more appropriate to speak of a magnetic (or optical-dipole) lens.}, which is capable of producing similar results within a substantially shorter time \cite{Chu_1986,Ammann_1997}. The basic idea is to release the atoms, let the cloud expand for some time and then switch on again the trapping potential for a short time. During this short period the kinetic energy of the atoms gets converted into potential energy, and by adjusting its duration appropriately, a significant fraction of the original kinetic energy can be removed (with the corresponding decrease of the momentum spread). A related method consists in letting the atom cloud expand in a shallower trap (compared to the originally confining one) and switch it off at the right time, when most kinetic energy has been converted into potential energy \cite{Cornell_1991,Kovachy_2014}.

DKC can be applied to both BECs and thermal clouds, but employing BECs has further advantages, as we will see. For nondegenerate thermal clouds, and assuming an isotropic harmonic trap for simplicity, the product of position and momentum widths, which is preserved by the time evolution according to Liouville's theorem, satisfies the inequality $2\, \sigma_p\, \sigma_x / \hbar \gtrsim N^\frac{1}{3}$. For an atom number $N \sim 10^6$, the right-hand side of the inequality is already of order $10^2$; moreover, for an ensemble far from degeneracy the product of the widths is actually much larger than $N^\frac{1}{3}$. This means that in order to achieve very narrow momentum widths one needs to have relatively large cloud sizes.
In contrast, for BECs one has $\sigma_p\, \sigma_x / \hbar \sim 1$. This means that even for a trapped condensate one can have a fairly narrow momentum distribution of order $\sigma_p \sim \hbar / R_\text{TF}$, and where the Thomas-Fermi radius $R_\text{TF}$  for a given trap frequency grows with the atom number. The use of DKC is still necessary in order to achieve very narrow momentum distributions because when the BEC is released from the trap, the nonlinear interaction energy is converted into kinetic energy and the expansion rate of the BEC increases significantly. Nevertheless, given the relation between the position and momentum widths in this case, a given momentum spread can be achieved with a much smaller size 
(by a factor $N^\frac{1}{3}$ or more) than for a thermal cloud.%
\footnote{For large atom numbers (e.g.~$N \sim 10^6$) the effect of nonlinear interactions can be non-negligible, even for condensate sizes of hundreds of micrometers, when momentum widths comparable to the Heisenberg limit are reached. In those cases, the final momentum widths attainable may be not so much better than those achieved by working with thermal clouds evaporatively cooled close to quantum degeneracy but with a small condensate fraction \cite{Kovachy_2014}.}
Hence, BECs are an ideal candidate for high-precision measurements with long-time interferometry where a combination of a smaller cloud size and a very low expansion rate can be achieved.
This helps to minimize the unwanted effects associated with wave-front distortions and reduces the loss of integrated contrast due to rotations and gravity gradients; moreover, the mitigation strategy put forward in Ref.~\cite{Roura_2014} is particularly effective for BECs (compared to thermal clouds).

BECs have recently been employed in precision measurements with atomic fountains \cite{Debs_2011,Altin_2013}, their use for atom interferometry in microgravity environments has been demonstrated in drop-tower experiments \cite{Zoest_2010,Muentinga_2013} and they are a key ingredient in plans for a dedicated space mission capable of performing tests of the equivalence principle that would improve the current bounds by several orders of magnitude \cite{Aguilera_2014,ste-quest_general}. Their combination with DKC has been implemented in drop-tower experiments \cite{Muentinga_2013}, where cloud sizes of just $1\, \text{mm}$ were achieved after $2\, \text{s}$ of expansion time, and in ground experiments \cite{Mcdonald_2013b}.

\subsection{Previous phase-shift calculations}

A path-integral approach was employed in Ref.~\cite{Storey_1994} to calculate the wave-function propagator and to obtain the phase shift for an atom interferometer in terms of the action evaluated along an appropriate classical trajectory for each branch. This reference mainly focused on the effects of rotations and uniform gravitational fields, but the same approach was later used to calculate the contribution of gravity gradients to the phase shift \cite{Wolf_1999,Peters_2001}. By considering an expansion in powers of time (up to a sufficiently high order) of the exact solution for the classical trajectories, the result was further extended to include higher-order contributions to the phase shift due to uniform gravitational fields, gravity gradients and rotations as well as cross terms \cite{Bongs_2006}.

A second approach analogous to the so-called ABCD formalism in optics was developed in refs.~\cite{Borde_1992,Borde_2001,Borde_2002} to obtain the propagator in position representation associated with a quadratic Hamiltonian. The propagator was then employed to calculate the evolution of a Gaussian wave packet corresponding to the dynamics of the atoms between laser pulses (and of a basis of Gauss-Hermite wave packets too). Finally, this was combined with the phases acquired from the interaction with the laser pulses to calculate the phase shift between the two interferometer branches governing the oscillations of the integrated particle number at each exit port.  Making use of these results, a general formula for the phase shift valid for arbitrary pulse sequences and including the effects of uniform gravitational fields, rotations, gravity gradients and weak gravitational waves was later derived \cite{Antoine_2003a,Antoine_2003}.
More recently, this question was revisited in Ref.~\cite{Borde_2008}, where the evolution of the massive matter particles was described in terms of massless particles propagating in a five-dimensional spacetime following an approach similar to Kaluza's theory \cite{Kaluza_1921}. The advantage of a formulation based on massless particles is that the phase of the associated waves (propagating in five-dimensional spacetime) is constant along the null rays determined by the classical trajectories, which are orthogonal to the spacetime hypersurfaces of constant phase, and this was exploited in refs.~\cite{Borde_2008,Borde_2013} to provide an elegant explanation for the nontrivial cancellation between different contributions to the phase shift found in previous calculations.

A related  procedure was carried out in Ref.~\cite{Hogan_2008}, where the evolution between laser pulses was described, for quadratic Hamiltonians, in terms of a (symmetric) centered wave packet with vanishing position and momentum expectation values together with position and momentum displacement operators acting on it and whose time-dependent arguments are given by classical trajectories, as well as a phase factor involving their associated classical action. By combining this result with the phases acquired from the interaction with the laser pulses and the additional phase shift that arises when the two wave packets interfering at each exit port have different central positions and momenta, a derivation was provided of the basic formula that had served as the starting point for the phase shift calculations in Ref.~\cite{Bongs_2006}.
Furthermore, by identifying the classical action with the proper time integrated along a worldline and working with Fermi normal coordinates to establish the connection with the nonrelativistic result, the basic formula for the phase shift was later extended to freely falling atoms in curved spacetimes \cite{Dimopoulos_2008} as long as the size of the wave packets remained small compared to the curvature radius. The resulting formalism was employed to analyze the possible application of atom interferometry to tests of general relativistic effects and the measurement of certain PPN parameters \cite{Dimopoulos_2008}. It was also a key ingredient in the calculation of a coordinate-invariant result for the differential phase shift of a pair of atom interferometers separated by a long baseline and sharing common lasers, and assessing its sensitivity to gravitational waves \cite{Dimopoulos_2008b}.

A third approach which has also proved to be very useful is based on working at the operator level. By accounting for the kicks from the laser pulses with momentum displacement operators, it was shown in Ref.~\cite{Audretsch_1994} that the action of the evolution operators between pulses can be translated into a linear transformation, in terms of position and momentum operators, of the exponent of those momentum displacement operators. This was then exploited to calculate the probabilities associated with each exit port for Ramsey-Bordé and Mach-Zehnder interferometers including the effects of constant acceleration and uniform forces, time-independent gravity gradients (up to first order) and rotations with constant angular velocity described in the rotating frame (up to quadratic order). These results were further extended to multiloop configurations with additional intermediate $\pi$ pulses in Ref.~\cite{Marzlin_1996}, where the possibility of canceling the effects of time-independent accelerations and the lowest-order contributions of rotations by adjusting the time separations between the intermediate pulses was analyzed.

More recently, there has been renewed interest in the use of operator methods for studying the phase shift in light-pulse atom interferometers. A representation-free derivation, based on operator algebra, of the phase shift for a Mach-Zehnder configuration in a uniform gravitational field was provided in Ref.~\cite{Schleich_2013a}. And the main results that will be obtained in the remaining sections of the present paper can be regarded as a generalization to quadratic potentials of such operator-algebra methods for deriving the detection probabilities at each exit port, together with a general treatment of rotations and non-inertial effects.
Furthermore, a representation-free description for the state evolution in interferometers with general quadratic Hamiltonians has been presented in Ref.~\cite{Roura_2014}; see also Ref.~\cite{Zeller_2015}. There the evolution of the interfering wave packets along each branch of the interferometer was described in terms of centered wave packets which characterized their expansion and shape evolution, as well as displacement operators which characterized their motion and whose arguments were given by classical phase-space trajectories including the kicks from the laser pulses. The main emphasis was on the key features of the fringe pattern arising in the  density profile of ``open interferometers'' (for which the trajectories associated with the different branches do not close in phase space after the last beam splitter), how this can lead to a loss of contrast in the oscillations of the integrated particle number at each exit port, and an efficient mitigation strategy to overcome such loss of contrast when due to gravity gradients.
In addition, a simple derivation of the general expression for the phase shift, based on the recursive use of the composition formula for the product of two displacement operators, was provided. It extends the general formula for the phase shift obtained by Antoine and Bordé \cite{Antoine_2003a,Antoine_2003} to the case of (possibly) branch-dependent forces.  The result is further generalized to anharmonic potentials, but locally harmonic (within regions of the size of the wave packets), in Ref.~\cite{Zeller_2015}. Furthermore, the formalism in Ref.~\cite{Roura_2014} has recently been exploited to propose a novel scheme \cite{Roura_2015} capable of simultaneously overcoming the two challenges associated with gravity gradients in tests of UFF: the loss of contrast and the initial co-location problem (the need to control very precisely the relative central position and momentum of the initial wave packets for the two atomic species).

There have also been interesting derivations based on the phase-space description of quantum mechanics in terms of Wigner functions. The time evolution of the Wigner function and the corresponding phase shift for interferometers with three and four laser pulses, including the effect of Earth's gravity gradient and rotation, was obtained in Ref.~\cite{Dubetsky_2006}. In some sense the method can be regarded as the phase-space counterpart of the representation-free description of the state evolution in Refs.~\cite{Roura_2014,Zeller_2015}. The Wigner function at the exit ports can be directly compared with the result for the Wigner function obtained in Ref.~\cite{Zeller_2015}, which is derived by expressing the state evolved with the representation-free approach in terms of the phase-space representation.
In turn, the derivation in Ref.~\cite{Giese_2014} can be regarded as the phase-space counterpart of the representation-free calculation of Ref.~\cite{Schleich_2013a}.
The results obtained in Refs.~\cite{Dubetsky_2006} apply to Raman-based interferometers, but they can be straightforwardly extended to Bragg-based ones.

Finally, it should be pointed out that all the analytic calculations described so far in this subsection made use of idealized laser pulses entirely described by instantaneous momentum displacement operators associated with planar laser wave fronts. Neither wave-front distortions, excitations of off-resonant diffraction orders, dispersion effects (including velocity selectivity) nor the evolution of the atom's COM during the finite duration of the laser pulse were taken into account. This might be a reasonable approximation under certain conditions (e.g.~small wave packets compared to the wave-front curvature, narrow momentum distributions and short laser pulses). However, in general these effects need to be taken into account for a sufficiently accurate description. The corrections due to the finite duration of the laser pulses were investigated in refs.~\cite{Cheinet_2008,Jansen_2008} and the effects of COM evolution during the laser pulse have been studied analytically including the effects of the gravitational acceleration \cite{Laemmerzahl_1995,Antoine_2006} and rotations \cite{Antoine_2007}. Dispersion effects can be investigated analytically for single Raman diffraction \cite{Moler_1992}. On the other hand, the effects due to excitations of off-resonant diffraction orders can be studied analytically in the \emph{quasi-Bragg} regime \cite{Mueller_2008b,Giese_2013} and dispersion effects can be approximately studied in the deep Bragg regime, but a numerical treatment (such as that of Ref.~\cite{Szigeti_2012}) is in general necessary. Algorithms making use of a suitable semi-analytical approach can be devised to generate codes that take these effects into account and can simulate full interferometer sequences with arbitrarily long interferometer times and no increase in computational costs.
As for the effects of wave-front distortions, the usual treatments are applicable to thermal clouds \cite{Fils_2005,Louchet_2011,Schkolnik_2014}, but an accurate description of their effects on large BECs requires new analytical tools currently under development.

%
%
%
%
%
%
%
%
\section{Matter-wave interferometry in quadratic potentials}
\label{sec_MWI_quadratic_potentials}
Matter-wave interferometers are composed of
a sequence of beam splitters which determines the specific interferometer geometry. With the help of the presented %
formalism we are able to efficiently describe arbitrary geometries.

We assume a two-level system coherently manipulated by beam splitters in the presence of 
an external quadratic potential. 
Note that our compact description of interferometry is also applicable to more complicated level structures, for instance necessary 
in double Bragg diffraction~\cite{Giese_2013}, by modifying the 
presented interaction model (beam splitter). 
In particular, internal level structures reducible to effective two-level systems, 
for instance Raman transitions in atom interferometers, are covered 
in an exact way. We illustrate our formalism without loss of generality by means of light-pulse atom interferometers.
\subsection{Total Hamiltonian}
\label{ssec_total_hamiltonian}

A two-level atom in a semiclassical electric field is a standard and often discussed problem in quantum optics~\cite{Exploring_the_Quantum_Haroche_2006, Atom-Photon_Interaction(Cohen-Tannoudji_1998), Scully_1997}.
Suppose we have given a two-level atom with the internal ground state $\ket{0}$ and the internal excited state $\ket{1}$ an electric dipole transition with frequency 
$\omega_\a$ will drive Rabi oscillations. 
The total Hamiltonian is given by
  \begin{align}\label{total_Hamiltonian}
     \hat{H}(t) &= \frac{\hat{\vec p}_\a^2}{2m} +  \hbar \omega_\a \frac{\hat{\sigma}_3}{2}  - ~ \vec{\hat{d}} \cdot \vec E(t,\hat{\vec x}) +  V_{\e}(t, \hat{\vec x}) 
  \end{align}
in which we have additionally included a time-dependent external potential $V_{\e}$ to the standard atom-field Hamiltonian of quantum optics. 
%
%
%
%
The atomic part consists of the kinetic energy of the atom (momentum $\vec p$; atomic mass $m$) and the energy term 
describing the internal level separation by the energy $\hbar \omega_\a$. 
The third term describes the linear coupling of the dipole operator and the semiclassical 
electric field (dipole approximation), where 
the dipole operator $\vec{\hat{d}}= \vec d \ \hat{\sigma}_+  + \vec d^* \hat{\sigma}_- $ includes the atomic excitation creation/annihilation operators $\hat{\sigma}_{+} = \ \ketbra{1}{0}$ and $\hat{\sigma}_{-} = \ \ketbra{0}{1}$,
and the dipole transition element $\vec d$.
%
%
\subsection{State description}
\label{ssec_state_description}
The (atomic) matter wave is described by a state with internal 
as well as external degrees of freedom
 \begin{align}
    \ket{\Psi} = \frac{1}{\sqrt{2}} \left(\ket{\psi^{(0)}} \ket{0}\, + \ket{\psi^{(1)}} \ket{1}  \right).
 \end{align}
In general, it is a superposition of atoms being in the internal ground state $\ket{0}$ and the 
internal excited state $\ket{1}$ with the corresponding external state $\ket{\psi^{(0)}}$ 
and $\ket{\psi^{(1)}}$, respectively. 
The internal states satisfy the orthonormality relation
 $ \braket{i}{j} = \delta_{ij}~(i,j \in \{0,1\})$
and form a complete set spanning up a two-dimensional Hilbert space.
Additionally, the state is assumed to be normalized: $ \braket{\Psi}{\Psi} = 1$.
%
%
\subsection{Interferometer sequence}
\label{ssec:Interferometer_sequence}
\begin{figure}[h]
    \centering
	\includegraphics[width=0.7\textwidth]{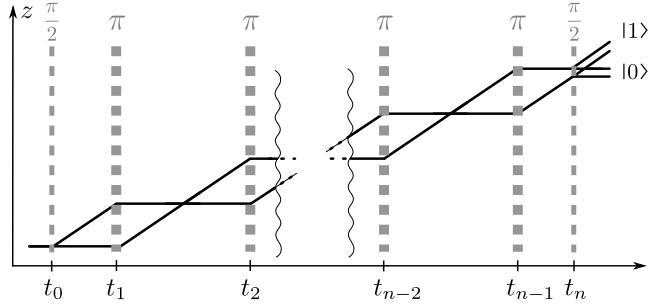}
    \caption[Multi-loop geometry]{Multi-loop geometry: In between two $\pi/2$-pulses (at time $t_0$ and $t_n$), which create a superposition of the internal ground state 
$\ket{0}$ and the internal excited state $\ket{1}$, ($n\!-\!1$) $\pi$-pulses redirect the atom. The atomic center-of-mass motion (black solid lines) is entangled with the 
internal states $\ket{0}$ and $\ket{1}$ via the laser pulses (dashed gray lines).}
    \label{fig:multi-loop geometry}
  \end{figure}

The simplest interferometer sequence, including all effects for a straightforward generalization to arbitrary interferometer
 geometries, is called the Mach-Zehnder pulse sequence (see \fig\ref{fig.MZI}). 
Pointing out the  quintessence, which includes the key elements of our formalism, this special geometry shall serve us 
in Section~\ref{sec:Compact_description:MZ_geometry} as a ``learning platform'' 
for a general theory describing more advanced geometries, for instance, a multi-loop geometry 
(see \fig\ref{fig:multi-loop geometry}). 

\begin{figure}[h]
    \centering
	\includegraphics[width=0.375\textwidth]{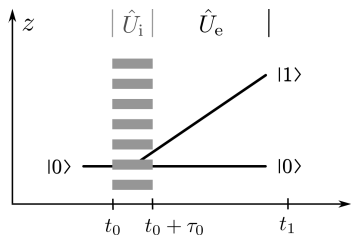}
    \caption[Different zones in an interferometer]{Detail of an interferometer sequence: The ``interaction zone'' 
(dashed gray region; operator $\hat{U}_\i$) of a laser field and an atom 
(black solid line = center-of-mass motion), initially in the internal ground state $\ket{0}$, causes a superposition 
of $\ket{0}$ and $\ket{1}$. 
The superposition weight is governed by the interaction time $\tau_0$. 
Afterwards, a ``free-propagation zone'' follows (external dynamics; operator $\hat{U}_\e$). ``Free-propagation'' is meant 
to include the external potential but no laser interaction is present.}
    \label{fig.interaction+free_evolution}
  \end{figure}

Below we split the interferometer pulse sequence in its elementary parts. In this sense, the interferometer can be described 
by individual zones which will correspond to unitary operators:
In an interferometer on every laser pulse (unitary time evolution $\hat{U}_{\i}$) a free time evolution $\hat{U}_{\e}$ 
follows (see \figs\ref{fig:multi-loop geometry} and \ref{fig.interaction+free_evolution}). Thus, for the final state we get the sequence  
  \begin{align}\label{separation_ansatz}
      \ket{\Psi(t_n\!+\!\tau_n)} &= \hat{U}_{\i}(t_n\!+\!\tau_n,t_n)~ ... ~ \hat{U}_{\e}(t_1,t_0\!+\!\tau_0) \, \hat{U}_{\i}(t_0\!+\!\tau_0,t_0)  \ket{\Psi_0}.
  \end{align}
This separation ansatz of $\hat{U}_\i$ (internal dynamics) and $\hat{U}_\e$ (external dynamics) is valid
for a sufficiently small atom-laser interaction time $\tau$ in order to neglect the effects of an 
external potential $V_{\e}(\vec x)$ during the internal dynamics. 
A detailed analysis of the beam-splitter process with a finite interaction time $\tau$ in the presence of various 
external potentials can be found in \cite{Antoine_2006}. For the limit of a quasi-instantaneous laser pulse we model 
the internal dynamics in Section~\ref{sec:Compact_description_of_interferometry} by a simple beam-splitter matrix while neglecting the 
external potential. 

Finally, an arbitrary interferometer can be composed by successively applying unitary operators for the internal as well as the 
external dynamics. 
Moreover, the interferometer phase shift emerges as a consequence of the commutation relations of such operators. \\

First, we discuss the time evolution of a matter wave within an external potential. Following this, the internal dynamics is studied.
%
%
%
\subsubsection{Free-propagation zone (external dynamics)}
\label{Free_propagation_zone}
The ``free-propagation zone'' is concerned with the time evolution within the external potential $V_{\e}$. 
Note that ``free-propagation'' is meant to include the effects of the external potential but no laser 
interaction is present. 
%
%
%
\paragraph{External Hamiltonian}
The time evolution between two laser pulses is governed by the time-dependent external Hamiltonian
  \begin{align}\label{Hamiltonian_ext_time-dependent}
      \hat{H}_{\e}(t) &= \frac{\hat{\vec p}_\a^2}{2m} + V_\e(t, \hat{\vec x}),
  \end{align}
where the first term corresponds to the kinetic energy of the matter wave, the second to the potential energy, which are both 
part of the total Hamiltonian~\eqref{total_Hamiltonian}. 
The external Hamiltonian \eqref{Hamiltonian_ext_time-dependent} only acts on the external degrees of freedom, i.e. on 
$\ket{\psi^{(0)}}$ and $\ket{\psi^{(1)}}$, respectively. 
%
%
%
\paragraph{Time evolution (external)}
The integrated version of the familiar Schrödinger equation maps the initial state $\ket{\Psi(t_0)}$ onto the final state $\ket{\Psi(t)}$ via the time-evolution 
operator $\hat{U}_\e(t,t_0)$ 
  \begin{align}\label{time_evolution}
     \ket{\Psi(t)} = \hat{U}_{\e}(t,t_0) \ket{\Psi(t_0)}.
  \end{align}

So far, we have dealt with the external dynamics. We consider next the internal one.
%
%
%
\subsubsection{Interaction zone (internal dynamics)}
\label{sec:Interaction_zone}
We have already considered the total Hamiltonian (\ref{total_Hamiltonian}) describing both the internal and the external dynamics of the matter wave.
In the following we will neglect the atomic center-of-mass motion in the semiclassical laser field which 
was already assumed for the separation ansatz~\eqref{separation_ansatz}. 
%
%
%
%
\setcounter{paragraph}{0}
\paragraph{Internal Hamiltonian}
Since the dynamics of the center-of-mass motion can be disregarded, the kinetic energy and the external 
potential in Eq.~\eqref{total_Hamiltonian} can be set to zero in order to get the internal Hamiltonian
  \begin{align}\label{internal_Hamiltonian}
      \hat{H}_{\i} (t) =  \hbar \omega_\a \frac{\hat{\sigma}_3}{2} -\vec{\hat{d}} \cdot \vec E(t,\hat{\vec x}).
  \end{align}
Moreover, $\hat{H}_{\i} (t)$ couples the internal states $\ket{0}$ and $\ket{1}$ via the dipole operator 
$\vec{\hat{d}} = \vec d \!\ketbra{1}{0} + ~ \vec d^* \! \ketbra{0}{1}$ 
to its center-of-mass motion ($\hat{\vec x} \equiv$~external degree of freedom). 
%
%
%
\paragraph{Time evolution (internal)}
The time evolution is given by \newline
$     \ket{\Psi(t)\!} = \hat{U}_{\i}(t,t_0) \ket{\Psi(t_0)}$,
where the unitary operator
\begin{align}\label{equ:time_evolution_int}
 \hat{U}_{\i}(t,t_0) = T ~ \e^{-\frac{\i}{\hbar} \int_{t_0}^t ~ \dif t' \hat{H}_\i(t') }
\end{align}
is governed by the time-dependent internal Hamiltonian~\eqref{internal_Hamiltonian}; $T$ is the time-ordering operator.
%
%
%
%
%
\subsection{Probability}
The probability of detecting the atom in the ground state $\ket{0}$ at position $\vec x$ after a sequence of laser pulses and 
free propagations reads%
  \begin{align}
    |\psi^{(0)}(t_n,\vec x)|^2 = |\bra{\vec x}\!\!\!\braket{0}{\Psi(t_n)}|^2.
  \end{align}
Here, the state $\ket{\Psi(t_n)}$ after the $n$-th laser pulse is given by the sequence (\ref{separation_ansatz}). In addition, we 
omit the pulse length $\tau_n$ in the notation. 
The ground-state detection probability irrespective of position becomes
  \begin{align}\label{equ_probabiltiy}
      P_{\,\indket{0}}(t_n)    &= \!\! \int\limits_{\mathbb{R}^{3}}  ~\dif^{3} x ~ |\psi^{(0)}(t_n,\vec x)|^2. 
  \end{align}
Note that any further free propagation $\hat{U}_\e(t_{n+1},t_n)$ within the external potential after the last beam splitter 
does not alter the probability. 
This can be easily checked by calculating the probability $P_{\,\indket{0}}(t_{n+1})$ with the 
state $ \ket{\Psi(t_{n+1})} = \hat{U}_\e(t_{n+1},t_n) \ket{\Psi(t_n)}$. 
In other words, the probability to find the atom in the internal ground state after a given pulse sequence 
is determined right after the last beam splitter at time $t_n$.
%
%
%
\subsection{External potential}
\label{ssec:External_potential}
%
In this section, we specify the external potential $V_\e(t,\vec x)$ and answer the question: What are the external potentials that 
matter-wave interferometers would naturally like to probe? 

On the one hand, clearly the most intuitive answer is: the gravitational potential. Indeed, gravity directly couples to 
the (gravitational) mass of matter waves. 
On the other hand, one can also think of other coupling potentials. For instance effective potentials due to the interaction between the (atomic) 
spin in a magnetic field or electric charges in electric fields. However, this would require additional intrinsic 
properties of matter waves.  
Thus, we restrict ourselves to the first mentioned and most natural one, the coupling 
of (gravitational) mass to the gravitational field. But keep in mind that the presented formalism holds true for various 
physical problems subject to the same mathematical model. \\

In Appendix~\ref{app:gravitational_potential}, we discuss the gravitational potential and 
expand it up to second order (harmonic approximation) 
	\begin{align}\label{harm_grav_potential}
	      V_\g(t,\vec x )  \approx	&~  V_\g(\vec\rho(t)) - m\, \vec g^\T(t)\left[\vec x - \vec\rho(t)\right] + \frac{1}{2} m \left[\vec x - \vec\rho(t)\right]^\T \Gamma(t)\left[\vec x -  \vec\rho(t)\right].
	\end{align}
This approximation is well-suited for high precision measurements at present as well as in future as long as 
the expanding point $\vec\rho(t)$ is chosen sufficiently close to the atomic center-of-mass motion. 
Depending on the chosen reference frame, $\vec g(t)$ can denote the local gravitational acceleration and/or accounts 
for arbitrary time-dependent inertial forces (e.\,g. due to vibrations). Moreover, $\Gamma(t)$ stands for the gravity gradient. 
%
%
\subsection{General quadratic Hamiltonians and their dynamical behavior}
\label{ssec:General_quadratic_Hamiltonians}
In order to provide a compact description of dynamics in matter-wave interferometers subject to a general quadratic Hamiltonian, 
we study in some detail the underlying symplectic structure in Appendix~\ref{app:math_excursus_Sympl_group}, 
especially the standard symplectic group. As a result we arrive at the well-known canonical description of quantum mechanics for a general quadratic 
Hamiltonian where the time-evolution matrix is symplectic.
%
%
%
\subsubsection{The general quadratic Hamiltonian}
\label{ssec:The_general_quadratic_Hamiltonian}
The most general form of a quadratic Hamiltonian is given by
   \begin{align}\label{quad_Hamilt}
      \hat{H} = H(\hat \xi) =  \mathcal{F}(t) + \vec{\mathcal{G}}^\T\!(t)\, \hat{\vec \xi} + \frac{1}{2}\, \hat{\vec \xi}^\T\, \mathcal{H}(t)\, \hat{\vec\xi}\,.
   \end{align}
Here, we have defined the phase-space vector operator $\hat{\vec \xi} = (\hat{\vec x}, \hat{\vec p})^\T$ 
which is a six-dimensional vector consisting of the three-dimensional position operator $\hat{\vec x}$ 
and the three-dimensional momentum operator $\hat{\vec p}$.
The time-dependent first- and second-order coefficients $\vec{\mathcal{G}}(t)$ 
and $ \mathcal{H}(t)$ are a six-dimensional vector and a $6 \times 6$ matrix, respectively. The scalar function $\mathcal{F}(t)$ only imprints a physically irrelevant 
energy offset to the Hamiltonian, for convenience, we set it to zero. 
%
%
%
\subsubsection{Heisenberg picture and the representation-free description}
%
So far, we were considering states and operators in the Schrödinger picture. Now, we go to the Heisenberg picture so that 
time evolution is accounted for by time-dependent operators. 
A representation-free description of matter-wave 
interferometry will circumvent the interpretation problem of the origin of the interferometer phase shift in 
representation-dependent approaches (such as in position or momentum representation)~\cite{Schleich_2013a}. 
In this way, we purely concentrate on the time evolution of operators and 
only use operator algebra methods. \\  

The phase-space vector operator $\hat{\vec\xi}$, which corresponds to the Schrödinger picture, reads in the Heisenberg picture
   \begin{align}
	  \hat{\vec\xi}_\H = \hat{\vec\xi}_\H(t) = \hat{U}^\dagger(t,t_0) \, \hat{\vec \xi} \, \hat{U}(t,t_0)\,.
   \end{align}
For the rest of the paper, we omit the argument ``$t$'' in the notation since time is implicitly included in the subscript 
``H'' for the Heisenberg picture.

Position and momentum operators fulfill the canonical commutation relations, which translate into the compact form
   \begin{align}\label{commutation_relation}
	  \left[ \hat{\vec\xi}_{\H,i} , \hat{\vec\xi}_{\H,k} \right] = \i \hbar \,\mathcal{J}_{ik}\,.
   \end{align}
Here, $\mathcal{J}_{ik}$ with $i,k \in \{1,2,3,4,5,6\}$ is the common symplectic form, see Eq.~\eqref{symplectic_form}. 
Since position and momentum do not commute, the canonical description of quantum mechanics naturally brings in 
the symplectic form $\mathcal{J} \in \mathbb{R}^{6 \otimes 6}$.  \\
%
%
\subsubsection{Heisenberg equation of motion}
%
The general quadratic Hamiltonian (\ref{quad_Hamilt}) in Heisenberg picture \newline
$	  \hat{H}_\H  	= \hat{U}^\dagger(t,t_0)\, H(\hat{\vec \xi})\, \hat{U}(t,t_0) 
			= H(\hat{\vec\xi}_\H)$
immediately yields the Heisenberg equation of motion 
\begin{subequations}\label{Heisenberg_equation}
   \begin{align}\label{Heisenberg _equ}
	  \frac{d\hat{\xi}_{\H,i}}{dt}	&= \frac{\i}{\hbar} \left[\hat{H}_\H, \hat{\xi}_{\H,i}\right] \nonumber \\ 
					&= \sum_k \mathcal{J}_{ik}\, \vec{\mathcal{G}}_k(t) + \sum_{k,l} \mathcal{J}_{ik}\, \mathcal{H}_{kl}(t)\,  \hat{\xi}_{\H,l}\,,
   \end{align}
where we made use of the symmetry of the second-order coefficient $\mathcal{H}_{kl} = \mathcal{H}_{lk}$.
Hence, the Heisenberg equation of motion is an inhomogeneous, linear differential equation
  \begin{align}\label{equ_of_motion}
	  \frac{d\hat{\vec\xi}_{\H}}{dt}	&= \mathcal{J}\, \mathcal{H}(t)\, \hat{\vec\xi}_\H + \mathcal{J}\, \vec{\mathcal{G}}(t)\,.
  \end{align}
\end{subequations}
The formal solution of the Heisenberg equation of motion will be determined next. 
%
\subsubsection{General solution}
%
The general solution of Eq.~\eqref{Heisenberg_equation} consists of a homogeneous solution and a particular solution. 
\setcounter{paragraph}{0}
\paragraph{Homogeneous solution}
%
The (formal) time-dependent homogeneous solution reads
  \begin{align}\label{hom_sol}
	  \hat{\vec\xi}_{\H}^{(h)}	&= \mathcal{T}(t,t_0)\, \hat{\vec\xi}\, ,
  \end{align}
where we have introduced the time-evolution matrix $\mathcal{T}(t,t_0) \in \mathbb{R}^{6 \otimes 6}$.
For general quadratic Hamiltonians with time-dependent second-order coefficient matrix $\mathcal{H}(t)$ only a numerical 
determination of $\mathcal{T}(t,t_0)$ is possible. Note that $\mathcal{T}$ satisfies the homogeneous equation of 
motion of Eq.~\eqref{Heisenberg_equation} and is a symplectic matrix (see Appendix~\ref{app:tau_is_sympl}). 
%
%
\paragraph{Particular solution}
%
We are left with the particular solution of the Heisenberg equation of motion
which follows from the method of variation of constants
  \begin{align}\label{particular_sol}
	  \hat{\vec\xi}_{\H}^{(p)}	&= \mathcal{T}(t,t_0) \!\! \int\limits_{t_0}^t ~\dif t' ~ \mathcal{T}^{-1}(t',t_0)\,  \mathcal{J}\,  \vec{\mathcal{G}}(t')\, .
  \end{align}
%
%
%
\paragraph{General solution}
%
The general (time-dependent) solution of the Heisenberg equation of motion 
is given by the sum of Eqs.~\eqref{hom_sol} and \eqref{particular_sol}
  \begin{align}\label{general_solution_of_Heisenberg}
	  \hat{\vec\xi}_{\H}	&= \mathcal{T}(t,t_0) \left[ \hat{\vec\xi} + \!\! \int\limits_{t_0}^t ~\dif t' ~ \mathcal{T}^{-1}(t',t_0)\,  \mathcal{J}\,  \vec{\mathcal{G}}(t')\right].
  \end{align}
In conclusion, we have studied (in a rather formal way) the main ingredients on which any interferometer is based. In particular, 
we have discussed the dynamics in general quadratic potentials. \\

%
%
%
%
\subsubsection{Perturbative approach for time-dependent Hamiltonians}
\label{sssec:perturbative_approach}
In this section, we provide a recursive formula for the time-evolution matrix $\mathcal{T}(t,t_0)$. 
We deal with the time-dependent second-order coefficient 
 by decomposing  $\mathcal{H}(t)=\mathcal{H}_0+\lambda \mathcal{H_\I}(t)$ in an unperturbed part $\mathcal{H}_0$
and a small perturbation $\lambda\mathcal{H}_\I(t)$ which takes into account time-dependent (gravitational, magnetic, etc.) 
gradients in general. \\

First of all, the time-evolution matrix $\mathcal{T}$ defined via 
$\hat{\vec\xi}^{(h)}_\H = \mathcal{T} (t) \, \hat{\vec\xi}$ has to fulfill the homogeneous part of the equation of motion~\eqref{equ_of_motion}. 
Thus, the homogeneous equation of motion reads
  \begin{align}\label{tau_hom}
	  \frac{d\, \mathcal{T}(t)}{dt}	&= \mathcal{J} \left[\mathcal{H}_0 + \lambda \mathcal{H}_\I(t)\right]\, \mathcal{T}(t) \,.
  \end{align}
To solve this differential equation, we write the time-evolution matrix as a power series
  \begin{align}\label{series_tau}
	  \mathcal{T}	&= \mathcal{T}^{(0)}\! + \lambda \mathcal{T}^{(1)}\! + \lambda^2 \mathcal{T}^{(2)} + \ldots \,
  \end{align}
where $\lambda$ denotes a small perturbation parameter. Substituting $\mathcal{T}$ in Eq.~\eqref{tau_hom} yields
  \begin{align}
	  \frac{d}{dt} \left(\mathcal{T}^{(0)}\! + \lambda \mathcal{T}^{(1)}\! +  \ldots\right)	
		      &= \mathcal{J} \left[\mathcal{H}_0 + \lambda \mathcal{H}_\I(t)\right] \left(\mathcal{T}^{(0)}\! + \lambda \mathcal{T}^{(1)}\! +  \ldots\right)\,.
  \end{align}
The unperturbed time evolution, which corresponds to $\lambda=0$, is given by 
$\mathcal{T}^{(0)}(t,t_0) = \text{exp}\{\mathcal{J} \mathcal{H}_0 (t-t_0)\}$. Hence, 
the linear independence of orders in $\lambda$ induces the following recursive formula 
(for $\lambda,\lambda^2, ...$ with $\lambda \neq 0$)
  \begin{align}
	  \frac{d}{dt} \mathcal{T}^{(n)}(t,t_0) - \mathcal{J} \mathcal{H}_0 \mathcal{T}^{(n)}(t,t_0) &= \mathcal{J} \mathcal{H}_\I(t) \mathcal{T}^{(n-1)}(t,t_0)\,,
  \end{align}
which reads in terms of an integrated recursive formula for $n>0$
  \begin{align}\label{recursive_int}
	  \mathcal{T}^{(n)}(t,t_0)	&=  \!\! \int\limits_{t_0}^t ~\dif t' ~ \mathcal{T}^{(0)}(t,t')\,  \mathcal{J} \mathcal{H}_\I(t') \, \mathcal{T}^{(n-1)}(t',t_0)\,.
  \end{align}\\
The solution of the previous equation enables us to deal with time-dependent Hamiltonians in a perturbative way. Indeed, the recursive formula~\eqref{recursive_int} takes 
into account quadratic perturbations $\mathcal{H}_\I(t)$ to the unperturbed Hamiltonian. Moreover, with the time-evolution matrix 
$\mathcal{T}$ at hand, we can also include arbitrary time-dependent local accelerations as well as inertial forces (e.\,g. due to 
vibrations) via the first-order coefficient $\vec{\mathcal{G}}(t)$ by means of Eq.~\eqref{particular_sol}. 
Following these lines, we arrive at a perturbative solution for Eq.~\eqref{general_solution_of_Heisenberg}. \\

As an example of a time-dependent Hamiltonian, we will see in Section~\ref{Non-inertial_frame} that 
the description of interferometry in non-inertial frames requires time-dependent gradients and therefore the recursive 
formula, Eq.~\eqref{recursive_int}.

Needless to say, the perturbative approach can also be applied for constant Hamiltonians. For instance, the time-evolution matrix 
$\mathcal{T}$ for the case of constant gradients can be alternatively calculated in such a way. 
The perturbative treatment as well as the exact analytical solution are presented in Section~\ref{Time-dependent_phases_+_displ}.\\

Next, we put the previous results in concrete terms and outline our general approach modeling matter-wave interferometry. 
We are especially interested in a compact description combining both the ``free-propagation zone'' 
and the ``interaction zone''.
%
%
\section{Generalized beam splitter}
\label{sec:Compact_description_of_interferometry}
The purpose of the current section is to provide a more concrete but still versatile description of matter-wave interferometry. 
Therefore, the atom-laser interaction is modeled in a semiclassical way and yields a simple beam-splitter matrix. 
It turns out that the combination of the (atom-laser) ``interaction zone'' and the ``free-propagation zone'' can be described by 
a generalized (time-dependent) beam-splitter matrix. 
%
%
\subsection{Generalized beam-splitter matrix}
\label{sec:Beam_splitter}
In Section~\ref{sec_MWI_quadratic_potentials}, we have already discussed the internal and external dynamics. 
We have split their contribution to the total evolution into an atom-field interaction (``interaction zone'') and a time 
evolution in an external potential (``free-propagation zone''). Now, we are interested in finding an explicit expression 
for the combination of both.
\setcounter{paragraph}{0}
%
\subsubsection{Atom-laser interaction}
\label{Atom-laser_interaction}
We consider a two-level system in the semiclassical electric field of two counter-propagating lasers where the internal dynamics 
is governed by the Hamiltonian~\eqref{internal_Hamiltonian}. Since we can neglect the (atomic) center-of-mass motion 
and assume a constant electric amplitude during the atom-laser interaction, the Hamiltonian becomes 
time-independent. Hence, the time-evolution operator~\eqref{equ:time_evolution_int} becomes the standard quantum optic 
operator
    $\hat{U}_\i (\Theta, \vec n) = \text{exp}\{-\i \,\Theta  \, \hat{\vec n}\, \hat{\vec \sigma}/2 \}$ with
the Bloch vector 
       $ \hat{\vec n} = \left(   \cos [\vec k \hat{\vec x} + \varphi],
							       \sin [\vec k \hat{\vec x} + \varphi],
							      0 
						      \right) $, 
the Pauli vector $\hat{\vec\sigma} = (\hat{\sigma}_1,\hat{\sigma}_2,\hat{\sigma}_3)^\T$ and
the pulse area $\Theta$ which depends on the individual pulse lengths $\tau_n$. For pedagogical reasons we neglect detuning. 
However, it can be easily included by taking the corresponding Bloch vector and an effective pulse area 
\cite{Exploring_the_Quantum_Haroche_2006, Atom-Photon_Interaction(Cohen-Tannoudji_1998), Scully_1997}. \\

As a result, the unitary time evolution $\hat{U}_\i$ can be written as the following matrix (basis $\{\ket{0}, \ket{1}\}$)
    \begin{align}\label{beam_splitter_matrix}
	\hat{S}^{(\Theta)}_n 	=& \begin{pmatrix} 
					\cos \frac{\Theta}{2} & -\i \sin \frac{\Theta}{2} ~ \e^{-\i[\vec k_n \hat{\vec x} + \varphi_n]}  \\[0.05cm]
					-\i \sin \frac{\Theta}{2} ~ \e^{+\i[\vec k_n \hat{\vec x} + \varphi_n]} & \cos \frac{\Theta}{2}
				   \end{pmatrix}\, ,
    \end{align}
where the subscript $n$ stands for the $n$-th ``interaction zone''; $\vec k_n$ and $\varphi_n$ corresponds to the wave vector and the phase of
the $n$-th laser pulse, respectively.  
Moreover, the pulse area $\Theta$ can be arbitrarily chosen. Note that we have changed the notation from $\hat{U}_\i$ to 
$\hat S^{(\Theta)}$ to make clear that the time scale on which $\hat{U}_\i $ acts is small compared 
to the external dynamics $\hat{U}_\e$. Thus, the internal dynamics is a quasi-instantaneous process solely described by 
the time-independent beam-splitter matrix $\hat S^{(\Theta)}_n$, Eq.~\eqref{beam_splitter_matrix}. \\

For a straightforward generalization of the beam-splitter matrix it is useful to 
introduce the displacement operator (see also Appendix~\ref{app:displacement_operator})
\begin{align}\label{displ_op}
   \hat{D}(\vec\chi) = \hat{D}(\vec\chi_x,\vec\chi_p) = \e^{\frac{\i}{\hbar}[\vec\chi_p \hat{\vec x} - \vec\chi_x \hat{\vec p}]},
\end{align}
which accounts for a displacement in phase space by the displacement vector $\vec\chi = (\vec\chi_x,\vec\chi_p)^\T$. 

Hence, we get for the beam-splitter matrix~\eqref{beam_splitter_matrix} of the $n$-th ``interaction zone''
    \begin{align}\label{beam_splitter_matrix_displ_0}
	\hat{S}^{(\Theta)}_n 	=& \begin{pmatrix} 
					\cos \frac{\Theta}{2} & -\i \sin \frac{\Theta}{2} ~ \e^{-\i\varphi_n}\hat{D}(-\bar{\vec\chi}_n)   \\[0.05cm]
					-\i \sin \frac{\Theta}{2} ~ \e^{+\i\varphi_n} \hat{D}(\bar{\vec\chi}_n) & \cos \frac{\Theta}{2}
				   \end{pmatrix}\, ,
    \end{align}
where $\bar{\vec\chi}_n=(\vec 0, \hbar \vec k_n)^\T$ is the phase-space displacement vector corresponding to the photon recoil 
$\hbar \vec k_n$ of the $n$-th laser pulse.
%
%
%
%
%
%
%
\subsubsection{Heisenberg picture}
The aim of this subsection is to transform the beam-splitter matrix into Heisenberg picture in order to deal with the 
``free-propagation zone'' in a convenient way. \\
%
%
%
\paragraph{The beam-splitter matrix becomes time-dependent}
We take into account the external dynamics of the atom in the gravitational field by means of the free-evolution operator 
$\hat{U}_\e$ introduced in Eq.~\eqref{time_evolution}. Thus, the beam-splitter matrix~\eqref{beam_splitter_matrix_displ_0} 
in Heisenberg picture formally reads 
    \begin{align}\label{bsmh}
	\hat{S}_{\H,n}^{\Theta} &=  \hat{U}_\e^\dagger(t_n,t_0) ~ \hat{S}^{\Theta}_n ~ \hat{U}_\e(t_n,t_0).
    \end{align}
We immediately see from Eq.~\eqref{beam_splitter_matrix_displ_0} that the displacement operator $\hat D$ will become time-dependent 
due to the transformation~\eqref{bsmh}. For a detailed discussion of the displacement operator in Heisenberg picture we refer to 
Appendix~\ref{app:displacement_operator}. Next, we only present the main ideas in order to derive a 
compact expression for the beam-splitter matrix in Heisenberg picture (generalized beam splitter). \\

We already know that the external evolution $\hat{U}_\e(t_n,t_0)$ can be split into: (i) the time evolution $\mathcal{T}(t_n,t_0)$, 
which corresponds to the homogeneous part of the Heisenberg equation of motion~\eqref{Heisenberg_equation}, and 
(ii) the time evolution which accounts for the inhomogeneity, Eq.~\eqref{particular_sol}. It is convenient to introduce the (time-dependent) 
generalized phase 
    \begin{align}\label{phi_n}
	    \Phi_n	&=	  \varphi_n + \frac{1}{\hbar} \! \int\limits_{t_0}^{t_n}\, \dif t' \, \vec\chi^\T_{n}(t',t_n)\, \vec{\mathcal{G}}(t')\, ,
    \end{align}
which includes the latter time evolution and can be interpreted as a generalization of the 
constant laser phase $\varphi_n$ in Eq.~\eqref{beam_splitter_matrix_displ_0}. 

In the same sense, the displacement operator $\hat{D}(\bar{\vec\chi}_n)$ in Eq.~\eqref{beam_splitter_matrix_displ_0} becomes time-dependent  
via the transformation~\eqref{bsmh}. Indeed, when we write the displacement operator in the following way
  \begin{align}\label{displ_op_2}
      \hat{D}(\vec\chi_n) = \e^{-\frac{\i}{\hbar}  \vec\chi_n^\T  \mathcal{J} \hat{\vec \xi}} \, ,
  \end{align}
we see that the phase-space displacement vector becomes time-dependent
  \begin{align}\label{chi_n(t)}
    \vec\chi_n &=  \mathcal{T}(t_0,t_n) \,\bar{\vec\chi}_n
  \end{align}
and accounts for the time-evolution of the homogeneous solution~\eqref{hom_sol}. 
Here, we recall the symplectic form $\mathcal{J}$, Eq.~\eqref{symplectic_form}, and 
$\bar{\vec\chi}_n = ( \vec 0 , \hbar \vec k_n )^\T $ which corresponds to the momentum kick $\hbar \vec k_n$ 
of the photon absorbed by the atom at time $t_n$. Note that the arguments of $\mathcal{T}(t_0,t_n)$ 
are exchanged with respect to the time evolution of $\hat{\vec \xi}_\H^{(h)} = \mathcal{T}(t_n,t_0)\, \hat{\vec \xi}$
because we have shifted the time-dependence from $\hat{\vec \xi}_\H^{(h)}$ to $\bar{\vec\chi}_{n}$ via 
Eq.~\eqref{change_relation}. \\

Finally, the beam-splitter matrix~\eqref{beam_splitter_matrix_displ_0} of the $n$-th ``interaction zone'' reads 
after the transformation~\eqref{bsmh}
    \begin{align}\label{BSM_H_2}
	    \hat{S}^{(\Theta)}_{\H,n} 	&= \begin{pmatrix} 
					      \cos \frac{\Theta}{2} &\!\!\! -\i \sin \frac{\Theta}{2} \, \e^{-\i\Phi_n} \, \hat{D}(-\vec\chi_n) \\
					      -\i \sin \frac{\Theta}{2}  \, \e^{+\i\Phi_n} \, \hat{D}(\vec\chi_n) &\!\!\! \cos \frac{\Theta}{2} 
					   \end{pmatrix}\, .
    \end{align}

Additionally, we can include state-dependent external potentials as well as state-dependent laser interactions when 
we introduce different displacement vectors $\vec\chi_n^{\pm}$ and phases $\Phi_n^\pm$ for each transition element \cite{Roura_2014}. 
On the one hand, a state-dependent external potential modifies the time-evolution of $\vec\chi_n$ and $\Phi_n$ for 
the internal states differently. For instance, this can be used to implement (external) anharmonic potentials \cite{Zeller_2015} although our 
approach is solely based on quadratic Hamiltonians (locally quadratic for each interferometer branch). On the other hand, 
state-dependent laser interactions can be implemented by state-dependent momentum kicks, where $\hbar \vec k_n^{+}$ describes 
the momentum kick corresponding to the transition from the internal ground state $\ket{0}$ to the internal excited state 
$\ket{1}$ and $\hbar\vec k_n^{-}$ the vice versa process. Finally, we call such a beam splitter an asymmetric beam splitter 
and take into account all these effects by the general beam-splitter matrix~\eqref{BSM_H_2} and the substitution: 
$\hat{D}(\pm\vec\chi_n) \rightarrow \hat{D}(\pm\vec\chi_n^{\pm})$ and $\pm\Phi_n \rightarrow \pm\Phi_n^\pm$. \\

In summary, we have modeled the atom-laser interaction by the beam-splitter matrix (\ref{beam_splitter_matrix_displ_0}). 
Going to the Heisenberg picture enables the combination of internal as well as external dynamics. We have 
introduced the time-dependent displacement operator~\eqref{displ_op_2} and the generalized phase~\eqref{phi_n} in order to include 
the effects of an external potential. 
The final beam splitter in Heisenberg picture, Eq.~\eqref{BSM_H_2}, serves as a building block for our 
interferometer description and enables a compact description of both the atom-field interaction and the external dynamics governed by a 
general quadratic Hamiltonian. We call Eq.~\eqref{BSM_H_2} a generalized beam splitter to emphasize 
that the beam splitter accounts for the common mixing of the internal states but also for a phase shifter 
due to the external dynamics.  \\
%
%
%
%
\paragraph{Beam splitter and mirror}
\label{Beam_splitter_and_mirrors}
%
%
%
%
By setting the pulse area $\Theta = \pi/2$, the matrix \eqref{BSM_H_2} becomes a fifty-fifty beam splitter creating an 
equally weighted superposition of the internal states
$\ket{0} = (1,0)^\T$ and $\ket{1} = (0,1)^\T$
  \begin{subequations}\label{S_H_pi/2_ground}
    \begin{align}
	\hat{S}^{(\frac{\pi}{2})}_{\H,n} \ket{0}	&= \frac{\ket{0} - \i\, \e^{+\i\Phi_n} \hat{D}( \vec\chi_n)  \ket{1} }{\sqrt{2}}\ , \\
	\hat{S}^{(\frac{\pi}{2})}_{\H,n} \ket{1}	&= \frac{\ket{1} - \i\, \e^{-\i\Phi_n} \hat{D}(-\vec\chi_n)  \ket{0} }{\sqrt{2}}\, .
    \end{align}
  \end{subequations}
Hence, the internal states accumulate the time-dependent relative phase  $-\i\,\e^{\pm\i\Phi_n}$. Additionally, 
the $\pi/2$-pulse leads to a displacement $\hat{D}(\pm\vec\chi_n)$ inducing the 
atomic center-of-mass motion by the time-dependent displacement vector $\vec\chi_n$. \\

A pulse area $\Theta = \pi$ inverts the populations of the internal states (mirror) while imprinting an additional phase and 
displacement
  \begin{subequations} \label{S_H_pi_ground_excited}
    \begin{align}
	\hat{S}^{(\pi)}_{\H,n} \ket{0}	&= -\i \e^{+\i\Phi_n} \hat{D}( \vec\chi_n)  \ket{1}\, ,  \\
	\hat{S}^{(\pi)}_{\H,n} \ket{1}	&= -\i \e^{-\i\Phi_n} \hat{D}(-\vec\chi_n)  \ket{0}\, .
    \end{align}
  \end{subequations}

These two pulses \eqref{S_H_pi/2_ground} and \eqref{S_H_pi_ground_excited}, but especially the $\pi/2$-pulse, 
are necessary for implementing interferometry. Therefore, it was important to understand 
their effect on internal states leading to additional phases and displacements. Moreover, we will see that these 
time-dependent phases and displacements generate the total interferometer phase shift and determine the visibility.
%
%
%
%
%
\subsection{Generalized phases and displacements for constant coefficients}
\label{Time-dependent_phases_+_displ}
The purpose of the present section is to provide explicit expressions for $\Phi_n$ and $\vec\chi_n$ in the 
presence of a time-independent, quadratic potential of the form (\ref{harm_grav_potential}). With respect to the general quadratic 
Hamiltonian (\ref{quad_Hamilt}) this implies the following second-order coefficient matrix
    \begin{align}\label{second_coeff_gravity}
	  \mathcal{H}^{(\Gamma)} =  \left(\begin{matrix}  m \Gamma & 0 \\  0 & \frac{1}{m} \mathds{1}_3 \end{matrix}\right) = const. \in \mathds{R}^{6 \otimes 6}.
    \end{align}
We use the superscript ``$\,\Gamma\,$'' to distinguish it from the most general coefficient matrix $\mathcal{H}(t)$. \\

The local acceleration naturally comes in via the first-order coefficient of Eq.~\eqref{quad_Hamilt}
    \begin{align}\label{first_coeff_gravity}
	  \vec{\mathcal{G}} = \left(\begin{matrix} - m \vec g \\ \vec 0 \end{matrix}\right) = const. \in \mathds{R}^6\, .
    \end{align}
For an exact analytical treatment we assume for a moment a time-independent acceleration $\vec g$ and a constant gradient~$\Gamma$. \\
%
%
%
%
%
%
%
\paragraph{Time-evolution matrix}
%
The homogeneous solution of the Heisenberg equation of motion (\ref{Heisenberg_equation})
is simply given by the phase-space operator $\hat{\vec\xi}^{(h)}_\H = \mathcal{T}(t,t_0)\, \hat{\vec\xi}^{~}$,
where the (symplectic) time-evolution matrix reads
  \begin{align}\label{sympl_matrix_gravity}
	 \mathcal{T}(t,t_0) = \left( \begin{matrix}  \cos{(\sqrt{\Gamma}[t-t_0])} & \frac{\sin(\sqrt{\Gamma}[t-t_0])}{m \sqrt{\Gamma}} \\ 
								-\sqrt{\Gamma} m ~\sin(\sqrt{\Gamma}[t-t_0]) &  \cos{(\sqrt{\Gamma}[t-t_0])} \end{matrix}  \right).
  \end{align}
This matrix is just the time evolution in phase space for the standard harmonic oscillator with frequency $\sqrt{\Gamma}$.
Note that $\Gamma$ is a matrix. So, if cosine and sine in Eq.~\eqref{sympl_matrix_gravity} are defined by means of the spectral decomposition of $\Gamma$, 
the matrix character of $\Gamma$ is unproblematic. In particular, if one of the eigenvalues of $\Gamma$ is negative, sine and cosine 
become the corresponding hyperbolic function. 
%
%
\paragraph{Perturbative treatment}
%
Despite the fact that a constant gradient does not require the perturbative treatment introduced in Section~\ref{sssec:perturbative_approach}, 
surely it can be done. Thus, we show how to get the previous result, Eq.~\eqref{sympl_matrix_gravity}, 
in order to come familiar with our perturbative approach. 

The time evolution for a vanishing gradient ($\Gamma=0$) is given by the
time-evolution matrix
  \begin{align}\label{tau_0}
	 \mathcal{T}^{(0)}(t,t_0) &=  \begin{pmatrix}  \mathds{1}_3 & \frac{t-t_0}{m} \\ 
								0 &  \mathds{1}_3      
				     \end{pmatrix}.
  \end{align}
The perturbation parameter is supposed to be the gravity gradient $\Gamma$, so that the second-order coefficient~\eqref{second_coeff_gravity} can be decomposed in the 
following way
\begin{align}\label{H_gamma_perturbative}
    \mathcal{H}^{(\Gamma)} 	&=  \mathcal{H}_0 + \Gamma\, \mathcal{H_\I}
			 =  \begin{pmatrix}  0 & 0 \\ 
					     0 &  \frac{1}{m}\mathds{1}_3
			    \end{pmatrix}
			 +  \Gamma\, \begin{pmatrix} m & 0 \\
						   0 & 0
			           \end{pmatrix}
			 = const.
\end{align}
Finally, the recursive formula~\eqref{recursive_int} yields the time-evolution matrix up to any order in $\Gamma$
  \begin{align} \label{tau_perturbative_gamma_const}
	 \mathcal{T}(t,t_0)   &= \mathcal{T}^{(0)}(t,t_0) + \Gamma\, \mathcal{T}^{(1)}(t,t_0) + \mathcal{O}[\Gamma^2]  \nonumber \\
 			      &=  \begin{pmatrix}  \mathds{1}_3 & \frac{t-t_0}{m} \\ 
								0 &  \mathds{1}_3      
				     \end{pmatrix}
			      +  \Gamma \begin{pmatrix} -\frac{1}{2}(t-t_0)^2  &  -\frac{1}{6m}(t-t_0)^3 \\[0.1cm]
						          -m(t-t_0)             & -\frac{1}{2}(t-t_0)^2
			           \end{pmatrix}
			      + \mathcal{O}[\Gamma^2]\,.
  \end{align}
%
%
%
%
%
%
%
\subsubsection{Displacement vector for a single laser pulse}
\begin{figure}[h]
    \centering
	\includegraphics[width=0.65\textwidth]{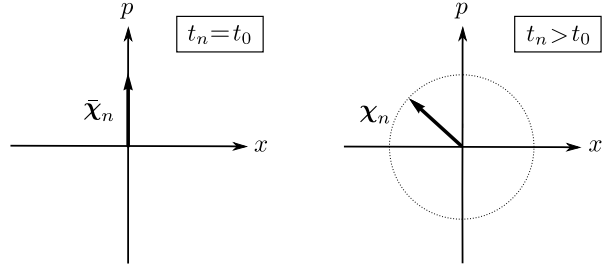}
    \caption[Displacement vector]{The displacement vector $\vec\chi_n =\mathcal{T}(t_0,t_n)\,  \bar{\vec\chi}_{n}$ (plotted for $m\sqrt{\Gamma}=1$) is rotating 
in counter-clockwise direction in phase space (trajectory of a harmonic oscillator) 
in the presence of the gradient $\Gamma$. The displacement vector describes the time-dependent displacement in position $x$ and momentum $p$ of the internal excited state 
relative to the ground state.}
    \label{fig:chi_MZ_T_variabel}
  \end{figure}

The time evolution of the displacement vector is given by the symplectic matrix (\ref{sympl_matrix_gravity})
\begin{align}\label{chi_n_time-dependent}
  \vec \chi_n &= \mathcal{T}(t_0,t_n)\,  \bar{\vec\chi}_{n}\, .
\end{align}
Note that $\mathcal{T}(t_0,t_n)$ is propagating the 
displacement vector $\bar{\vec\chi}_n = (\vec 0, \hbar \vec k_n)^\T$ backwards in time to the initial time~$t_0$. 
Finally, we get the expression
  \begin{align}\label{chi_n_gradient}
	 \vec \chi_n  &= \left( \begin{matrix}  - \frac{\sin(\sqrt{\Gamma}[t_n-t_0])}{m \sqrt{\Gamma}} ~ \hbar \vec k_n \\  \cos{(\sqrt{\Gamma}[t_n-t_0])} ~ \hbar \vec k_n \end{matrix}  \right).
  \end{align}
Thus, the spatial part of the displacement vector does not vanish any longer. The time evolution in the presence of the gradient 
$\Gamma$ rotates the displacement vector in phase space (see \fig\ref{fig:chi_MZ_T_variabel}). In other words, 
the time-dependent displacement vector (\ref{chi_n_gradient}) follows the (classical) trajectory of the atomic center-of-mass motion. 
%
%
%
%
%
%
%
\subsubsection{Generalized phase for a single laser pulse}
Since the time-dependent phase (\ref{phi_n}) depends on the displacement vector calculated above, we can now determine $\Phi_n$. 
Employing the displacement vector (\ref{chi_n_gradient}) and the first-order coefficient (\ref{first_coeff_gravity}), 
the generalized phase for a single laser pulse reads after integration
    \begin{align}\label{phi_n_4}
 	    \Phi_n        	&= \varphi_n - \vec g^\T \left[ \frac{ \cos{(\sqrt{\Gamma}[t_n-t_0]}) - 1 }{\Gamma} \right]  \vec k_n\, .
    \end{align}
It linearly depends on the local acceleration $\vec g$ and 
the wave vector $\vec k_n$ in the presence of an external field (in harmonic approximation). 
Moreover, the gradient $\Gamma$ modulates the phase in a non-linear way. \\
%
%
%
%
\paragraph{Vanishing gradient}
Assuming $\Gamma = 0$, we arrive at
    \begin{align}\label{phi_n_Gamma=0}
 	    \Phi_n	&= \varphi_n + \frac{1}{2}\, \vec k_n^\T\, \vec g\,  [t_n-t_0]^2.
    \end{align}
Hence, the laser phase as well as the scalar product of the local acceleration and the wave vector appear in a linear combination. 
The second term vanishes if the lasers stand perpendicular to $\vec g$. 
In other words, the interferometer phase is sensitive to the projection of the wave vector $\vec k_n$ on the local acceleration 
$\vec g$.   
%
%
\subsection{Perturbative treatment for time-dependent coefficients}
%
In the previous section, we assumed time-independent first- and second-order coefficients $\vec{\mathcal{G}}$ and $\mathcal{H}$, 
respectively. In particular, we have shown how to deal with 
a constant gradient in a perturbative way; see Eqs.~\eqref{tau_0}-\eqref{tau_perturbative_gamma_const}. Of course, 
our perturbative approach is valid for time-dependent coefficients 
in general. At this point, we want to recall that the general scheme of handling time-dependent Hamiltonians in a 
perturbative way is presented in Section~\ref{sssec:perturbative_approach}. As a result, the time-evolution matrix is given 
as a series expansion, Eq.~\eqref{series_tau}, determined by the recursive formula~\eqref{recursive_int}. The time-evolution matrix $\mathcal{T}$ at hand, 
it is straightforward to calculate all relevant quantities, for instance the displacement 
vector~\eqref{chi_n(t)} or the generalized phase~\eqref{phi_n}. 
To come familiar with the perturbative treatment, we apply our approach to different 
scenarios: (i) a constant gravity gradient as a simple example for interferometry in inertial reference frames 
(see the previous section and especially Eqs.~\eqref{tau_0}-\eqref{tau_perturbative_gamma_const}), (ii) a general 
time-dependent gradient and (iii) 
the case of an orbiting observer (non-inertial frame), see Section~\ref{Non-inertial_frame}. 
%
\paragraph{Generalized phase}
%
For time-dependent (gravitational, magnetic, ...) gradients the second-order coefficient can be written as
\begin{align}\label{H_gamma(t)_perturbative}
    \mathcal{H}(t) 	&=  \mathcal{H}_0 + \mathcal{H_\I}(t)
			 =  \begin{pmatrix}  0 & 0 \\ 
					     0 &  \frac{1}{m}\mathds{1}_3
			    \end{pmatrix}
			 +  \begin{pmatrix} m \Gamma(t) & 0 \\
						   0 & 0
			     \end{pmatrix}.
\end{align}
Here, the unperturbed time-evolution matrix $\mathcal{T}^{(0)}$, Eq.~\eqref{tau_0}, corresponds to $\mathcal{H}_0$. 
In addition, the recursive formula~\eqref{recursive_int} yields correction terms due to (small) effects of 
the gradient $\Gamma(t)$. Hence, we arrive at the 
generalized phase (up to first order in $\Gamma$)
    \begin{align}\label{phi_n_gamma(t)}
	  \Phi_n &=  \varphi_n +  \frac{1}{\hbar} \! \int\limits_{t_0}^{t_n}\, \dif t' \, \left[\mathcal{T}^{(0)}(t',t_n)\,\bar{\vec\chi}_{n} + \mathcal{T}^{(1)}(t',t_n)\,\bar{\vec\chi}_{n} + \dots \right]^\T   \vec{\mathcal{G}}(t')\nonumber \\ 
		 &=  \varphi_n - \! \int\limits_{t_0}^{t_n}\, \dif t' \, \vec k_n^ \T\, \vec g(t')\, [t'-t_n]    
		      +   \! \int\limits_{t_0}^{t_n}\, \dif t' \! \int\limits_{t_n}^{t'}\, \dif t'' \vec k_n^ \T\, \Gamma(t'') \, \vec g(t')\,[t'-t''][t''-t_n]  +  \dots
    \end{align}
valid for an arbitrary time-dependent gradient $\Gamma(t)$ and local acceleration $\vec g(t)$.

%
%
%
%
%
%
%
%
\section{Compact description of interferometry}
\label{sec:Compact_description:MZ_geometry}
So far we have developed a compact description of the internal and external dynamics of a two-level system in the presence 
of an external quadratic potential. The result was the generalized beam-splitter matrix~\eqref{BSM_H_2} describing the beam-splitter 
process in the Heisenberg picture. Now, we show that a sequence of such beam-splitter matrices allow the description of 
any interferometer geometry. With no loss of generality we choose the Mach-Zehnder interferometer as paradigmatic example in 
this section. Based on these considerations, we derive in Section~\ref{Probability_of_ground_state_detection} results 
which are even valid for more advanced geometries, e.\,g. the multi-loop geometry.  \\
  \begin{figure}[h]
    \centering
	\includegraphics[width=0.45\textwidth]{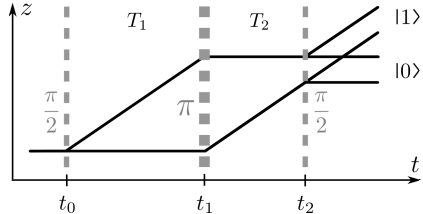}
    \caption[Mach-Zehnder interferometer]{Mach-Zehnder interferometer: An atomic wave function initially in the internal ground state $\ket{0}$ is coherently split by a $\pi/2$-pulse 
(dashed gray line at time $t_0$; atom-field interaction zone) ending in a superposition of $\ket{0}$ and $\ket{1}$. Influences coming from a potential coupling to the 
external degree of freedom (center-of-mass motion) cause a state-dependent phase accumulation of the coherently superimposed states $\ket{0}$ and $\ket{1}$ 
(time interval $T_1$; free-evolution zone = no laser field). Afterwards, a $\pi$-pulse (dashed gray line at time $t_1$) redirects the two atomic paths (black lines). After a second 
free-evolution zone (time interval $T_2$) a final $\pi/2$-pulse at time $t_2$ coherently recombines the internal states.}
    \label{fig.MZI}
  \end{figure}

\fig\ref{fig.MZI} sketches a Mach-Zehnder geometry commonly used in (light-pulse) atom interferometry. The initial 
state $\ket{\Psi(t_0)} = \ket{\psi^{(0)}(t_0)}\ket{0}$ is equally split by a $\pi/2$-pulse into a coherent 
superposition of the internal ground state $\ket{0}$ and the internal excited state $\ket{1}$.
Since every internal atomic transition coincides with a momentum kick (recoil of an absorbed/emitted photon with 
momentum $\hbar \vec k_n$), the center-of-mass motion is state-dependent. We have already mentioned this internal-external 
coupling in the context of the interaction Hamiltonian in Section~\ref{sec_MWI_quadratic_potentials}. A $\pi$-pulse 
subsequently inverts the populations of $\ket{0}$ and $\ket{1}$ and redirects the (classical) atomic paths. 
A final $\pi/2$-pulse coherently recombines the internal states. 
%
%
%
\subsection{Mach-Zehnder pulse sequence in an external potential}
The spatially separated paths of the states $\ket{0}$ and $\ket{1}$ get a state-dependent 
phase accumulation in the presence of an external potential. Therefore, the relative phase between the ground and the exited state 
after a Mach-Zehnder pulse sequence measures the influence of the external potential. \\

Let us compose the Mach-Zehnder interferometer sequence via the separation ansatz~\eqref{separation_ansatz} and the 
corresponding matrices $\hat{S}_n^{(\Theta)}$ and $\hat{U}_\e(t_{n+1},t_n)$ in the Schrödinger picture:
    \begin{align}\label{final_MZI_state_1}
	\ket{\Psi(t_2)}	 &=  \hat{S}^{(\frac{\pi}{2})}_{2}  ~ \hat{U}_\e(t_2,t_1) ~ \hat{S}^{(\pi)}_{1} ~ \hat{U}_\e(t_1,t_0) ~ \hat{S}^{(\frac{\pi}{2})}_{0} ~ \ket{\Psi_0}\, .
    \end{align}
The final state (at time $t_2$) after a Mach-Zehnder pulse sequence consists of the individual zones: \\
\vspace{-0.25cm}
\begin{center}
    \begin{tabular}{rl}
	$\hat{S}_0^{(\frac{\pi}{2})}$:			&	{\small $\frac{\pi}{2}$-pulse at $t_0$ (wave vector $\vec k_0$, laser phase $\varphi_0$),}\\[0.1cm]
	$\hat U_\e(t_1,t_0)$:				&	{\small ``Free propagation'' from $t_0$ to $t_1$ in $V_\e$,} \\[0.1cm]
	$\hat{S}_1^{(\pi)}$:				&	{\small $\pi$-pulse at $t_1$ (wave vector $\vec k_1$, laser phase $\varphi_1$),}\\[0.1cm]
	$\hat U_\e(t_2,t_1)$:				&	{\small ``Free propagation'' from $t_1$ to $t_2$ in $V_\e$,} \\[0.1cm]
	$\hat{S}_2^{(\frac{\pi}{2})}$:			&	{\small $\frac{\pi}{2}$-pulse at $t_2$ (wave vector $\vec k_2$, laser phase $\varphi_2$).} \\[0.2cm]
    \end{tabular}
\end{center} 
For the rest of the paper we omit the subscript ``$\,\e\, $''; in particular $\hat{U}$ will indicate free propagation in an 
external quadratic potential.
%
%
%
%
\subsection{Final state}
Recalling the transformation \eqref{bsmh} of the beam-splitter matrix into the Heisenberg picture,
we can rewrite the expression (\ref{final_MZI_state_1}) for the final state and arrive at
    \begin{align}\label{final_state}
	\ket{\Psi(t_2)} &=  \hat{U}(t_2,t_0) \, \hat{U}_{\MZ}  \ket{\Psi_0}\, .
    \end{align}
At this point, we have introduced the Mach-Zehnder operator
    \begin{align}\label{MZ_operator}
	\hat{U}_{\MZ} = \hat{S}^{(\frac{\pi}{2})}_{\H,2} ~ \hat{S}^{(\pi)}_{\H,1} ~ \hat{S}^{(\frac{\pi}{2})}_{\H,0}\, ,
    \end{align}
which fully characterizes the Mach-Zehnder geometry. 
In this sense, any interferometer can be understood as a series of generalized beam-splitter matrices. Knowing the interferometer 
geometry, the calculation of the characteristic interferometer operator (here $\hat{U}_{\MZ}$) is straightforward 
(see Appendix~\ref{app:Mach-Zehnder_operator}).
%
%
%
%
\section{Probability of ground-state detection}
\label{Probability_of_ground_state_detection}
The probability of detecting the matter-wave in the ground state $\ket{0}$ after a Mach-Zehnder pulse sequence is given by 
  \begin{align}\label{prob_MZ_1}
      \tilde{P}_{\,\indket{0}}(t_2) &= \trace_{\!\! \i,\e}\{\hat\rho(t_2)  \ket{0}\bra{0} \}\,,
  \end{align}
where the trace runs over the internal as well as the external degrees of freedom and 
$\hat\rho(t_2) = \hat{U}(t_2,t_0) ~ \hat{U}_{\MZ} ~ \hat\rho(t_0) ~ \hat{U}^{\dagger}_{\MZ} ~ \hat{U}^{\dagger}(t_2,t_0) $
is the density-matrix operator.
When we assume a state initially starting in the ground state $\ket{0}$ and take the trace over the internal degrees of freedom, 
we arrive at
  \begin{align}\label{ground_state_probability}
      P_{\,\indket{0}}(t_2) &= \trace_{\!\! \e}\{ \hat{\mathcal{O}}_\e \, \hat\rho_\e(t_0) \, \hat{\mathcal{O}}^{\dagger}_\e \}\,.
  \end{align}
Here, we have introduced the operator $\hat{\mathcal{O}}_\e = \bra{0}  \hat{U}_{\MZ}  \ket{0}$ which 
contains the evolution of the external degrees of freedom while passing a Mach-Zehnder pulse sequence. Note, 
since we have already taken the trace over the internal degree of freedom, $\hat{\mathcal{O}}_\e$ has to be an operator only 
acting on the external degrees of freedom. 
Furthermore, the free-evolution operator $\hat{U}(t_2,t_0)$ does not alter the probability regardless whether interferometer pulse-sequence 
was running before. Indeed, since the free-evolution operator only acts on external degrees of freedom, $\hat{U}(t_2,t_0)$ 
and the projector $\ket{0}\bra{0}$ commute and we can change their order. Taking advantage 
of the unitarity relation  $\hat{U}(t_2,t_0) \,\hat{U}^{\dagger}(t_2,t_0)= \mathds{1}$ we get the 
result~\eqref{ground_state_probability}.
%
%
%
%
\subsection{Characteristic operators of the Mach-Zehnder geometry}
\label{Characteristic_operators_MZI}
When we consider the ground-state detection probability, we recognize that the details of the interferometer 
are included in the operators $\hat{\mathcal{O}}_\e$ and $\hat{U}_{\MZ}$. Thus, we study these operators in more detail.
%
%
%
\subsubsection{Mach-Zehnder operator}
The Mach-Zehnder operator $\hat{U}_{\MZ}$ is supposed to be the starting point of our discussion. Its detailed calculation is 
given in Appendix~\ref{app:Mach-Zehnder_operator}. The result is the matrix Eq.~\eqref{app:U_MZ}
including the generalized phase $\Phi_n$, Eq.~\eqref{phi_n}, as well as the displacement operator 
$\hat{D}(\vec\chi_n)$ defined in Eq.~\eqref{displ_op_2}. Due to its generality, we get a rather lengthy expression for the Mach-Zehnder 
operator. However, it includes the full information about the geometry: 
It provides the phase shifts and displacements and takes into account all possible paths 
of a Mach-Zehnder interferometer. More precisely, every matrix element corresponds 
to the propagation of the initial (external) state. For instance, the matrix element 
$ (\hat{U}_{\MZ})_{ij} = \bra{i}  \hat{U}_{\MZ}  \ket{j} $ with $i,j\in\{0,1\}$ 
describes the propagation of the external state corresponding to the scenario in which we initially start in the internal 
state  $\ket{j}$ and end in $\ket{i}$. 
%
%
%
%
\subsubsection{Vertex rule}
%
Let us have a closer look at vertices (interactions) 
connecting the internal states.
Whenever an interaction, described by the generalized beam splitter~\eqref{BSM_H_2}, is stimulating internal transitions ($\ket{0}\! \rightarrow \, \ket{1}$ or $\ket{1} \! \rightarrow \, \ket{0}$) 
an additional phase $-\i \e^{\pm\i\Phi_n}$ and a displacement $\hat{D}(\pm \vec\chi_n)$ is accumulated; 
see also Eq.~\eqref{S_H_pi_ground_excited}.
The displacement operator accounts for the photon recoil and the center-of-mass motion 
whereas the phase shift contains, amongst others, the laser phase $\varphi_n$ imprinted by the $n$-th interaction. In addition, 
every 50:50 node ($\pi/2$-pulse) corresponds to the additional factor $1/\sqrt{2}$, see Eq.~\eqref{S_H_pi/2_ground}.    
We summarize these rules in TABLE~\ref{tab:additional_phases_displ}. \\
  \begin{table}[h]
    \centering
	\begin{tabular}{m{4cm}|m{3cm}}
	\toprule[0.5pt] 
	\begin{center}\vspace{-0.3cm}  phases/displacements \end{center}	& 	\begin{center}\vspace{-0.3cm} $n$-th vertex \end{center} \\ [-0.4cm]  \midrule[0.5pt]   
	\centering$-\i \e^{+\i\Phi_n} \hat{D}(+ \vec\chi_n)$			&	\begin{center}\includegraphics[width=0.1\textwidth]{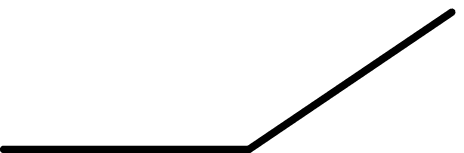}	\end{center}  \\  
	\centering$-\i \e^{-\i\Phi_n} \hat{D}(-\vec\chi_n)$			& 	\begin{center}\includegraphics[width=0.1\textwidth]{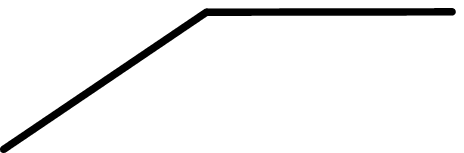} \end{center}  \\  \midrule[0.5pt]  
        \begin{center}\vspace{-0.3cm}	prefactor \end{center}			&	\begin{center}\vspace{-0.3cm} node (50:50)  \end{center} \\ [-0.4cm] \midrule[0.5pt] 
	\centering$\frac{1}{\sqrt{2}}$						& 	\begin{center}\vspace{0.15cm} \textbullet \end{center} \\ 
	\end{tabular}
    \caption[Vertex rule]{Vertex rule: Imprinted phases and displacements by the generalized 
beam splitter. Whenever a transition between initial states takes place, the state accumulates the phase $-\i \e^{\pm\i\Phi_n}$ 
and the displacement  $\hat{D}(\pm \vec\chi_n)$. For 50:50 nodes ($\pi/2$-pulse) we get an additional prefactor $1/\sqrt{2}$.}
    \label{tab:additional_phases_displ}
  \end{table} 
  \begin{figure}[h]
    \centering
	\includegraphics[width=0.45\textwidth]{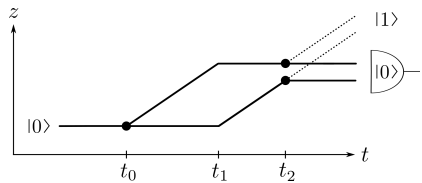}
    \caption[Displacement and phases of a Mach-Zehnder geometry]{Mach-Zehnder geometry: When we start in the ground state $\ket{0}$ 
and go along the upper or the lower path (paths are sketched by solid lines), we get for every vertex, which we will cross, 
the term $-\i \e^{\pm\i\Phi_n} \hat{D}(\pm\vec\chi_n)$. In addition, 
every node corresponds to the factor $1/\sqrt{2}$. The resulting displacement operator sequence causes a displacement in position $z$ as well as in momentum $p$ of the final state $\ket{0}$ (parallel solid lines) as well as 
the final excited state $\ket{1}$ (parallel dotted lines). The ground state is detected at the end.}
    \label{fig:two_paths}
  \end{figure} 

By means of the Mach-Zehnder interferometer sketch (\fig\ref{fig:two_paths}), we can easily get the final state, corresponding 
to the ground-state detection, via the ``vertex rule'', TABLE~\ref{tab:additional_phases_displ}:
The upper path consists of two vertices at the times $t_0$ and $t_1$ and two nodes. Thus, the final upper state for the ground-state 
detection reads
    \begin{align}\label{path_I}
	    \ket{\psi^{(0)}(t_2)}_\text{upper} &= \frac{1}{\sqrt{2}} \left(-\i \e^{-\i\Phi_1} \hat{D}(- \vec\chi_1)\right) \times  \nonumber \\
					      & \qquad \times \left(-\i \e^{+\i\Phi_0} \hat{D}(+ \vec\chi_0)\right)  \frac{1}{\sqrt{2}}  \ket{\psi^{(0)}(t_0)}\,.
    \end{align}
Indeed, every vertex brings in one phase factor and one displacement. Each node corresponds to the factor $1/\sqrt{2}$.

The final state for the lower path shows the same total number of vertices and nodes but at different times
    \begin{align}\label{path_II}
	    \ket{\psi^{(0)}(t_2)}_\text{lower} &= \frac{1}{\sqrt{2}} \left(-\i \e^{-\i\Phi_2} \hat{D}(- \vec\chi_2)\right) \times   \nonumber \\
					      & \qquad \times \left(-\i \e^{+\i\Phi_1} \hat{D}(+\vec\chi_1)\right)  \frac{1}{\sqrt{2}}  \ket{\psi^{(0)}(t_0)}\,.
    \end{align}

We remark that impure $\pi$- or $\pi/2$-pulses result in slightly different pulse areas $\Theta$.
Therefore, a vertex always appears with a node. However, the node no longer corresponds to the factor of $1/\sqrt{2}$, but to 
the sine and cosine prefactors given by the generalized beam-splitter matrix~\eqref{BSM_H_2}. \\

Finally, the superposition of the upper path, Eq.~\eqref{path_I}, and the lower path, Eq.~\eqref{path_II}, yields the final state 
(for the ground-state detection) after a Mach-Zehnder pulse sequence
    \begin{align}\label{final_superpostion}
	    \ket{\psi^{(0)}(t_2)}   &=      - \frac{1}{2} \left\{  \e^{+\i[\Phi_0-\Phi_1]} ~ \hat{D}(-\vec\chi_1)\, \hat{D}(\vec\chi_0) + \right. \nonumber \\
				    &\,~\qquad    \left.    +\e^{+\i[\Phi_1-\Phi_2]} ~ \hat{D}(-\vec\chi_2)\, \hat{D}(\vec\chi_1) \right\}  \ket{\psi^{(0)}(t_0)} \nonumber \\
				    &= \hat{\mathcal{O}}_\e \ket{\psi^{(0)}(t_0)} \, .
    \end{align}
In the last step, we connect our result (based on the ``vertex rule'') with the calculation of the matrix element 
$\hat{\mathcal{O}}_\e = \bra{0}  \hat{U}_{\MZ}  \ket{0}$ of the matrix~\eqref{app:U_MZ}, done in Appendix~\ref{app:Mach-Zehnder_operator}. 
Analogously, the ``vertex rule'' also provides the other three matrix elements $\bra{i}  \hat{U}_{\MZ}  \ket{j}$; $i,j \in \{0,1\}$.

%
%
%
%
%
\subsubsection{Generalized displacement operator}
So far, we permitted all possible initial as well as final states ($\ket{0}$, $\ket{1}$) which yields the four 
matrix elements $\bra{i}  \hat{U}_{\MZ}  \ket{j}$; $i,j \in \{0,1\}$. Next, we consider 
the standard scenario in which we initially start in the internal state $\ket{0}$: We get the time evolution
  \begin{align}\label{U_MZ_0}
     \hat{U}_{\MZ}  \ket{0}	&=\frac{1}{2}		\left( \!\! \begin{array}{c} 
							      - \e^{+\i[\Phi_1-\Phi_2]} ~ \hat{D}(-\vec\chi_2)\, \hat{D}(\vec\chi_1)  ~ \{1  + \mathfrak{\hat D}_{\MZ} \} \\[0.1cm]
  							      - \i ~ \e^{+\i\Phi_1} \hat{D}(\vec\chi_1) ~ \{1 - \mathfrak{\hat D}_{\MZ} \}
							\end{array} \!\!\right) , 
  \end{align}
motivated by the previous section and exactly calculated in Appendix~\ref{app:Mach-Zehnder_operator}.
Here, we have factored out the phases and displacements corresponding to the 
lower path. Moreover, we have introduced the generalized Mach-Zehnder displacement operator defined as
  \begin{align}\label{gen_displ_1}
     \mathfrak{\hat D}_{\MZ}		&=   \e^{+\i[\Phi_0 - 2 \Phi_1 + \Phi_2]}  ~ \hat{D}(-\vec\chi_1)\, \hat{D}(\vec\chi_2)\, \hat{D}(-\vec\chi_1)\, \hat{D}(\vec\chi_0)\, ,
  \end{align}
which describes the relative phase and displacement (in phase space) of the upper path with respect to the lower path.
Note that the expression (\ref{U_MZ_0}) still includes both exit ports. Indeed, the first and second vector components correspond 
to the atom finally being in the ground state $\ket{0}$ and in the excited state $\ket{1}$, respectively. 

The ``sandwich rule''
\begin{align}\label{sandwich_rule_main}
  \hat{D}(-\vec\chi_1)\, \hat{D}(\vec\chi_2) \, \hat{D}(-\vec\chi_1) = \hat{D}(\vec\chi_2 - 2 \vec\chi_1)
\end{align}
as well as the ``composition rule''
\begin{align}\label{Composition_Rule}
   \hat D(\vec\chi_1) \, \hat D(\vec\chi_0) = \hat D(\vec\chi_1 + \vec\chi_0) ~ \e^{\frac{\i}{2\hbar}\vec\chi_0^\T  \mathcal{J} \vec\chi_1}\, ,
\end{align}
both derived in Appendix~\ref{app:displacement_operator}, can be employed to combine the product of displacement operators 
into just one effective displacement operator accompanied by an additional phase term 
  \begin{align}\label{gen_displ_2}
     \mathfrak{\hat D}_{\MZ}\!		&=   \e^{\!+\i[\Phi_0\! - 2 \Phi_1\! + \Phi_2] +\! \frac{\i}{2\hbar} \vec\chi_0^\T  \mathcal{J} (\vec\chi_0\!-2\vec\chi_1\!+\vec\chi_2)}  \hat{D}(\vec\chi_0 \!-\! 2 \vec\chi_1\! +\! \vec\chi_2).
  \end{align}

This phase correction stems from the composition of $ \hat{D}(\vec\chi_0)$ and $\hat{D}(\vec\chi_2 - 2 \vec\chi_1)$. 
The physical reason is the non-commutative property of the canonical operators $\hat{\vec x}$ and $\hat{\vec p}$ in the displacement operators. Hence, we get a 
quantum mechanical correction to the phase which follows from the Baker-Campbell-Hausdorff relation. 
In contrast, the ``sandwich rule'' does not show any phase corrections due to the bilinear form: The phase correction coming from 
the product of the displacement operator $\hat{D}(-\vec\chi_1)$ on the left cancels out with the correction coming from 
the displacement operator $\hat{D}(-\vec\chi_1)$ on the right.  
These statement is crucial for all loop geometries whether closed or not as long as we assume symmetric momentum kicks 
for the upper and the lower interferometer path; see the discussion of Eq.~\eqref{BSM_H_2}. Otherwise, the left and the right 
displacement operators show different displacement vectors $\vec\chi_1^+$ and $\vec\chi_1^-$. 
As a result, for every displacement operator we get an additional phase correction à la ``composition rule''~\eqref{Composition_Rule}. \\

By defining the generalized Mach-Zehnder phase 
\begin{align}\label{def_MZ_phase}
  \Phi_{\MZ} = \Phi_0 - 2 \Phi_1 + \Phi_2
\end{align} 
and the generalized Mach-Zehnder displacement vector 
\begin{align}\label{def_MZ_displacement}
   \vec\chi_{\MZ} = \vec\chi_0 - 2  \vec\chi_1 +  \vec\chi_2\, ,
\end{align}
we can write the generalized displacement operator in a compact form:
  \begin{align}\label{gen_displ_3}
     \mathfrak{\hat D}_{\MZ}		&=   \e^{+\i\Phi_{\MZ} + \frac{\i}{2\hbar} \vec\chi_0^\T  \mathcal{J} \vec\chi_{\MZ}} ~ \hat{D}(\vec\chi_{\MZ})\,.
  \end{align}
We already anticipate that the phases present in the generalized displacement operator will contribute to the total 
interferometer phase. Additionally, we note that $\mathfrak{\hat D}_{\MZ}$ is a  unitary operator. 
In contrast, the external operator  $\hat{\mathcal{O}}_\e$, which we will consider now, is not a unitary one.
%
%
%
\subsubsection{External operator}
With the time evolution (\ref{U_MZ_0}) at hand we find for the matrix element of the Mach-Zehnder operator 
  \begin{align}\label{ext_operator_MZ}
     \hat{\mathcal{O}}_\e 	&= 	\bra{0}  \hat{U}_{\MZ}  \ket{0} \nonumber \\
				&=	- \frac{1}{2} ~ \e^{+\i[\Phi_1-\Phi_2]} ~ \hat{D}(-\vec\chi_2)\, \hat{D}(\vec\chi_1)\, \{1 +  \mathfrak{\hat D}_{\MZ} \}\, .
  \end{align}
Clearly, the operator $\hat{\mathcal{O}}_\e$ only includes displacement operators $\hat{D}(\pm\vec\chi_n)$ and phases $\Phi_n$. Since displacement operators affect the 
position and momentum of the atom (external degrees of freedom), we call $\hat{\mathcal{O}}_\e$ an external operator.
%
%
%
\subsection{General probability}
\label{General_probability_MZ}
When we employ the previous result for the external operator $\hat{\mathcal{O}}_\e$, the expression for the ground-state detection probability (\ref{ground_state_probability}) reads
  \begin{align}\label{probability_MZ_1}
      P_{\,\indket{0}}(t_2)	&= 	\trace_{\!\! e}\{ \hat{\mathcal{O}}_\e \, \hat\rho_\e(t_0)\,  \hat{\mathcal{O}}^{\dagger}_\e \} \nonumber \\
					&=	\frac{1}{2} \left[1+ \Re\left(\trace_{\!\! e}\{ \mathfrak{\hat D}_{\MZ} \, \hat\rho_\e(t_0) \}\right)\right],
  \end{align}
where $\Re(c)$ indicates the real part of $c \in \mathds{C}$. 
The trace over the external degrees of freedom corresponds to the characteristic function 
\cite{Ferraro_2005}
    \begin{align}
      \eta[\hat\rho_\e(t_0)](\vec\chi_{\MZ})	&=	\trace_{\!\! e}\{ \hat{D}(\vec\chi_{\MZ})\, \hat\rho_\e(t_0) \}\, ,
    \end{align}
which is defined in Appendix~\ref{app:characteristic_function} and depends here on the initial density operator $\hat\rho_\e(t_0)$ and the displacement vector $\vec\chi_{\MZ}$. 
Hence, the ground-state probability reads in terms of the characteristic function
  \begin{align}\label{prob_MZ_charact_func}
	P_{\,\indket{0}}(t_2)	&=	\frac{1}{2} \left[1  \! +  \Re \!\left(   \e^{+\i\Phi_{\MZ} + \frac{\i}{2\hbar} \vec\chi_0^\T  \mathcal{J} \vec\chi_{\MZ}} \,  \eta[\hat\rho_\e(t_0)](\vec\chi_{\MZ})\right)\right]\!.
  \end{align}
As we will see, the characteristic function determines the visibility of the interferometer and brings in an additional term 
to the total interferometer phase. For an introduction into the characteristic function and its relation to the Wigner 
function we refer to Appendix~\ref{app:Wigner_func}. \\

When we express the characteristic function in terms of its corresponding Wigner function, which is a real function, via
 \begin{align}
        \eta[\hat\rho_\e(t_0)](\vec\chi_{\MZ}) = \!\!\int\limits_{\mathbb{R}^{2\N}}  \dif^{2\N} \xi ~ \e^{-\frac{\i}{\hbar} \vec\chi_{\MZ}^\T \mathcal{J} \vec\xi} ~ \text{W}[\hat\rho_\e(t_0)](\vec\xi)\, ,
 \end{align}
the above ground-state probability can be brought into the most convenient, final form 
  \begin{align}\label{prob_MZ_3}
      P_{\,\indket{0}}(t_2)	&=	\frac{1}{2} \left[1+ \!\!\int\limits_{\mathbb{R}^{2\N}}  \dif^{2\N} \xi ~ \text{W}[\hat\rho_\e(t_0)](\vec\xi) \, \cos  \left( \Delta \phi_{\MZ}(\vec\xi) \right)  \right] .
  \end{align}
Here, we have introduced the total Mach-Zehnder phase
  \begin{align}\label{Phi_total}
     \Delta \phi_{\MZ}(\vec\xi)	&=	\Phi_{\MZ} + \frac{1}{\hbar} \left[ \frac{\vec\chi_0}{2} + \vec\xi \right]^\T \!\!\mathcal{J}  \vec\chi_{\MZ}
  \end{align}
consisting of the Mach-Zehnder phase (\ref{def_MZ_phase}) and a phase contribution coming from the non-closed interferometer 
geometry ($\vec \chi_{\MZ} \neq 0$).

In conclusion, we arrived at one of the main results, Eq.~\eqref{prob_MZ_3}, which is the most general expression for the 
ground-state detection probability for a Mach-Zehnder pulse sequence (in the presence of an external quadratic potential), 
if the initial state was the internal ground state.
%
%
%
%
%
%
%
\subsection{Probability for Gaussian initial states}
\label{Probability_for_Gaussian_initial_states}
The previously obtained, most general expression for the ground-state detection probability $P_{\,\indket{0}}(t_2)$ 
significantly  simplifies if we restrict ourselves to the class of Gaussian states. 
In fact, this restriction is in accordance with thermal cloud experiments, see for example \cite{Peters_2001} or \cite{Kasevich_1991}, 
but also with BEC experiments using Mach-Zehnder pulse sequences. 
Thereby, the atoms initially trapped in a harmonic potential are described by the density operator $\hat\rho$. Afterwards, they are launched adiabatically and perpendicular to Earth's surface. At the moment of 
turning off the trapping potential, the atomic cloud is given by $\hat{D}(\vec\chi)\,  \hat\rho\, \hat{D}^{\dagger}(\vec\chi)$, 
which corresponds to a state at the mean position $\vec\chi_{x}$
with the mean initial momentum $\vec\chi_{p}$. \\

At the end of this subsection, we will additionally show that a final non-vanishing displacement 
$\vec\chi_\MZ$ results in a decreasing visibility of our interferometer.
%
%
%
\subsubsection{Wigner function}
The Wigner function of a general Gaussian state $\hat\rho_\e^{(\G)}$ is given by
 \begin{align}\label{wignerfunction_MZ}
    \text{W}[\hat\rho_\e^{(\G)}](\vec\xi)  &=  \frac{1}{\sqrt{(2\pi)^{6} \text{det} \Sigma}} ~ \e^{-\frac{1}{2} (\vec\xi-\langle \vec\xi\rangle)^\T \Sigma^{-1} (\vec\xi-\langle \vec\xi\rangle)}
 \end{align}
where the mean value of the Gaussian distribution reads
 $  \langle \vec\xi\rangle = \trace \{ \hat\rho_\e^{(\G)} \vec\xi \}$ 
and the corresponding positive definite covariance matrix is defined as
 $  \Sigma_{ik} = \frac{1}{2} \langle \xi_i \xi_k + \xi_k \xi_i\rangle -  \langle \xi_i \rangle \langle \xi_k \rangle; ~i,k \in \{1,...,6\}. $
Note that the choice of $\langle \vec\xi\rangle$ and $\Sigma$ fully determines the Gaussian state and therefore the 
Wigner function. Such Gaussian states can be used for thermal states, but also for coherent or 
squeezed states. 
%
%
%
%
\subsubsection{Characteristic function}
The corresponding characteristic function for Gaussian states follows from the (symplectic) Fourier transform 
of the Wigner function  
 \begin{align}
    \eta[\hat\rho_\e^{(\G)}](\vec\chi_{\MZ}) &=    \!\!\int\limits_{\mathbb{R}^{2\N}}  \dif^{2\N} \xi ~ \e^{-\frac{\i}{\hbar} \vec\chi_{\MZ}^\T \mathcal{J} \vec\xi} ~  \text{W}[\hat\rho_\e^{(\G)}](\vec\xi)\, .
 \end{align}
The characteristic function for Gaussian states can be rewritten in the following way (for details see Appendix~\ref{app:Wigner_func}): 
 \begin{align}\label{char_func_gauss_1}
    \eta[\hat\rho_\e^{(\G)}](\vec\chi_{\MZ})  &=   \e^{-\frac{1}{2\hbar^2} (\mathcal{J}\vec\chi_{\MZ})^\T \Sigma_0 (\mathcal{J}\vec\chi_{\MZ})+ \frac{\i}{\hbar}  \langle \vec\xi_0\rangle^\T (\mathcal{J}\vec\chi_{\MZ}) },
 \end{align}
where its time-dependence is included in the displacement vector $\vec\chi_{\MZ}$. 
The initial mean value $\langle \vec\xi_0\rangle$ and the initial covariance matrix $\Sigma_0$ are determined by the initial 
Gaussian state $\hat\rho_\e^{(\G)} = \hat\rho_\e^{(\G)}(t_0)$.
We recognize that the first term of the exponent is a Gaussian function in the 
displacement vector $\vec\chi_{\MZ}$ while the second term brings 
in the additional phase $ \langle \vec\xi_0\rangle^\T (\mathcal{J}\vec\chi_{\MZ}) /\hbar$. These two terms will give us the 
visibility and an additional contribution to the total interferometer phase, respectively.
%
%
%
\subsubsection{Probability (Gaussian initial states)}
\begin{figure}[h]
    \centering
	\includegraphics[width=0.4\textwidth]{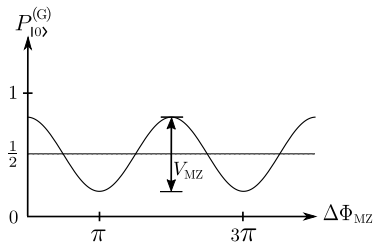}
    \caption[Probability]{Probability of ground-state detection: Tuning the total Mach-Zehnder phase $\Delta\Phi_{\MZ}$ will 
oscillate the probability of finding the atom in the ground state $\ket{0}$ after a Mach-Zehnder pulse sequence. Thereby, 
the visibility $V_{\MZ}$ determines the amplitude.}
    \label{fig:probability}
  \end{figure}

With the characteristic function (\ref{char_func_gauss_1}) at hand, the ground-state detection probability (\ref{prob_MZ_charact_func}) 
for initial Gaussian states can be written in the very compact and common form, known from Ramsey interferometry,
  \begin{align}\label{prob_MZ_compact_form}
	P^{(\G)}_{\indket{0}}(t_2)	&=	\frac{1}{2} \left[ 1 + V_{\MZ} \cos\left( \Delta\Phi_{\MZ}   \right)\right]\,.
  \end{align}
Here, we have introduced the visibility $V_\MZ$ and the total Mach-Zehnder phase $\Delta \Phi_\MZ$ for Gaussian states, defined and 
discussed next below. \\
%
%
%
%
%
%
\paragraph{Visibility}
The bilinear form in the exponent of the visibility 
  \begin{align}\label{Visibility_MZ}
	V_{\MZ}	&=	\e^{-\frac{1}{2\hbar^2} (\mathcal{J}\vec\chi_{\MZ})^\T \Sigma_0 (\mathcal{J}\vec\chi_{\MZ})}
  \end{align}
is a Gaussian originally coming from the characteristic function (\ref{char_func_gauss_1}).
The visibility decreases ($V_\MZ<1$, see \fig\ref{fig:probability}) for non-closed interferometers, which means for a non-vanishing 
displacement vector ($\vec\chi_{\MZ}\neq 0$), and depends on the covariance matrix $\Sigma_0$. By squeezing the initial state 
appropriately we can modify the covariance matrix and therefore improve the visibility. For the loss of visibility due to the effects of 
gradients and rotations (see discussion of the displacement vector $\vec\chi_\MZ$, Eq.~\eqref{chi_MZ_gradient}, 
in Section~\ref{sec_characteristic_quantities}), we can choose suitable time-asymmetric interferometer pulse sequences to achieve 
nearly closed interferometers ($\vec\chi_\MZ \approx 0$) and therefore also improve the visibility~\cite{Roura_2014}. \\
%
%
%
%
%
\paragraph{Total phase}
Besides the visibility, the second term in the exponent of the characteristic function \eqref{char_func_gauss_1} yields 
a correction to the Mach-Zehnder phase $\Phi_{\MZ}$. 
Hence, the total phase shift for a Mach-Zehnder pulse sequence is given by
  \begin{align}\label{total_ph_MZ}
	\Delta\Phi_{\MZ}		&=	\Phi_{\MZ} +  \frac{1}{\hbar} \left[ \frac{\vec\chi_0}{2} + \langle \vec\xi_0\rangle \right]^\T \!\! \mathcal{J}  \vec\chi_{\MZ}\, .
  \end{align}
There are three terms contributing to the total phase: 
(i) The first term is the generalized Mach-Zehnder phase (\ref{def_MZ_phase}) which takes into
account the effect of a linear external potential (e.g.\,local acceleration $\vec g$).  
(ii) The second term is a correction coming from the composition of displacement operators: In this sense, it is a 
Baker-Campbell-Hausdorff correction by composing exponential functions/displacement operators including non-commuting 
canonical operators~\cite{Wilcox_1967}. 
(iii) The last term is the additional phase coming from the characteristic function (\ref{char_func_gauss_1}): 
It takes into account the initial position as well as the initial momentum of the wave packet. Indeed, the initial phase-space coordinate $\langle \vec\xi_0\rangle$ only arises for a non-vanishing displacement vector, 
which is the case for gradients or rotations. Hence, different initial conditions will lead to different state-dependent 
phase accumulations. 

Finally, we emphasize that all phase terms depend on the specific interferometer geometry, which is formally included in the 
subscript ``MZ'' of $\Phi_{\MZ}$ and $\vec\chi_{\MZ}$. But more 
importantly, the total interferometer phase is a consequence of the non-commutative property of canonical operators: On the one 
hand the time-evolution, included in $\Phi_{\MZ}$ and 
$\vec\chi_{\MZ}$, stems from the Heisenberg equation of motion $\eqref{Heisenberg_equation}$ and thus from the 
commutator of the Hamiltonian $\hat H_\H$ with the operator $\hat{\vec\xi}_\H$; on the other hand the composition of displacement operators
leads to commutation relations in the context of Baker-Campbell-Hausdorff corrections, which enter in the total phase.
In conclusion, the phase shift measured by the interferometer can be understood as a quantum mechanical effect based on 
commutation relations of canonical observables.    
%
%
%
%
\subsection{Multi-loop geometry}
\label{ssec:Multi-loop_geometry}
Analogously to the Mach-Zehnder geometry, see Eqs.~\eqref{final_state} and \eqref{MZ_operator}, we can write the final state of a 
multi-loop geometry (\fig\ref{fig:multi-loop geometry}) in the following form
    \begin{align}\label{final_state_I}
	\ket{\Psi(t_n)}	 &=  \hat{U}(t_n,t_0) \, \hat{U}_{\I}  \ket{\Psi_0}.
    \end{align}
Here, the interferometer operator
\begin{align}\label{interferometer_op}
 \hat{U}_{\I} = \hat{S}^{(\frac{\pi}{2})}_{H,n} ~ \hat{S}^{(\pi)}_{H,n-1} ~ \cdots ~ \hat{S}^{(\pi)}_{H,1} ~ \hat{S}^{(\frac{\pi}{2})}_{H,0}
\end{align}
accounts for the $\pi/2$-pulse at the beginning (time $t_0$) and at the end (time $t_n$) of the multi-loop geometry as well as the $(n\!-\!1)$ $\pi$-pulses 
in between.\\

Next, we can substitute the corresponding generalized beam splitters~\eqref{BSM_H_2} into Eq.~\eqref{interferometer_op} 
and arrive at an explicit expression for the interferometer matrix $\hat{U}_\I$. Afterwards, we can determine the ground-state 
detection probability. However, we want to derive the probability on another way. We use the ``vertex rule'' in order to get the characteristic 
quantities $\Phi_\I$ and $\vec\chi_\I$ of the multi-loop geometry. Therefore, we recall that every vertex corresponds to a 
phase $-\i \e^{\pm\i\Phi_n}$ and a displacement $\hat{D}(\pm \vec\chi_n)$ (``vertex rule''). Thus, each interferometer 
geometry belongs to a specific combination of displacements and phases. For the case of the Mach-Zehnder geometry this yields the characteristic quantities 
$\Phi_\MZ$, Eq.~\eqref{def_MZ_phase}, and $\vec\chi_\MZ$, Eq.~\eqref{def_MZ_displacement}. The combination of both quantities determines 
the generalized displacement operator $\hat{\mathfrak{D}}_\MZ$, Eq.~\eqref{gen_displ_3}. 

When we apply the ``vertex rule'' to the multi-loop geometry, it is obvious by induction that the characteristic quantities are given by the generalized interferometer phase 
  \begin{align}\label{Phi_I}
    \Phi_\I &= \Phi_0 \!+ 2 \left[- \Phi_1 +  \Phi_2 -\! ... + \! (-1)^{n-1}  \Phi_{n-1} \right]\! +\! (-1)^n \Phi_n
  \end{align}
and the interferometer displacement vector
  \begin{align}\label{chi_I}
      \vec\chi_\I &= \vec\chi_0 \! + 2 \left[ - \vec\chi_1 + \vec\chi_2   -\! ... +\! (-1)^{n-1}  \vec\chi_{n-1}\right]\! +\! (-1)^n\vec\chi_n\,.  \end{align}
Indeed, since a multi-loop geometry starts with the same vertices as the Mach-Zehnder geometry, the first and the second term in 
Eqs.~\eqref{def_MZ_phase}, \eqref{Phi_I},  and Eqs.~\eqref{def_MZ_displacement}, \eqref{chi_I} has to be the same. Moreover, every 
$\pi$-pulse corresponds to two vertices (upper and lower path), which yields the additional factor of $2$. Since subsequent 
beam splitters always change the internal state and therefore the momentum, the sign for each term alternates. \\

Since the interferometer geometry only explicitly enters in the characteristic quantities $\Phi_\I$ and $\vec\chi_\I$, 
we can combine the sequence of phases, Eq.~\eqref{Phi_I}, and displacements, Eq.~\eqref{chi_I}, in the generalized displacement operator  
  \begin{align}\label{gen_displ_I}
     \mathfrak{\hat D}_{\I}		&=   \e^{+\i\Phi_{\I} + \frac{\i}{2\hbar} \vec\chi_0^\T  \mathcal{J} \vec\chi_{\I}} ~ \hat{D}(\vec\chi_{\I})
  \end{align}
by analogy to the Mach-Zehnder one, Eq.~\eqref{gen_displ_3}. 

Moreover, the results derived for the Mach-Zehnder geometry also holds true for the multi-loop case 
(except for an alternating sign in the equations for the probability; 
compare Eq.~\eqref{prob_MZ_compact_form} with \eqref{prob_I_final} and Eq.~\eqref{probability_MZ_1} with \eqref{probability_I}).
Indeed, since Eqs.~\eqref{probability_MZ_1}-\eqref{total_ph_MZ} include the quantities 
$\Phi_\MZ$, $\vec\chi_\MZ$ and $\hat{\mathfrak{D}}_\MZ$ only as parameters, we can substitute them by 
the corresponding multi-loop ones, which are given by Eqs.~\eqref{Phi_I}-\eqref{gen_displ_I}.
Therefore, we get via the external operator $\hat{\mathcal{O}}_\e 	= 	\bra{0}  \hat{U}_{\I}  \ket{0}$ 
the ground-state detection probability for a multi-loop geometry 
  \begin{align}\label{probability_I}
      P_{\,\indket{0}}(t_n)	&= 	\trace_{\!\!e}\{ \hat{\mathcal{O}}_\e \, \hat\rho_\e(t_0)\,  \hat{\mathcal{O}}^{\dagger}_\e \} \nonumber \\
					&=	\frac{1}{2} \left[1+  (-1)^n \,\Re\left(\trace_{\!\! e}\{ \mathfrak{\hat D}_{\I} \, \hat\rho_\e(t_0) \}\right)\right],
  \end{align}
which is finally determined by the displacement operator~\eqref{gen_displ_I}. 
The alternating sign $(-1)^n$ stems from the different number of vertices for each path of a multi-loop geometry, \fig~\ref{fig:multi-loop geometry}. 
Every vertex corresponds to an additional factor of ``$-\i$'', 
see TABLE~\ref{tab:additional_phases_displ}, and the number of vertices per path determines the final phase of each path. 
Thus, for an odd number of $\pi$-pulses (e.g. Mach-Zehnder geometry) there is no 
difference in the number of vertices per path and therefore no difference in the number of ``$-\i$''. However, for an 
even number of $\pi$-pulses one path always shows two vertices more than the other one and we get the relative phase shift 
$(-\i)^2=-1$. In conclusion, for every additional $\pi$-pulse in the 
interferometer sequence, the sign alternates in the ground-state detection probability~\eqref{probability_I}.

Analogously to Eq.~\eqref{prob_MZ_compact_form}, the ground-state detection probability for initial Gaussian states reads
  \begin{align}\label{prob_I_final}
	P^{(G)}_{\indket{0}}(t_n)	&=	\frac{1}{2} \left[ 1 + (-1)^n ~ V_{\I} \cos\left( \Delta\Phi_{\I}   \right)\right]
  \end{align}
with the total interferometer phase shift
  \begin{align}\label{total_ph_I}
	\Delta\Phi_{\I}		&=	\Phi_{\I} +  \frac{1}{\hbar} \left[ \frac{\vec\chi_0}{2} + \langle \vec\xi_0\rangle \right]^\T \!\! \mathcal{J}  \vec\chi_{\I}
  \end{align}
and the visibility 
  \begin{align}\label{visibility_interferometer}
	V_{\I}	&=	\e^{-\frac{1}{2\hbar^2} (\mathcal{J}\vec\chi_{\I})^\T \Sigma_0 (\mathcal{J}\vec\chi_{\I})}\, .
  \end{align}
The phase shift and the visibility include the characteristic interferometer quantities $\Phi_\I$, Eq.~\eqref{Phi_I}, and/or $\vec\chi_\I$, Eq.~\eqref{chi_I}. 

Finally, we remark that multi-loop geometries arise from appropriate chosen time separations $T_n$ for the laser pulses. 
Then, the upper and the lower path cross each other between subsequent laser pulses. However, since we can 
arbitrarily choose the time intervals $T_n$ (and in addition account for asymmetric momentum kicks in the beam splitters), 
our formalism allows a description of general multi-pulse geometries. But in practice, multi-loop geometries 
(with equal loop areas) are used in order to benefit from cancellation effects in the total interferometer 
phase and the visibility; for example to measure in first order gradients of the external potential.  
%
%
%
\section{Mach-Zehnder \& Butterfly geometries}
\label{sec_characteristic_quantities}
\subsection{Mach-Zehnder interferometer: Characteristic quantities}
\label{ssec:MZ_characteristic}
The ground-state detection probability is the typical quantity of interest in interferometer experiments. 
For the Mach-Zehnder interferometer we have derived a convenient form~\eqref{prob_MZ_compact_form} in the previous section.
We are left with the problem of the explicit calculation of the characteristic interferometer 
quantities $V_{\MZ}$ and $ \Delta\Phi_{\MZ}$. The goal of the present section is 
to determine the total phase $\Delta\Phi_{\MZ}$ and the visibility $V_{\MZ}$ of the 
Mach-Zehnder interferometer. We will consider the Sagnac effect by taking into account 
rotating lasers (with respect to an inertial reference frame), and the effect emerging 
from different time intervals between the successive laser pulses (time-asymmetric pulse sequence; see \fig\ref{fig.MZI}). 
Thereby, we use the results of Section~\ref{sec:Compact_description_of_interferometry} for the time-dependent phases and displacement vectors so that we will arrive at 
a full description of the Mach-Zehnder interferometer including all (three) major effects: 
Sagnac effect, time-asymmetry and the effects of gravity gradients.

Now, we take into account a uniform rotation rate of our lasers (Sagnac effect with time-independent axis of rotation) 
in the presence of a quadratic gravitational potential. We describe the effects of rotation by a rotated wave vector for 
the $n$-th laser pulse
\begin{align}\label{roation_k}
  \vec k_n 	&=	 	\cos(\Omega\, t_n)\, \vec k_0 + \sin(\Omega\, t_n) (\vec n \times \vec k_0) 
			      + [1-\cos(\Omega\, t_n)](\vec n \cdot \vec k_0)\,\vec n\, .
\end{align}
This is an active rotation of the wave vector $\vec k_0$ (first pulse) around the axis $\vec n = \vec \Omega/|\vec \Omega|$ 
with the rotation angle $\Omega\,  t_n$. 
%
%
%
\subsubsection{The generalized Mach-Zehnder phase}
The generalized Mach-Zehnder phase 
  $    \Phi_{\MZ} = \Phi_0 - 2\, \Phi_1 + \Phi_2$
explicitly includes the interferometer geometry. When we use the time-dependent phases (\ref{phi_n_4}) for the $n$-th laser pulse, 
we arrive at the general expression
    \begin{align}\label{phi_{MZ}_gamma}
	    \Phi_{\MZ} 	&= \varphi_{\MZ} +   2\, \vec g^\T \left[  \frac{ \cos{(\sqrt{\Gamma}[t_1-t_0]}) - 1 }{\Gamma} \right]  \vec k_1 
					 - \vec g^\T \left[\frac{ \cos{(\sqrt{\Gamma}[t_2-t_0]}) - 1 }{\Gamma} \right] \vec k_2
   \end{align}
for the generalized Mach-Zehnder phase in the presence of an external quadratic potential.
Here, we recall the definition of the Mach-Zehnder laser phase $\varphi_{\MZ} = \varphi_0 - 2 \varphi_1 + \varphi_2$, 
where $\varphi_2,~\varphi_1$ and $\varphi_0$ are the individual laser phases of each laser pulse. \\

The pure effect of a gradient will be achieved by setting the wave vectors equal ($\vec k_0 = \vec k_1 = \vec k_2$)
as well as the time intervals ($T=T_2=T_1$ with $T_2=t_2-t_1$ and $T_1=t_1-t_0$).
Table \ref{tab:phase_MZ_different_effects} summarizes all pure effects (to first order) of gradient, time-asymmetry and the Sagnac effect
for the generalized Mach-Zehnder phase $\Phi_{\MZ}$. Note that pure effect means without mixed terms.
For an expansion including mixed terms too, we refer to Appendix~\ref{The total Mach-Zehnder phase}, in which the first 
three lines of the total Mach-Zehnder phase (\ref{Delta_Phi_MZ}) correspond to the expansion of Eq.~\eqref{phi_{MZ}_gamma}.
  \begin{table}[h]
    \centering
	\begin{tabular}{lll}
	\toprule[1pt]\addlinespace[0.05cm] 
				&	\multicolumn{1}{c}{zeroth order}		& \multicolumn{1}{c}{first order}	\\   \midrule[1pt] \addlinespace[0.1cm]
	Standard case		&	$\varphi_{\MZ} + \vec k_0^\T \vec g  T^2$	& \multicolumn{1}{c}{--}		\\ \addlinespace[0.1cm] 
	Time-asymmetry		&	\multicolumn{1}{c}{$\varphi_{\MZ}$}		& $ + \vec k_0^\T \vec g [- T_1^2 + \frac{1}{2} (T_2+T_1)^2 ]$		\\ \addlinespace[0.1cm] 
	Sagnac effect		&	$\varphi_{\MZ} + \vec k_0^\T \vec g  T^2$	& $ + 3  (\vec\Omega \times \vec k_0)^\T \vec g   T^3 $		\\ \addlinespace[0.1cm] 
	Gradient		&	$\varphi_{\MZ} + \vec k_0^\T \vec g  T^2$	& $ - \frac{7}{12}\vec k_0^\T \Gamma  \vec g  T^4 $ \\ \addlinespace[0.1cm] \bottomrule[0.5pt] 
	\end{tabular}
    \caption{Generalized Mach-Zehnder phase $\Phi_{\MZ}$ for pure effects (to first order) of rotating lasers (Sagnac effect), gradient and time-asymmetric pulse sequences.}
    \label{tab:phase_MZ_different_effects}
  \end{table}
%
%
%
\subsubsection{The Mach-Zehnder displacement vector}
\label{ssec:MZ_Displacement_vector}
We have defined the generalized Mach-Zehnder displacement vector by 
    $    \vec \chi_{\MZ} = \vec\chi_0 - 2\, \vec\chi_1 +  \vec\chi_2$.
Consequently via Eq.~\eqref{chi_n_gradient}, the generalized Mach-Zehnder displacement vector is given by  
    \begin{align}\label{chi_MZ_gradient}
	    \vec \chi_{\MZ} &= \left( \begin{matrix} 0\\\hbar \vec k_0\end{matrix}\right) 
				- 2 \left( \begin{matrix} -\frac{\sin(\sqrt{\Gamma}[t_1-t_0])}{m \sqrt{\Gamma}} \\  \cos{(\sqrt{\Gamma}[t_1 - t_0])}  \end{matrix} \right) \hbar \vec k_1 \nonumber \\[0.2cm]
			    &	\qquad +  \left( \begin{matrix} - \frac{\sin(\sqrt{\Gamma}[t_2-t_0])}{m \sqrt{\Gamma}} 
				\\      \cos{(\sqrt{\Gamma}[t_2-t_0])}    \end{matrix}  \right) \hbar \vec k_2\,.
    \end{align}
Analogously to what we have done for $\Phi_\MZ$, we summarize in Table~\ref{tab:chi_MZ} the pure effects up to first order.
We see that time-asymmetry will never produce a displacement in momentum since the first order term is zero and 
no higher order terms are present. \\

Finally, we emphasize the importance of the displacement vector in the context of the visibility.
The knowledge of $\vec \chi_{\MZ}$ and the initial covariance matrix $\Sigma_0$ fully determines the visibility, Eq.~\eqref{Visibility_MZ}. 
Since rotations, gradients or time-asymmetric pulse sequences yields a non-vanishing displacement vector, Eq.~\eqref{chi_MZ_gradient}, 
the visibility immediately decreases ($V_\MZ<1$, see \fig\ref{fig:probability}). However, an appropriate choice of the time intervals $T_1$ and $T_2$ 
(time-asymmetric pulse sequence) can be used to get a nearly vanishing displacement vector ($\vec\chi_\MZ\approx0$; 
see Table~\ref{tab:chi_MZ}) and therefore improve the visibility~\cite{Roura_2014}.  
  \begin{table}[h]
    \centering
	\begin{tabular}{lcc}
	\toprule[1pt] \addlinespace[0.05cm] 
				&~displacement~ 		&\multicolumn{1}{c}{first order}								\\ \midrule[1pt] \addlinespace[0.1cm] 
	Time-asymmetry		&$\chi_{\vec x}$	&$  +\frac{\hbar \vec k_0}{m} \{T_1 -  T_2 \} $				\\ 
				&$\chi_{\vec p}$	&$  \vec 0$											\\  \midrule[1pt] \addlinespace[0.1cm] 
	Sagnac effect		&$\chi_{\vec x}$	&$  -\frac{2\hbar }{m}(\vec \Omega \times \vec k_0) T^2 $			\\ 
				&$\chi_{\vec p}$	&$  \vec 0$											\\ \midrule[1pt] \addlinespace[0.1cm] 
	Gradient		&$\chi_{\vec x}$	&$  + \Gamma \frac{\hbar \vec k_0}{m} T^3$										\\ 
				&$\chi_{\vec p}$	&$  - \Gamma \hbar \vec k_0 T^2$												\\ \midrule[1pt]
  \end{tabular}
    \caption{The components of the generalized Mach-Zehnder displacement vector $\vec\chi_{\MZ} = (\chi_{\vec x}, \chi_{\vec p})^\T$ are shown. We take into account pure effects (to first order) of rotating lasers 
(Sagnac effect), gradient $\Gamma$ and time-asymmetry.}
    \label{tab:chi_MZ}
  \end{table}
%
%
%
%
%
\subsubsection[Total Mach-Zehnder phase]{The total Mach-Zehnder phase}
The total Mach-Zehnder phase shift $\Delta\Phi_{\MZ}$, given by Eq.~\eqref{total_ph_MZ}, is fully determined by the above generalized 
Mach-Zehnder phase \eqref{phi_{MZ}_gamma}, the generalized displacement vector \eqref{chi_MZ_gradient} and the initial condition 
$\langle \vec\xi_0\rangle$, and leads to an exact analytical treatment of the Mach-Zehnder interferometer in the presence of 
rotations and gradients. However, in most of the experiments a time-symmetric pulse sequence is chosen
while rotations and gradients are naturally coming in by imperfections. 
Hence, we are usually interested in effects of 
small rotations and gradients and get for the total Mach-Zehnder phase shift the expansion formula \eqref{Delta_Phi_MZ} 
obtained in Appendix~\ref{The total Mach-Zehnder phase}.	
%
%
%
%
\subsection{Butterfly interferometer}
\label{ssec:BUI}
\begin{figure}[h]
    \centering
	\includegraphics[width=0.6\textwidth]{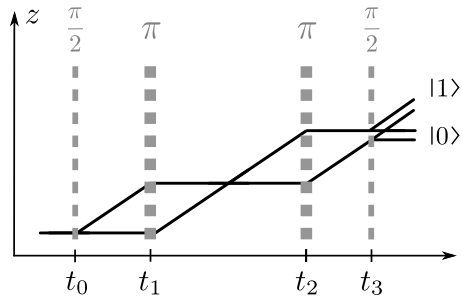}
    \caption[Butterfly interferometer]{Butterfly interferometer: An atomic wave function initially in the internal ground state $\ket{0}$ is coherently split by a $\pi/2$-pulse 
(dashed gray line at time $t_0$; atom-field interaction zone) ending in a superposition of $\ket{0}$ and $\ket{1}$. 
Influences coming from a potential coupling to the external degree of freedom (center-of-mass motion) cause a state-dependent phase accumulation of the coherently superimposed states $\ket{0}$ and $\ket{1}$ 
(free-evolution zone = no laser field).
Meanwhile, two $\pi$-pulses (dashed gray lines at time $t_1$ and $t_2$) redirect the two atomic paths (black lines). 
The final $\pi/2$-pulse at time $t_3$ coherently recombines the internal states.}
    \label{fig:BUI}
  \end{figure}

In the previous section, we discussed the characteristic quantities of the Mach-Zehnder geometry. 
Now, we will concentrate on the so-called  Butterfly or figure-eight geometry (\fig\ref{fig:BUI}). In analogy to the Mach-Zehnder interferometer, 
the Butterfly interferometer is another special case of a multi-loop/pulse geometry discussed in 
section~\ref{ssec:Multi-loop_geometry}. 
Therefore, it is straightforward to determine the generalized phase $\Phi_\BU$ as well as the displacement vector $\vec\chi_\BU$, 
which fully characterize the Butterfly geometry.  

The Butterfly interferometer \cite{Canuel_2006,Clauser_1988,Dubetsky_2006,Gustavson_2000,LEDUC_2004,Leveque_2010,Marzlin_1996,McGuirk_2002,Takase_2008,Tonyushkin_2008,Wu_2007,Stockton_2011}
is a two loop geometry. We will show that a configuration consisting of two symmetric loops directly measures gradients and 
is not sensitive to local accelerations (in first order). 

In comparison to the Mach-Zehnder geometry (\fig\ref{fig.MZI}), the Butterfly pulse sequence shows an additional $\pi$-pulse 
(see \fig\ref{fig:BUI}). Thus, it starts with a $\pi/2$-pulse creating an equally 
weighted superposition of the atomic ground $\ket{0}$ and excited $\ket{1}$ state. But then, two $\pi$-pulses redirect 
the two atomic paths. In-between, the superimposed states usually cross each other. A final $\pi/2$-pulse coherently 
recombines the internal states. Note that the crossing of the upper and the lower path is not necessary in general 
(our formalism allows for arbitrary time-intervals $t_n - t_0$). But in practice, it benefits to have 
loops equally in area with the result of a nearly closed geometry, and thus the full capability of cancellation effects 
based on a two loop geometry.

In the following, we present the characteristic quantities $\Phi_\BU$ and $\vec\chi_\BU$ in analogy to what we have done for the 
Mach-Zehnder geometry.
%
%
%
%
\paragraph{Generalized Butterfly phase}
The generalized Butterfly phase is given by 
   $  \Phi_{\BU} = \Phi_0 - 2\, \Phi_1 + 2\, \Phi_2 - \Phi_3  $ 
as a special case of the multi-loop geometry, see Eq.~~\eqref{Phi_I}. Via the explicit expression, Eq.~\eqref{phi_n_4}, 
for the n-th laser pulse, we arrive at the general result 
    \begin{align}\label{generalized_phi_BU}
	    \Phi_{\BU} &= \varphi_{\BU} 	+   2\, \vec g^\T \left[  \frac{ \cos{(\sqrt{\Gamma}[t_1-t_0]}) - 1 }{\Gamma} \right]  \vec k_1 
						-   2\, \vec g^\T \left[\frac{ \cos{(\sqrt{\Gamma}[t_2-t_0]}) - 1 }{\Gamma} \right] \vec k_2 \nonumber \\[0.1cm]
		       &\qquad			+     \vec g^\T \left[\frac{ \cos{(\sqrt{\Gamma}[t_3-t_0]}) - 1 }{\Gamma} \right] \vec k_3\, .
    \end{align}
The generalized phase $ \Phi_{\BU}$ explicitly includes arbitrary time-intervals ($t_n-t_0$) in-between successive laser pulses 
and the effect of a constant gradient $\Gamma$. Moreover, the Sagnac effect is implicitly 
included in the vector notation of the n-th wave vector $\vec k_n$, see Eq.~\eqref{roation_k}. 
The individual laser phases $\varphi_n$ are abbreviated by the Butterfly laser phase 
$\varphi_{\BU} = \varphi_0  - 2 \varphi_1 + 2 \varphi_2 - \varphi_3$. \\

The pure effect of the gradient is achieved by setting the wave vectors equal ($\vec k_0 = \vec k_1 = \vec k_2 = \vec k_3$)
and choosing the in experiments common time intervals: $T=T_1=T_3=T_2/2$ and $T_n=t_n-t_{n-1}$ (for equal loop areas). 
Table~\ref{tab:phase_BU_different_effects} summarizes all pure effects (to first order) of gradient, time-asymmetry and 
the Sagnac effect for the generalized Butterfly phase $\Phi_{\BU}$. The series expansion of $\Phi_{\BU}$, also including mixed terms, 
is presented in Appendix~\ref{app:Butterfly}.
  \begin{table}[h]
    \centering
	\begin{tabular}{lll}
	\toprule[1pt]\addlinespace[0.05cm] 
				&	\multicolumn{1}{c}{zeroth order}	\hspace{-0.4cm}	& \multicolumn{1}{c}{first order}	\\   \midrule[1pt] \addlinespace[0.1cm]
	Standard case		&	\multicolumn{1}{c}{$\varphi_{\BU} $}	\hspace{-0.4cm}	& \multicolumn{1}{c}{--}		\\ \addlinespace[0.1cm] 
	Time-asymmetry		&	\multicolumn{1}{c}{$\varphi_{\BU} $}	\hspace{-0.4cm}	& $+ \vec k_0^\T \vec g [- T_1^2 \! +\! (T_2\! +\! T_1)^2\! - \!\frac{1}{2} (T_3 \!+\! T_2 \!+\! T_1)^2 ]$		\\ \addlinespace[0.1cm] 
	Sagnac effect		&	\multicolumn{1}{c}{$\varphi_{\BU} $}	\hspace{-0.4cm}	& $- 6 (\vec \Omega \times \vec k_0)^\T \vec g T^3 $		\\ \addlinespace[0.1cm] 
	Gradient		&	\multicolumn{1}{c}{$\varphi_{\BU} $}	\hspace{-0.4cm}	& $+ 4 \vec k_0^\T  \Gamma  \vec g  T^4 $ \\ \addlinespace[0.1cm] \bottomrule[0.5pt] 
	\end{tabular}
    \caption{Generalized Butterfly phase $\Phi_{\BU}$ for pure effects (to first order) of rotating lasers (Sagnac effect), gradient and time-asymmetric pulse sequences. The 
time intervals are denoted by $T_n = t_n - t_{n-1}$.}
    \label{tab:phase_BU_different_effects}
  \end{table}
%
%
%
%
%
\paragraph{Butterfly displacement vector}
The Butterfly displacement vector is given by 
	$    \vec \chi_{\BU} = \vec\chi_0 - 2\, \vec\chi_1 + 2\, \vec\chi_2 - \vec\chi_3,$ see Eq.~\eqref{chi_I},
and therefore reads
    \begin{align}\label{chi_BU_gradient}
	    \vec \chi_{\BU} &= \left( \begin{matrix} 0\\\hbar \vec k_0\end{matrix}\right) 
				- 2 \left( \begin{matrix} -\frac{\sin(\sqrt{\Gamma}[t_1-t_0])}{m \sqrt{\Gamma}} \\  \cos{(\sqrt{\Gamma}[t_1 - t_0])}  \end{matrix} \right) \hbar \vec k_1 
			    	+ 2 \left( \begin{matrix} - \frac{\sin(\sqrt{\Gamma}[t_2-t_0])}{m \sqrt{\Gamma}} 
				\\      \cos{(\sqrt{\Gamma}[t_2-t_0])}    \end{matrix}  \right) \hbar \vec k_2 \nonumber \\[0.2cm]
			    & \qquad - 2 \left( \begin{matrix} - \frac{\sin(\sqrt{\Gamma}[t_3-t_0])}{m \sqrt{\Gamma}} 
				\\      \cos{(\sqrt{\Gamma}[t_3-t_0])}    \end{matrix}  \right) \hbar \vec k_3 \, ,
    \end{align}
where we have used Eq.~\eqref{chi_n_gradient} valid for constant gradients. Finally, Table~\ref{tab:chi_BU} summarizes 
the pure effects (in first order) of the gradient, time-asymmetric pulse sequence and an uniform rotation of the lasers. 
  \begin{table}[h]
    \centering
	\begin{tabular}{lcc}
	\toprule[1pt] \addlinespace[0.05cm] 
				&~displacement~ 		&\multicolumn{1}{c}{first order}								\\ \midrule[1pt] \addlinespace[0.1cm] 
	Time-asymmetry		&$\chi_{\vec x}$	&$+ \frac{\hbar \vec k_0}{m} \{ T_1 - T_2 + T_3     \}$				\\ 
				&$\chi_{\vec p}$	&$ \vec 0$											\\  \midrule[1pt] \addlinespace[0.1cm] 
	Sagnac effect		&$\chi_{\vec x}$	&$ \vec 0$			\\ 
				&$\chi_{\vec p}$	&$ \vec 0$											\\ \midrule[1pt] \addlinespace[0.1cm] 
	Gradient		&$\chi_{\vec x}$	&$- 2 \frac{\Gamma}{m} \vec k_0 T^3 $											\\ 
				&$\chi_{\vec p}$	&$ \vec 0$												\\ \midrule[1pt]
	\end{tabular}
    \caption{The components of the Butterfly displacement vector $\vec\chi_{\BU} = (\chi_{\vec x}, \chi_{\vec p})^\T$ are shown. 
We take into account pure effects (to first order) of rotating lasers (Sagnac effect), constant gradient $\Gamma$ and time-asymmetry.}
    \label{tab:chi_BU}
  \end{table}
%
%
%
%
\paragraph{The total Butterfly phase shift}
The total phase shift for an arbitrary interferometry is given by Eq.~\eqref{total_ph_I}. Thus, we arrive at the total Butterfly 
phase shift by substituting with the generalized interferometer phase, Eq.~\eqref{generalized_phi_BU}, and the 
displacement vector, Eq.~\eqref{chi_BU_gradient}, 
  \begin{align}\label{total_Phi_BU}
	\Delta\Phi_{\BU}		&=	\Phi_{\BU} +  \frac{1}{\hbar} \left[ \frac{\vec\chi_0}{2} + \langle \vec\xi_0\rangle \right]^\T \!\! \mathcal{J}  \vec\chi_{\BU}.
  \end{align}
The approximation of $\Delta \Phi_\BU$ for small, uniform rotations of the lasers and small gradients is presented in Appendix~\ref{app:Butterfly}.

%
%
\section{Non-inertial reference frames}
\label{Non-inertial_frame}
So far we have considered the atom dynamics in inertial frames. In particular,
the external potential was expanded around a fixed point $\vec \rho_0$ and any rotation effects corresponded to changes in the direction of the momentum transfer $\vec k_n$ for each laser pulse. 

In this section we will consider interferometers
in non-inertial frames where the expansion point as well as the potential are in general time dependent. 
We will start by reviewing the dynamics in non-inertial reference frames in order to establish the link to well-known 
results based on general quadratic Hamiltonians in rotating frames.
Next, we will introduce an equivalent (and general) description of interferometry in non-inertial frames, 
where the interferometer observables are calculated in a special reference frame co-moving with the interferometer device but non-rotating. In this way 
we will see that we can adapt our results obtained for inertial frames (with possibly rotating lasers) to this class of co-moving frames and derive in a simple way the interferometer phase shift and visibility for arbitrary non-inertial frames.  
As an example, we discuss at the end of this section interferometers in rotating frames 
(for instance experiments fixed on Earth) where we employ our alternative approach.
%
%
%
\subsection{General case}
\label{ssec:General_non-inertial_frame}
Here  we provide a general introduction to non-inertial frames needed for the classification of further calculations 
done in co-moving and co-rotating frames. 

  \begin{figure}[h]
    \centering
	\includegraphics[width=0.45\textwidth]{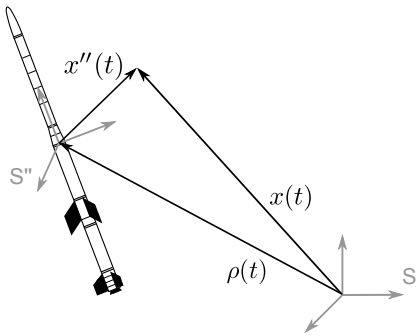}
    \caption[Rocket]{Rocket as an application for the most general case in non-inertial reference frames: The trajectory $\vec x''(t)$, seen by an observer flying in a rocket which follows 
the trajectory $\vec\rho(t)$ (non-inertial frame $\S''$), is connected with the trajectory $\vec x(t)$ (inertial frame $\S$)
by the relation $\vec x(t)= \vec\rho(t) + \mathcal{R}(t)\, \vec x''(t)$. $\mathcal{R}(t)$ accounts for an arbitrary time-dependent 
rotation of $\S''$ (with respect to $\S$). The vector notation in combination with primes is a short notation for the 
components measured in the corresponding reference frames.}
    \label{fig:rocket}
  \end{figure} 

An arbitrary trajectory in an inertial frame $\S$ shall be given by the time-dependent vector $\vec x(t)$. As in previous sections, 
we assume a general quadratic approximation, Eq.~\eqref{harm_grav_potential}, of the external potential $V$. It should be emphasized that $V$ 
does not necessarily have to be the gravitational potential, and can also be induced by magnetic or any other external fields. 

In many applications such as those involving freely falling capsules in drop towers, sounding rockets and dedicated satellite missions, it is convenient to consider a reference  frame $\S''$ co-moving and co-rotating with the interferometer set-up and the lasers (see \fig\ref{fig:rocket}). The origin of the non-inertial reference frame $\S''$ is taken as the expansion point $\vec \rho = \vec \rho(t)$ for the harmonic approximation of the potential.
Moreover, the coordinate axes of $\S''$ are in general arbitrarily rotating in time
(with respect to the inertial frame $\S$). Hence, the trajectory measured by an observer in the inertial frame $\S$ 
can be written as
    \begin{align}\label{general_coord_transformation}
	\vec x(t) &=  \vec \rho(t) + \mathcal{R}(t)\, \vec x''(t)\, ,
    \end{align}
where the vector $\vec x''(t)$ should be interpreted as the components of the trajectory relative to the coordinate axis of 
$\S''$ and $\mathcal{R}(t)$ takes into account an arbitrary time-dependent rotation of $\S''$ (with 
respect to $\S$).\\

Next, we show that effects due to non-inertial reference frames can be easily interpreted when the coordinate 
transformation, Eq.~\eqref{general_coord_transformation}, is performed in two steps: (i) the transformation to an 
accelerated reference frame, but without rotations of the coordinate axis 
(co-moving frame $\S'$; see \fig\ref{fig:co-moving_frame}) and (ii) the transformation to the frame $\S''$ 
which is assumed to be co-rotating with the interferometer device. 

In particular, we will point out in which sense such transformations concern the local acceleration as well as the gradient. 
%
%
%
%
\subsubsection{Dynamics in accelerated frames}
\label{ssec:Accelerations}
\begin{figure}[h]
    \centering
	\includegraphics[width=0.3\textwidth]{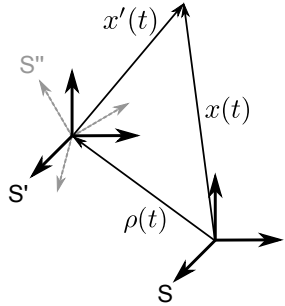}
    \caption[Co-moving and rotating frame]{Co-moving frame $\S'$ and rotating frame $\S''$: Both frames $\S''$ and $\S'$ possess the same time-dependent 
origin $\vec\rho(t)$, but just the co-moving frame $\S'$ shows aligned axes with respect to the
inertial frame $\S$. The position vector seen by an observer in $\S$ or $\S'$ is connected via $\vec x(t) = \vec\rho(t) + \vec x'(t)$.}
    \label{fig:co-moving_frame}
  \end{figure}

In this section, we consider interferometers in co-moving reference frames. Co-moving means arbitrary (translational) 
accelerations of $\S'$ while following the interferometer device, but aligned coordinate axes with respect to the inertial frame 
$\S$, i.e. $\mathcal R(t) =  \mathds{1}_3 $ (see also \fig\ref{fig:co-moving_frame}). 

The position vector in terms of the co-moving frame $\S'$ reads
    \begin{align}\label{vector_x}
	\vec x(t) = \vec \rho(t) + \vec x'(t)\, .
    \end{align}
The dynamics in the inertial frame $\S$ is determined by the following equation of motion 
    \begin{align}\label{dynamics_S}
	\ddot{\vec x}(t) = -\vec\nabla_{\vec x} V(t,\vec x(t))\, .
    \end{align}
However, for an observer in the co-moving frame $\S'$ the dynamics behaves differently due to fictitious forces 
arising in accelerated reference frames. We arrive at the equation of motion in $\S'$ by means of the coordinate 
transformation, Eq.~\eqref{vector_x}:
    \begin{align}\label{dynamics_S_tilde}
	\ddot{\vec x}'(t) = -\vec\nabla_{\vec x'} V'(t,\vec x'(t)) - \ddot{\vec \rho}(t)\, ,
    \end{align}
where we have introduced the potential $V'(t,\vec x') = V\big( t, \vec \rho(t) + \vec x'\big)$.
Hence, the dynamics in the co-moving frame $\S'$ depends on the second time derivative of $\vec\rho(t)$, 
which corresponds to the fictitious acceleration seen by an observer in the co-moving frame 
$\S'$   
    \begin{align}
	  \vec a_{\text{fi}}(t) = - \ddot{\vec \rho}(t)\,.
    \end{align}
The quadratic approximation of the external potential, Eq.~\eqref{harm_grav_potential}, yields the expression
    \begin{align}\label{acceleration'}
	\ddot{\vec x}'(t) \approx \vec g(\vec\rho(t)) - \Gamma(\vec\rho(t))\, \vec x'(t) + \vec a_{\text{fi}}(t)\, .
    \end{align}
%
%
\paragraph{Effective local acceleration and the time-dependence of the gradient}
When we define an effective, local acceleration
\begin{align}\label{g_tilde}
    \vec g'(t) := \vec g(t) + \vec a_{\text{fi}} (t)\,,
\end{align}
Eq.~\eqref{acceleration'} can be rewritten in quantities seen by an observer in the co-moving frame $\S'$
    \begin{align}\label{dynamics_S'}
	\ddot{\vec x}'(t) \approx \vec g'(\vec\rho(t)) - \Gamma(\vec\rho(t)) \,\vec x'(t).
    \end{align}
Here, 
 $	\vec g'(\vec\rho(t)) = \vec g(\vec\rho(t)) + \vec a_{\text{fi}} (\vec\rho(t)) $
is composed of the local acceleration $\vec g$ and the fictitious acceleration $\vec a_\text{fi}$, both evaluated at the time-dependent 
expansion point $\vec\rho(t)$.

In conclusion, the coordinate transformation from the inertial frame $\S$ to an arbitrary co-moving (but non-rotating) frame 
$\S'$ brings in the additional (fictitious) acceleration $\vec a_{\text{fi}}(\vec\rho(t))$. Moreover, the dynamics in $\S'$ is determined 
by an effective local acceleration and a gradient both time-dependent even when the external potential in the inertial frame is time-independent. \\

%
%
\subsubsection{Dynamics in rotating frames}
\label{ssec:}
In addition to origin translations, the general coordinate transformation in Eq.~\eqref{general_coord_transformation} also takes into account arbitrary time-dependent rotations 
of $\S''$ through the rotation matrix $\mathcal{R}(t)$. Hence, the second time derivative of this general coordinate transformation 
in combination with Eq.~\eqref{dynamics_S} gives 
    \begin{align}\label{}
	\ddot{\vec x}''(t) &= \mathcal{R}^\T(t) \left[-\vec\nabla_{\vec x''} V''(t,\vec x''(t)) - \ddot{\vec\rho}(t) - \ddot{\mathcal{R}}(t)\, \vec x''(t) - 2 \dot{\mathcal{R}}(t)\, \dot{\vec x}''(t)\right] \, ,
    \end{align}
where we have introduced the potential $V''(t,\vec x'') = V\big( t,\vec \rho(t) + \mathcal{R}(t)\, \vec x''\big)$.
This describes the dynamics of the atoms seen by an observer in the non-inertial frame $\S''$ where the first term comes from the external 
potential (which is already present in inertial frames), the second term is the acceleration arising due to 
the translational movement of $\S'$ and the last two terms 
correspond, respectively, to the centrifugal and the Coriolis accelerations.
Using the harmonic approximation of the external potential $V$, we finally get 
    \begin{align}\label{dynamics_arbitrary_rot}
	\ddot{\vec x}''(t) & \approx  \vec{g}''(\vec\rho(t)) - \Gamma''(\vec\rho(t)) \,\vec x''(t)  -  \mathcal{R}^\T(t) \left[\ddot{\mathcal{R}}(t)\, \vec x''(t) + 2 \dot{\mathcal{R}}(t)\, \dot{\vec x}''(t)\right]   
    \end{align}
where we have introduced the local acceleration $\vec g''(\vec\rho(t)) = \mathcal{R}^\T(t)\, \vec g'(\vec\rho(t))$ and the gradient 
$\Gamma''(\vec\rho(t)) = \mathcal{R}^\T(t)\,\Gamma(\vec\rho(t))\,\mathcal{R}(t)$ measured by an observer in $\S''$.
The corresponding Hamiltonian is given by 
    \begin{align}\label{Hamiltonian_rotating_frame}
	  H = \frac{{\vec p''}^2}{2m} - m\, {\vec{g}''}^\T\vec x''  + \frac{1}{2} m\, {\vec x''}^\T \Gamma'' \,\vec x'' - \vec\alpha^\T(t)\, (\vec x'' \times \vec p'')
    \end{align} 
where $\vec\alpha(t)$ denotes the time-dependent vector associated with the rotation matrix $\mathcal{R}(t)$.

%
%
\subsubsection{Hamiltonian in non-inertial frames}
\label{sssec:Equation of motion in non-inertial frames}
In the previous section we have seen that the dynamics in a non-inertial frame $\S''$ significantly differs from that in an inertial frame $\S$. We rewrite now the Hamiltonian in Eq.~\eqref{Hamiltonian_rotating_frame}, 
in a compact form and promote the phase-space variables to operators.
The general quadratic Hamiltonian
    \begin{align}\label{Hamiltonian_in_tilde_S_'}
	  \hat{H} = H(\hat{\vec \xi}) =  \mathcal{F}(t) + \vec{\mathcal{G}}^\T(t)\, \hat{\vec \xi} + \frac{1}{2} \hat{\vec \xi}^\T \, \mathcal{H}(t) \, \hat{\vec\xi}
    \end{align}
includes the first- and second-order coefficients $\vec{\mathcal{G}}(t)$ and $\mathcal{H}(t)$, respectively. 
Both get modified for the reference frame $\S''$ discussed above. According to Eqs.~\eqref{dynamics_arbitrary_rot} and \eqref{Hamiltonian_rotating_frame}, the second-order coefficient is now given by
    \begin{align}\label{Hamilt_matrix_gamma}
	  \mathcal{H}''(t) &=  \left(\begin{matrix}  m\, \Gamma''(\vec\rho (t)) & \vec\alpha(t) \cdot\vec\Lambda \\  -\vec\alpha(t) \cdot \vec\Lambda & \frac{1}{m} \mathds{1}_3 \end{matrix}\right)  \in \mathds{R}^{6 \otimes 6} .
    \end{align}
It is time dependent due to the gravity gradient $\Gamma''( \vec\rho (t))$ and the rotation $\mathcal{R}(t)$ characterized by the vector $\vec\alpha(t)$. In addition, we have introduced the matrix generator of an arbitrary time-dependent 
rotation, which is given by $\vec\alpha(t) \cdot \vec\Lambda$ (see Appendix~\ref{app:Rotation_group_SO(3)}). 
In turn, the first-order coefficient reads
    \begin{align}\label{linear_gravitational_contribution_tilde_S_'}
	 \vec{\mathcal{G}}''(t) &= \left(\begin{matrix}  - m \, \vec g''(\vec\rho (t)) \\ \vec 0 \end{matrix}\right)  \in \mathds{R}^6,
    \end{align}
where $\vec g''(\vec\rho(t)) = \mathcal{R}^\T(t) [\vec g(\vec\rho(t)) + \vec a_{\text{fi}}]$ comprises the local acceleration and the fictitious 
acceleration $\vec a_{\text{fi}}=-\ddot{\vec\rho}(t)$. \\

The dynamics, described in general by the inhomogeneous equation of motion (\ref{equ_of_motion}), cannot be calculated 
analytically for arbitrary time-dependent coefficients. Since the second-order coefficient, Eq.~\eqref{Hamilt_matrix_gamma}, is now time-dependent, we need to apply the perturbative approach introduced in Section~\ref{sssec:perturbative_approach}. 

The Hamiltonian presented in this section, as given by Eqs.~\eqref{Hamiltonian_in_tilde_S_'}-\eqref{linear_gravitational_contribution_tilde_S_'}, 
reduces to the standard description of interferometry in rotating frames~\cite{Dubetsky_2006,Audretsch_1994,Antoine_2003a,Antoine_2003} when the expansion 
point $\vec\rho(t)$ follows a circular motion determined by the same angular velocity as the coordinate axis, or 
more precisely when $\vec\rho(t)=\mathcal{R}(t)\, \vec\rho_0$. Thereby, one usually assumes that the lasers are at rest in the rotating 
frame $\S''$ but of course allows arbitrary time-dependent rotations of the frame and therefore considers the most general scenario.
However, below we pursue a different strategy which is based on calculations entirely done in the reference frame $\S'$ co-moving with the lasers (but non-rotating). Therefore, for the most general scenario we additionally have to include rotations of the lasers within 
$\S'$. In the next section we show in detail in which sense our strategy is more flexible and general. \\

%
%
%
\subsection{Description in co-moving frames (an alternative approach)}
\label{ssec:Interferometry_co-moving}
Our strategy to describe interferometers in arbitrary non-inertial frames is based on calculations entirely performed in the co-moving (but non-rotating)
frame $\S'$ following the interferometer device. 
In our approach we account for the rotation of the non-inertial reference frame $\S''$ 
by an (active) rotation (in the co-moving frame $\S'$) of the wave vectors $\vec k_n(t)$ for each laser pulse. 
For a discussion of rotations and in particular of 
passive and active transformations we refer to Appendix~\ref{app:Rotation_group_SO(3)}. 
Finally, we can adopt our results for the Sagnac effect obtained in inertial frames 
(see Section~\ref{sec_characteristic_quantities}) in order to get the phase shift and the visibility for interferometers in non-inertial frames. 
The main difference to the standard approach~\cite{Dubetsky_2006,Audretsch_1994,Antoine_2003a,Antoine_2003} is that we perform our 
calculations in the co-moving frame $\S'$. In this way we arrive 
at the same results as for an observer in the non-inertial frame $\S''$\footnote{Since phase shift and visibility are scalar observables and, therefore, frame-invariant.}. 
However, the big advantage is that our strategy is more general and flexible. 
Indeed, we can for example describe atomic fountains on Earth (which corresponds to the standard approach) 
but can also include easily independent rotations of the lasers 
and Earth's rotation (i.e. relevant for drop-tower experiments), or even more sophisticated orbital motions 
(for instance necessary for satellite missions).  

Since such complex motions within an external quadratic potential lead to inhomogeneous equations of motion 
including arbitrary time-dependent coefficients, a perturbative approach is necessary.
We have already applied our perturbative approach to interferometers in external potentials with 
constant, Eqs.~\eqref{tau_0}-\eqref{tau_perturbative_gamma_const}, as well as with time-dependent gradients, 
Eqs.~\eqref{H_gamma(t)_perturbative} and \eqref{phi_n_gamma(t)}. Consequently, it is clear how to calculate 
interferometer phases and the visibility when the approximated time-evolution matrix is given. For completeness, 
we recall the important steps now and apply in the following our approach to experiments in Earth's gravitational field.
%
%
\subsubsection{Phase shift and visibility in general}
In Section~\ref{sec:Compact_description:MZ_geometry} we have decomposed arbitrary interferometer geometries in elementary parts 
by generalized beam splitters. 
Since the generalized beam splitter consists of the phase $\Phi_n$ and the displacement vector $\vec\chi_n$, these are 
the crucial quantities to study but now in the context of non-inertial frames.

Our perturbative treatment for time-dependent Hamiltonians provides the time-evolution matrix as 
a series expansion, see Eq.~\eqref{series_tau}. Thus, we calculate the time-evolution matrix $\mathcal{T}$ 
by substituting the time-dependent coefficients in the recursive formula~\eqref{recursive_int} with the corresponding ones 
for the co-moving frame $\S'$, Eqs.~\eqref{Hamilt_matrix_gamma} and \eqref{linear_gravitational_contribution_tilde_S_'}.
This perturbative solution for $\mathcal{T}$ fully determines the displacement vector $\vec\chi_n$, see Eq.~\eqref{chi_n(t)}, and therefore 
the visibility~\eqref{visibility_interferometer} for a given interferometer geometry. 
Moreover, the displacement vector $\vec\chi_n$ and the expression for 
the generalized phase $\Phi_n$, Eq.~\eqref{phi_n_gamma(t)}, immediately yields the total phase shift 
$\Delta\Phi_\I$, Eq.~\eqref{total_ph_I}, seen by an observer in the co-moving frame $\S'$.
For the most general scenario discussed above (with lasers at rest in the rotating frame $\S''$)
we have to rotate the wave vectors for each laser pulse, which induces the following relation: 
\begin{align}\label{chi_n_0_non-inertial}
     \bar{\vec\chi}_n &=	\begin{pmatrix}   \vec 0  		   	\\  
						  \mathcal{R}(t_n)\, \hbar \vec k_0
				\end{pmatrix}
				 \in \mathds{R}^6\, .
\end{align}
However, this is formally equivalent to the calculations done for the Sagnac effect in inertial frames. The only difference now is 
that the external potential seen by an observer in the co-moving frame $\S'$ becomes time-dependent because of his motion within the external 
field (even when the field was initially time-independent). \\

In conclusion, we have decomposed the general coordinate transformation into elementary ones: (i) the transformation from 
the inertial frame $\S$ to the co-moving (but non-rotating) frame $\S'$, which brings in time-dependent local accelerations as well as time-dependent 
gradients (having an influence on the time-evolution matrix $\mathcal{T}$), and (ii) the rotation of $\S''$ (the frame co-rotating with the lasers) with respect to $\S'$. 
Therefore, we were able to introduce an alternative 
approach based on calculations entirely performed in the co-moving frame $\S'$ and where the rotation of $\S''$ is implicitly included in the rotation of the wave vectors; see Eq.~\eqref{chi_n_0_non-inertial}. 

Next, we illustrate the use of this alternative approach by applying it to the special case of experiments on Earth or in Earth's gravitational field.
%
%
%
\subsubsection{Experiments on Earth or in Earth's gravitational field}
\label{sssec:Experiments_on_Earth}

We already know from the Sagnac effect discussed in Section~\ref{sec_characteristic_quantities}  
how to implement arbitrary rotations of the lasers. Moreover, we have introduced in the preceding section all the tools 
necessary for a stepwise and straightforward calculation of the phase shift for 
arbitrary interferometer geometries in non-inertial frames. Now, we combine both and determine as an example the phase shift 
for a Mach-Zehnder interferometer seen by an observer following a circular motion 
(for instance an observer on Earth's surface). In particular, we provide the expressions for atomic fountains where the lasers are 
fixed on Earth (angular velocity of the lasers $\vec\Omega_k=\vec\Omega$) 
while the Earth is rotating with $\vec\Omega$. 
As an intermediate step we also provide the more general result for $\vec\Omega\neq\vec\Omega_k$, which for example corresponds 
to an observer in a satellite orbiting around the Earth (fixed altitude and constant $\vec\Omega$) 
while the lasers are rotating with a different angular velocity $\vec\Omega_k$.
  \begin{figure}[h]
    \centering
	\includegraphics[width=0.35\textwidth]{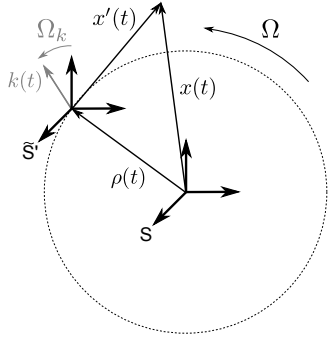}
    \caption[Circular motion]{Circular motion: The circular motion $\vec \rho(t) = \mathcal{R}_{\vec\Omega t} \,\vec \rho_0$ of a 
reference frame $\tilde{\S}'$ co-moving with the interferometer device
(for instance relevant for interferometers on Earth or as an example for orbiting satellites with constant altitude). 
The wave vector $\vec k(t)$ of each laser pulse is rotating with angular velocity $\vec\Omega_k$.}
    \label{fig:orbiting_frame}
  \end{figure}

The trajectory of such a co-moving frame $\tilde{\S}'$ (see \fig\ref{fig:orbiting_frame}) shall be given by the circular motion
    \begin{align}\label{rotation_rho_0}
	\vec \rho(t) = \mathcal{R}_{\vec\Omega t} \,\vec \rho_0\, ,
    \end{align}
where $\mathcal{R}_{\vec\Omega t}$ denotes the rotation matrix discussed in detail in Appendix~\ref{app:Rotation_group_SO(3)}. 
Further, we assume a uniform rotation around a time-independent axis with the rotation angle $\Omega t$.
%
%
%
%
%
%
\paragraph{Fictitious force becomes centrifugal force}
The dynamics in $\tilde{\S}'$ (and described by Eq.~\eqref{dynamics_S'} in general) depends on the local acceleration 
 $\vec g'(\mathcal{R}_{\vec\Omega t} \,\vec \rho_0) = \vec g(\mathcal{R}_{\vec\Omega t} \,\vec \rho_0) + \vec a_{\text{fi}} (\mathcal{R}_{\vec\Omega t} \,\vec \rho_0) $,
where the fictitious acceleration now becomes
    \begin{align}
	  \vec a_{\text{fi}} = - \ddot{\vec \rho}(t) = - \ddot{\mathcal{R}}_{\vec\Omega t} \vec \rho_0 - 2\dot{\mathcal{R}}_{\vec\Omega t} \dot{\vec \rho_0} - \mathcal{R}_{\vec\Omega t} \ddot{\vec \rho}_0\,.
    \end{align}
The first term in $\vec a_{\text{fi}}$ is governed by the second derivative of the rotation matrix, which brings in the 
centrifugal acceleration $\vec a_{\text{cf}}$, see Eq.~\eqref{second_derivative_R}. Since we assume a constant altitude $\vec\rho_0=const$, we arrive at
    \begin{align}\label{pseudo_acc}
	 \vec a_{\text{fi}} = -\ddot{\vec \rho}(t) = -\ddot{\mathcal{R}}_{\vec\Omega t} \vec \rho_0 = - \vec\Omega \times \left(\vec\Omega \times ( \mathcal{R}_{\vec\Omega t} \vec \rho_0 ) \right) .
    \end{align}
Next, we determine the effective local acceleration as well as the gradient for an external potential with spherical symmetry. As 
an example we choose the gravitational potential which corresponds to experiments on Earth or orbiting satellites within Earth's 
gravitational field (approximated by a spherically symmetric and time-independent configuration).
%
%
%
%
%
%
\paragraph{Effective local acceleration and the time-dependence of the gradient}
When we introduce the local gravitational acceleration for a spherically symmetric mass distribution, 
Eq.~\eqref{linear_gravitational_acceleration_2}, the circular motion of the expansion point~\eqref{rotation_rho_0} induces an effective rotation of the local 
gravitational acceleration
    \begin{align}\label{g_rotating}
	\vec g (\mathcal{R}_{\vec\Omega t}\, \vec\rho_0) &=  -\frac{G M_E [\mathcal{R}_{\vec\Omega t} \vec \rho_0]}{|\mathcal{R}_{\vec\Omega t} \vec \rho_0|^3} 
			      = \mathcal{R}_{\vec\Omega t}\, \vec g_0\, .
    \end{align}
In analogy, the gravity gradient, Eq.~\eqref{gravity_gradient_comp}, is also rotating with time
    \begin{align}\label{Gamma_rotating}
	\Gamma(\mathcal{R}_{\vec\Omega t}\, \vec\rho_0) = \mathcal{R}_{\vec\Omega t}\, \Gamma_0\, \mathcal{R}^\T_{\vec\Omega t}\, , 
    \end{align}
where $\Gamma_0 = \Gamma(\vec\rho_0)$ is the time-independent gravity gradient naturally chosen by an observer in the inertial frame~$\S$.

Finally, the dynamics in the co-moving frame $\tilde{\S}'$ is given by
    \begin{align}\label{dynamics_in_tilde_S_'}
	\ddot{\vec x}'(t) =  \mathcal{R}_{\vec\Omega t}\, \vec g'_0  -  \mathcal{R}_{\vec\Omega t}\, \Gamma_0 \, \mathcal{R}^{\T}_{\vec\Omega t}\, \vec x'(t) \,.
    \end{align}
Here, we used the general expression~\eqref{dynamics_S'} and 
$\vec g'_0 = \vec g_0 - \vec\Omega \times \left(\vec\Omega \times \vec \rho_0 \right)=\vec g_0 - \Omega^2 \vec \rho_0$ 
is the sum of the local gravitational acceleration and the centrifugal acceleration, both evaluated at $\vec\rho_0$. 
In the last term we have introduced the shorthand notation 
\begin{align}\label{eq:def_generator_rotation}
  \Omega\, \vec q \equiv (\vec\Omega \cdot \vec\Lambda)\, \vec q = \vec\Omega \times \vec q
\end{align}
for arbitrary vectors $\vec q \in \mathds{R}^3$, see also Appendix~\ref{app:Rotation_group_SO(3)} 
for the definition of $\vec\Lambda$ and further details. This notation allows a short and clear presentation of the following 
results that contain multiple vector products $\vec\Omega\times...(\vec\Omega\times...)\,$.

%
%
%

%
\paragraph{Time-evolution matrix}
In the following we assume that the gradient $\Gamma_0$ and the angular velocity $\vec\Omega$ are small parameters in order to apply the perturbative approach introduced in Section~\ref{sssec:perturbative_approach}. 
For the gravitational field of the Earth this is a very good approximation. If an even better approximation is needed (for 
instance to include anharmonicity effects), 
we have already mentioned the possibility of introducing a path-dependent potential in the generalized beam splitter 
in such a way that we expand the external potential for each branch of the interferometer separately \cite{Zeller_2015}. 

Hence, the rotation of the gradient, Eq.~\eqref{Gamma_rotating}, in combination with Eq.~\eqref{recursive_int} yields the 
series expansion (to first order in $\Gamma_0$ and $\Omega$)
\begin{align}\label{tau_expansion_co-moving_orbiting}
     \mathcal{T}(t) &= 	\begin{pmatrix}
			1\!-\! \frac{t^2}{2} \Gamma_0\! +\! \frac{t^3}{6} (\Gamma_0 \Omega\! -\! \Omega \Gamma_0)\! +\! \dots 
			  &\quad \frac{t}{m}\!  -\! \frac{t^3}{6 m}\Gamma _0 \! +\! \frac{t^4}{12 m}\left(\Gamma _0 \Omega\! -\! \Omega  \Gamma _0\right)\! +\! \dots \\[0.2cm]                 					
			    - m \Gamma _0  t\! + \!m \frac{t^2}{2}   \left( \Gamma _0 \Omega\! -\! \Omega \Gamma _0\right)\! +\! \dots
			  &\quad 1 \!-\! \frac{t^2}{2}\Gamma _0  \!+\! \frac{t^3}{3}\left( \Gamma _0 \Omega \!-\! \Omega \Gamma_0\right) \!+\! \dots
			\end{pmatrix} 
\end{align}
for the time-evolution matrix in the co-moving frame $\tilde{\S}'$. Note that the gradient $\Gamma_0\in\mathds{R}^{3\otimes3}$ 
as well as the generator of rotations $\Omega\in\mathds{R}^{3\otimes3}$, Eq.~\eqref{eq:def_generator_rotation}, are matrices.

%
\paragraph{Laser rotation}
So far we have not constrained the angular velocity of the lasers $\vec\Omega_k$. The corresponding displacement vector for 
the general scenario ($\vec\Omega_k\neq\vec\Omega$) is given via 
Eqs.~\eqref{chi_n(t)} and \eqref{chi_n_0_non-inertial} by
  \begin{align}\label{chi_n_orbiting}
     \vec\chi_n &= \mathcal{T}(t_0,t_n) \begin{pmatrix} \vec 0  		   	\\  
 				    \mathcal{R}_{\vec\Omega_k t_n} \hbar \vec k_0 
				\end{pmatrix}\, ,
  \end{align}
where $\mathcal{T}(t_0,t_n)$ denotes the perturbative solution~\eqref{tau_expansion_co-moving_orbiting}. 

\paragraph{General case}

The rotation of the effective local acceleration $\vec g' (t) = \mathcal{R}_{\vec \Omega t}\, \vec g'_0$ in combination with 
expression~\eqref{phi_n_gamma(t)}, valid for arbitrary time-dependent local accelerations and gradients, induces 
the following generalized phase 

\begin{align}\label{phi_n_different_omega}
 \Phi_n &= \varphi_n - \vec k^{\T}_0 \left\{ 
				      - \frac{1}{2}
				      + \frac{1}{6}\left[3 \Omega_k - \Omega_g \right](t_n-t_0) \right.\nonumber\\
		&\qquad\qquad\qquad		\left.      + \frac{1}{24} \left[\Gamma_0 + 4 \Omega_k \Omega_g   -  \Omega_g^2 - 6 \Omega_k^2 \right] (t_n-t_0)^2  \right. \nonumber\\
		&\qquad\qquad\qquad\qquad	      +  \frac{1}{120}  \left[ 3\Omega_{\Gamma} \Gamma_0  - 3 \Gamma_0 \Omega_{\Gamma}
                                      + \Gamma_0 \Omega_g + 5 \Omega_k \Omega_g^2  -   10 \Omega_k^2 \Omega_g  \right.\nonumber\\
		&\qquad\qquad\qquad\qquad\qquad	       - \Omega_g^3 - 5 \Omega_k \Gamma_0 
				                      \left. + 10 \Omega_k^3 \right] (t_n-t_0)^3 
		                                      \left.\vphantom{\frac{1}{2}}\right\} \vec g'_0\, (t_n-t_0)^2 \nonumber\\
		&\hphantom{= \varphi_n } + \mathcal{O}[(t_n-t_0)^6]\, .
\end{align}
Here, we distinguish between three different rotations: 
(i) the rotation of the gradient which corresponds to the generator $\Omega_\Gamma$ and comes in by 
the time-evolution matrix \eqref{tau_expansion_co-moving_orbiting}, (ii) the rotation of the effective local acceleration represented 
by $\Omega_g$, see Eq.~\eqref{g_rotating}, and (iii) the rotation of the wave vectors by
$\Omega_k$, see Eq.~\eqref{chi_n_orbiting}. As expected, for $\Omega_\Gamma=\Omega_g=0$ we reproduce the expressions for an 
observer in the inertial frame $\S$, in which only the lasers are rotating (Sagnac effect in Section~\ref{sec_characteristic_quantities}). 

Analogously to the calculations done for the inertial frame, we present  in Appendix~\ref{app:Phi_non-inertial} the \emph{general expression} for the total Mach-Zehnder 
phase shift $\Delta\Phi_\MZ$. 
The specific situation of atomic fountains 
on Earth, i.e.\ a Mach-Zehnder geometry with $\Omega=\Omega_k=\Omega_g=\Omega_\Gamma$, will be discussed next.
\paragraph{Particular case: atomic fountains on Earth}
For experiments on Earth the most natural reference frame is the one co-rotating with the Earth's surface. We have seen that our alternative 
approach presented above provides such special circular motions. Hence, the total Mach-Zehnder phase shift (measured in the rotating frame of the Earth or equivalently calculated by our 
alternative approach) is given by Eq.~\eqref{Phi_MZ_rotating} and $\Omega=\Omega_g=\Omega_\Gamma$. This 
result is a generalization for instance applicable to satellite or drop-tower experiments in which the lasers are rotating 
independently of the circular motion of the reference frame.
However, for atomic fountains, where the lasers are 
fixed on the Earth's surface ($\vec\Omega_k=\vec\Omega$), we get the rather simple expression
\begin{align}\label{phi_atomic_fountain}
 \Delta\Phi_\MZ &= \varphi_\MZ - \vec k^{\T}_0 \left[ 
				      -   1 
				      + 2  \Omega T
				      +   \frac{7}{12} \left(  \Gamma _0  -  3  \Omega^2  \right) T^2   \right. +   \frac{1}{2}  \left(2  \Omega ^3  -  \Omega  \Gamma _0 -   \Gamma _0 \Omega     \right)T^3 \nonumber\\
		&\qquad\qquad\qquad         +    \frac{31}{360} \left(3  \Omega ^2 \Gamma _0 + 3  \Gamma _0   \Omega ^2   
			              +   4  \Omega  \Gamma _0 \Omega - \Gamma _0^2  - 5  \Omega ^4 \right)T^4 \nonumber \\
		&\qquad\qquad\qquad\qquad	      + \frac{1}{20} \left(3 \Omega^5  -2 \Omega ^3 \Gamma_0  -  2 \Gamma_0 \Omega^3 - 3 \Omega \Gamma_0  \Omega^2 \right.  
			              - 3 \Omega ^2 \Gamma_0 \Omega \nonumber \\
		&\qquad\qquad\qquad\qquad\qquad\qquad        - 5 \Omega  \Gamma_0^2  - 5 \Gamma_0^2  \Omega  
		                               \left. +  13 \Gamma_0 \Omega \Gamma_0  \right)  T^5   + \dots  \left.\vphantom{\frac{1}{2}}\right] \vec g'_0\, T^2\nonumber\\
		&\hphantom{=}
		+\vec k^{\T}_0 \left[\vphantom{\frac{1}{2}}
			      \left( \Omega^2 -  \Gamma _0 \right) T^2 
			  +    \left(2  \Omega \Gamma_0 +    \Gamma_0   \Omega -   \Omega^3 \right)T^3 \phantom{\frac{a}{b}}\right.  \nonumber \\
		&\qquad\qquad         +   \frac{7}{12} \left( \Omega^4  -3 \Omega ^2  \Gamma_0 - \Gamma_0   \Omega^2 - 2 \Omega \Gamma_0  \Omega +  \Gamma_0^2 \right)T^4  \nonumber \\
		&\qquad\qquad\quad	  +   \frac{1}{4} \left(4  \Omega^3  \Gamma_0 +  \Gamma_0   \Omega^3 + 2 \Omega  \Gamma_0  \Omega^2 + 3  \Omega^2  \Gamma_0  \Omega -  2  \Omega \Gamma_0^2 \right. \nonumber \\
		&\qquad\qquad\quad\quad\quad\quad	   \left. - \Gamma_0^2  \Omega  - 2 \Gamma_0  \Omega  \Gamma_0 -  \Omega ^5 \right) T^5 
		+ \dots  \left.\vphantom{\frac{1}{2}}\right] \langle \vec x_0 \rangle \nonumber\\
		&\hphantom{=}
		+\vec k^{\T}_0 \left[ \vphantom{\frac{1}{2}}
			  -   2 \Omega T 
			  +   \left(3 \Omega^2 - \Gamma_0\right) T^2  +  \frac{7}{6} \left( \Omega \Gamma_0 + \Gamma_0 \Omega - 2 \Omega ^3  \right)T^3  \right. \nonumber \\
		&\qquad\qquad\quad	  +   \frac{1}{4}  \left(5 \Omega^4 -3 \Omega^2 \Gamma_0 - 4 \Omega \Gamma_0  \Omega   + \Gamma_0^2 - 3 \Gamma_0 \Omega ^2 \right) T^4 \nonumber\\
		&\qquad\qquad\quad\quad         -   \frac{31}{180} \left(3 \Omega ^5 + \Gamma_0  \Omega \Gamma_0 + \Gamma_0^2 \Omega + \Omega \Gamma_0^2 
			  -   2 \Gamma_0 \Omega^3 - 2 \Omega^3  \Gamma_0 
			  \right. \nonumber \\
		&\qquad\qquad\qquad\qquad\quad~  
			 \left.  -   3 \Omega^2 \Gamma_0 \Omega  -   3 \Omega \Gamma_0 \Omega^2 \right) T^5 
			  + \dots  \left.\vphantom{\frac{1}{2}}\right] T 
			 \left( \frac{\hbar \vec k_0}{2m} + \frac{\langle \vec p_0 \rangle}{m} \right).          
\end{align}
The comparison with the total phase shift measured in inertial frames, Eq.~\eqref{Delta_Phi_MZ}, shows that the additional rotation of the local acceleration
as well as the gradient modifies the rotation terms originally induced by the laser rotation seen in inertial frames. Finally, we 
arrive at analog contributions to the total phase shift but the prefactors become modified for observers in Earth's rotating frame (in comparison to an 
inertial frame where we only include laser rotations). 

Eq.~\eqref{phi_atomic_fountain} for the total phase shift in rotating frames can also be found in~\cite{Antoine_2003}\footnote{By comparison with \cite{Antoine_2003} 
we recognized that the last term in Eq.~(56) of the mentioned paper differs by a factor of two. 
Since Eq.~(56) depends on the expansion (54), which also shows this deviation in the last term, we expect a misprint in Eq.~(54) as 
well as in Eq.~(56).}. 
Moreover, references~\cite{Dubetsky_2006,Audretsch_1994,Antoine_2003a} study 
the same setup but either do not present explicit terms for the phase shift or expand their 
perturbation theory not to such high orders. However, all of them are using a Hamiltonian valid for rotating frames, 
which corresponds to the one introduced in Section~\ref{sssec:Equation of motion in non-inertial frames} and therefore 
inevitably includes the laser rotations (due to the rotation of the reference frame). 
In contrast, our alternative approach based on co-moving frames takes into account arbitrary translations as well as rotations and 
therefore provides a more general and flexible framework. To confirm this statement we showed for instance in 
Eq.~\eqref{Phi_MZ_rotating} one possible generalization of the well-known result, Eq.~\eqref{phi_atomic_fountain}. 
Additionally, the loss of visibility induced by non-inertial effects can be easily interpreted in the language of our approach 
by means of displacement vectors 
in phase space (see for example Eq.~\eqref{chi_n_orbiting}). In particular, we refer to Section~\ref{ssec:MZ_Displacement_vector} 
or Ref.~\cite{Roura_2014} for a discussion of the loss of contrast/visibility triggered by a non-vanishing displacement vector.

%
\section{Conclusion}
\label{sec_conclusion}
We have presented a straightforward and versatile method for determining the phase shift and the visibility for general interferometer geometries. 
We pursued a representation \!\!-free description in the context of light-pulse atom interferometry entirely based on operator algebra methods. 
In the course of modeling the internal and external dynamics of an effective two-level system in the presence of an external potential, we have taken into account 
local accelerations, gradients and rotations of the interferometer device.  
Assuming a two-level system kept the presentation as clear as possible while our formalism also allows more complicated level structures 
by exchanging the beam-splitter matrix by desired, suitable ones. 
In particular, internal level structures reducible to effective two-level systems, for instance Raman or Bragg transitions, are covered 
in an analytically exact way. For double diffraction schemes~\cite{Leveque_2009a,Giese_2013} most of our results simply needs to be 
modified by doubling the effective momentum transfer. 

We have analyzed in detail the time evolution in a general quadratic potential 
(e.g. induced by external magnetic or gravitational background fields) and described the internal dynamics by a simple 
beam-splitter matrix. In doing so, we quantized the internal as well as the external degrees of freedom. 
Since our interaction model is based on internal states associated with different internal energies, 
the effect of the internal atomic structure is also included. (However, the corresponding Rabi oscillations 
are described in the interaction picture where the frame is rotating with the laser frequency and therefore in our results 
this effect appears only after a back transformation.) 
For instance such internal energy splittings are the main contribution to the phase shift 
in microwave atomic clocks where the momentum transfer by the photon can be usually neglected. In this sense our approach 
provides a general framework combining both light-pulse atom interferometers and optical atomic clocks (with momentum transfer). 
In particular, the consecutive propagation in the ``interaction zone'' and the 
``free-propagation zone'' led to a time-dependent generalized beam-splitter matrix. 
By a sequence of these generalized beam splitters, we are able to construct any combination of interferometer pulse sequences. 

We described interferometers in an universal manner where the total interferometer phase shift and the visibility include 
the full information of the special interferometer geometry. By calculating these characteristic quantities, we arrived at a 
straightforward method for obtaining the ground-state detection probability for arbitrary loop geometries. We remark that 
instead of integrating over the external degrees of freedom after the interferometer sequence 
one can directly detect a fringe pattern in the spatial density profile at the exit ports (as long as a 
displacement of the final atomic wave packets is present, which can be achieved by a suitable timing of the pulses or naturally comes in by 
rotations or gradients). Hence, one can extract information 
on the visibility and the phase shift even for a single shot~\cite{Muentinga_2013,Sugarbaker_2013,Miller_2005}.

The Mach-Zehnder pulse sequence served as an example for our operator approach and established the connection to results already known in the literature, 
but also highlighted important features as the visibility (and its connection to relative displacements in phase space) 
or the phase shift as a consequence of commutation relations in the quantized dynamics. 
As a second example, we have studied the Butterfly geometry, where the phase shift is (to first order) directly sensitive to the gradient of the external quadratic potential. Moreover, 
in order to take into account corrections beyond our quadratic approximation, we have already mentioned the possibility 
of introducing a path-dependent potential for the generalized beam splitter 
in such a way that we locally expand the external potential for each branch of the interferometer separately \cite{Zeller_2015}. Hence, 
anharmonicity effects are included by a quadratic approximation but locally for each branch, 
which means within the size of the atomic wave packet.

Finally, we focused on interferometers embedded in non-inertial reference frames necessary not only for the 
description of state-of-the-art space missions (taking into account arbitrary accelerations and trajectories) 
but also for experiments in Earth's rotating frame. We have shown that the effects due to non-inertial reference frames 
can be easily interpreted when the general coordinate transformation is decomposed into elementary ones. 
In addition, we have introduced a simple way of dealing with such general situations in the context of co-moving (but non-rotating) frames.
Since the effects of external potentials and inertial forces were mainly studied in the context of the 
phase shift~\cite{Peters_2001,Audretsch_1994,Wolf_1999,Borde_2001,Borde_2002}, we have also emphasized the loss of contrast 
inevitably arising for non-vanishing displacements (in phase space)~\cite{Roura_2014} induced by these effects.

In conclusion, we have developed a compact and versatile formalism for the description of atom interferometers in inertial 
as well as in non-inertial frames well-suited for the next generation of high-precision measurements, particularly microgravity 
experiments performed in drop-tower facilities, sounding rockets or dedicated satellite missions.  
\newpage
%
\section*{Acknowledgments}
%
E.K. is grateful to J. Bolte, J. E. Sträng and A. Wolf for many fruitful
discussions during the early stages of this work.
S.K. and A.R. were supported by the German Space Agency (DLR) with 
funds provided by the Federal Ministry of Economics and Technology (BMWi) under grant number 50WM1136.
Moreover, W.P.S. thanks Texas A\&M University for a Texas A\&M Institute of Advanced Studies (TIAS) Faculty Fellowship.
%
%
%
%
\appendix
%
\renewcommand*{\thesection}{\Alph{section}}
%
%
%
%
\section{External potential}
\label{app:gravitational_potential}
Suppose we place a matter-wave interferometer in a satellite around the Earth (as e.g. proposed in the STE-QUEST mission of 
the European Space Agency~\cite{ste-quest_general}), the corresponding external potential energy for an atom with mass $m$ would (in first approximation) be given by the following gravitational one
    \begin{align}
      V_\g(\vec x) = -m\frac{GM_\E}{r}
    \end{align}
with the absolute value of the position vector $r= |\vec x|$, the Newtonian constant of gravity $G$  and the mass of the Earth 
$M_\E$. Here, the origin of the inertial frame is the Earth's center. When we additionally denote the trajectory 
of the satellite by $\vec\rho(t)$, we can expand the Earth's potential 
around $\vec\rho(t)$ and arrive for sufficiently small regions around the trajectory at 
	\begin{align}\label{V_g_approx}
	      V_\g(\vec x )  \approx	&~  V_\g(\vec\rho(t)) - m\, \vec g^\T(t)\left[\vec x - \vec\rho(t)\right]  + \frac{1}{2} m \left[\vec x - \vec\rho(t)\right]^\T \Gamma(t)\left[\vec x -  \vec\rho(t)\right].
	\end{align}
At this point, we have introduced the time-dependent local acceleration $\vec g(t)$ and the gradient matrix $\Gamma(t)$. 
In this harmonic approximation, quite sufficient for state-of-the-art experiments, the local acceleration 
as well as the gradient are primarily responsible for the effects on the external evolution of the matter wave. 
%
%
%
%
%
\paragraph{Local acceleration}
\label{app:uniform_grav_acc}
The Taylor series (\ref{V_g_approx}) includes the local acceleration which is the negative first derivative 
of the potential divided by $m$ and evaluated at $\vec\rho(t)$
		\begin{align}\label{linear_gravitational_acceleration_1}
		      \vec g (t) =  -\frac{\vec\nabla_{\vec x} V_\g(\vec x)}{m} \bigg\vert_{\vec x = \vec \rho(t)}\, .
		\end{align}
Via the absolute value of the local acceleration $g(t) = |\vec g (t)| =  G M_\E / |\vec \rho(t)|^2$
we find
		\begin{align}\label{linear_gravitational_acceleration_2}
		      \vec g (t) =  -\frac{G M_\E \vec \rho(t)}{|\vec \rho(t)|^3} = -\, g(t)\, \vec e_{\rho}\, .
		\end{align}
Hence, the local acceleration always points in the opposite direction as the trajectory $\vec \rho(t)$ where $\vec e_{\rho} = \vec \rho(t)/|\vec \rho(t)|$. 
%
%
%
%
%
\paragraph{Gradient}
\label{app:gradient}
The gradient is proportional to the second derivative of the potential evaluated at 
$\vec \rho(t)$
		\begin{align}\label{gravity_gradient_comp}
		      \Gamma_{ik}(t) = \frac{1}{m} \frac{\partial^2 V_\g(\vec x)}{\partial x_i \partial x_k} \bigg|_{\vec x = \vec \rho(t)}  = G M_\E \frac{\delta_{ik} |\vec \rho(t)|^2-3\rho_\i(t)\rho_k(t)}{|\vec \rho(t)|^5}.
		\end{align}
Moreover, it is an indefinite quadratic form with two positive eigenvalues 
\newline%
$G M_\E / |\vec\rho(t)|^3$ that correspond to the eigenvectors 
perpendicular to $\vec\rho(t)$ and one negative eigenvalue $-2 G M_\E / |\vec\rho(t)|^3$ for the eigenvector parallel to $\vec\rho(t)$.
%
\section{Symplectic group}
\label{app:math_excursus_Sympl_group}
In order to provide a compact description of the dynamics in matter-wave interferometers for general quadratic Hamiltonians,
we study the underlying symplectic structure, especially the standard symplectic group, and derive some useful properties. 
A more detailed analysis of group theory can be found in Ref.~\cite{Hall_2003} or~\cite{Arvind_1995}. \\

Besides the real orthogonal group and the complex unitary group, the symplectic group is one of the three major families of classical Lie groups. In both classical  and quantum mechanics
the real symplectic group plays an important role. The canonical formalism of classical dynamics as well as its counterpart in quantum mechanics naturally brings in the
symplectic structure. \\

We start with the definition of the symplectic matrix and will present some useful properties. 
\subsection{Definition of the symplectic matrix}
A matrix $S$ is a symplectic matrix, if and only if $S$ satisfies
      \begin{align}\label{def_sympl}
	    S^\T \mathcal{J} S = \mathcal{J}.
      \end{align}
Here, $\mathcal{J}$ is the anti-symmetric block matrix
  \begin{align}\label{symplectic_form}
    \mathcal{J} = \left(\begin{smallmatrix}  0 & \mathds{1}_\N \\  -\mathds{1}_\N & 0 \end{smallmatrix}\right) \in \mathds{R}^{2\N \otimes 2\N}
  \end{align}
which is generally referred to as the symplectic form; each block is of dimension $\text{N} \times \text{N}$. 
The matrices $S$ are elements of the symplectic group $\text{Sp}(\N)$. \\

For our purposes, the most important property is the anti-symmetry and the fact that its transpose 
is its inverse
  \begin{align}\label{property_J}
    \mathcal{J}^\T = -\mathcal{J} = \mathcal{J}^{-1}.
  \end{align}
Next, we show that the set of symplectic matrices $S$ indeed forms a group as claimed above.
%
%
%
%
\subsection{Symplectic matrices form a group}
The matrices $S$ satisfy the group axioms with the matrix multiplication as group operation:  
\begin{enumerate}
  \item[1.]The first axiom of closure can be easily proven by successively applying the definition of the symplectic matrix 
  \begin{align}
      (S_1 S_2)^\T \mathcal{J} (S_1 S_2) = S_2^\T S_1^\T \mathcal{J} S_1 S_2 = \mathcal{J}.
  \end{align}
  \item[2.]The second axiom (associativity) is immediately fulfilled by the associativity of matrices since the symplectic matrices form a 
	    sub-group of all matrices.
  \item[3.]The identity element is the identity block matrix $\mathds{1}_{2\N} \in \mathds{R}^{2\N \otimes 2\N}$ with
  \begin{align}      
      \left( \begin{matrix} \mathds{1}_\N & 0 \\ 0 & \mathds{1}_\N \end{matrix} \right)^\T \mathcal{J} \left(\begin{matrix} \mathds{1}_\N & 0 \\ 0 & \mathds{1}_\N \end{matrix}\right) = \mathcal{J}
  \end{align}
  \item[  ]and $\mathds{1}_{2\N} S = S \in \text{Sp}(\N)$.
  \item[4.]The last axiom leads to an explicit expression for the inverse
matrix $S^{-1}$. The definition of the symplectic matrix yields
  \begin{align}      
      (S^\T  \mathcal{J} S) \mathcal{J}^{-1} = \mathcal{J} \mathcal{J}^{-1} & \Leftrightarrow ~ S^\T (\mathcal{J} S \mathcal{J}^\T) = \mathds{1}_{2\N} ~ \nonumber \\
				    & \Leftrightarrow ~ (S^\T)^{-1} = \mathcal{J} S \mathcal{J}^\T.
  \end{align}
  \item[  ]Applying the property (\ref{property_J}) once again, we arrive at the inverse
  \begin{align}\label{S_inverse}      
       (S^\T)^{-1} = (S^{-1})^\T = (\mathcal{J} S^\T \mathcal{J}^\T)^\T ~ \Leftrightarrow ~ S^{-1} = \mathcal{J} S^\T \mathcal{J}^\T.
  \end{align}
\end{enumerate}

So far, we have seen that the symplectic matrices form a group. We conclude by showing that the 
transpose $S^\T$ is also an element of the symplectic group. 
%
%
%
%
\subsection{The transpose is symplectic as well}
%
%
If a given matrix $S$ is symplectic, the transpose $S^\T$ is symplectic as well. We proof this statement by recalling 
the definition of the symplectic matrix and take the inverse of both sides
    \begin{align}\label{S_trans_sympl}
	  S \in \text{Sp}(\N) ~ &\Leftrightarrow ~ S^\T \mathcal{J} S = \mathcal{J} \nonumber \\
				&\Leftrightarrow ~ S^{-1} \mathcal{J}^{-1} (S^\T)^{-1}	= \mathcal{J}^{-1}.
    \end{align}
Then, we use $\mathcal{J}^{-1} = -\mathcal{J}$ and arrive at
    \begin{align}
	S \in \text{Sp}(\N) ~   &\Leftrightarrow ~ S^{-1} \mathcal{J} (S^\T)^{-1}	= \mathcal{J} \nonumber \\
	\hphantom{S \in S} 	&\Leftrightarrow ~ \mathcal{J} = S \mathcal{J} S^\T \nonumber \\
				&\Leftrightarrow ~ S^\T \in \text{Sp}(\N)\,.
    \end{align}
%
%
%
%
\section{The time-evolution matrix $\mathcal{T}$ is a symplectic matrix} 
\label{app:tau_is_sympl}
First we define the matrix $Z(t,t_0) := \mathcal{T}^\T(t,t_0)\, \mathcal{J}\, \mathcal{T}(t,t_0)$ in order to 
prove the symplectic property of the time-evolution matrix $\mathcal{T}$. When we take the time derivative and recall that 
the time-evolution matrix satisfies the homogeneous part of the Heisenberg equation of motion \eqref{equ_of_motion}, we obtain 
  \begin{align}
      \frac{d Z}{dt}	
			&= (\mathcal{J} \mathcal{H}(t) \mathcal{T})^\T \mathcal{J} \mathcal{T} + \mathcal{T}^\T \mathcal{J} (\mathcal{J} \mathcal{H}(t) \mathcal{T})\,.
  \end{align}
Transposing the first bracket and recalling the relation (\ref{property_J}) yields
  \begin{align}
      \frac{d Z}{dt}	&= \mathcal{T}^\T \mathcal{H}^\T(t) \mathcal{J}^\T \mathcal{J} \mathcal{T} + \mathcal{T}^\T \mathcal{J}^2 \mathcal{H}(t) \mathcal{T} = 0\,.
  \end{align}
Hence, the matrix $Z(t,t_0)$ is constant in time. With the initial condition $\mathcal{T}(t_0,t_0) = \mathds{1}_6$ at hand we find 
  \begin{align}
      Z(t_0,t_0)	&= \mathcal{T}^\T(t_0,t_0)\, \mathcal{J}\, \mathcal{T}(t_0,t_0) = \mathcal{J}.
  \end{align}
Since $Z(t,t_0)$ is constant in time, we arrive at
  \begin{align}\label{T_trans_sympl}
      Z(t,t_0)	&= \mathcal{T}^\T(t,t_0)\, \mathcal{J}\, \mathcal{T}(t,t_0) = \mathcal{J}     \qquad     \forall ~ t \geq t_0\,.
  \end{align}
Indeed, this is the defining equation of the symplectic matrix (\ref{def_sympl}). Thus, the time-evolution matrix is an element of the symplectic group
$\text{Sp}(3)$, which means that the symplectic form is invariant under time transformations $\mathcal{T}$.
%
%
%
%
%
\section{Displacement operator}
\label{app:displacement_operator}
In this appendix, we first introduce the displacement operator and recall some useful properties and identities. 
At the end, we transform the displacement operator into the Heisenberg picture, which is necessary for the expression of the 
generalized beam-splitter matrix~\eqref{BSM_H_2}.
%
\subsection{Definition}
%
%
%
Let $\vec\chi_x,$ $\vec\chi_p \in \mathbb{R}^\N$ and $\hat{\vec{x}},$ $\hat{\vec{p}}$ be operators on $\text{L}^2 (\mathbb{R^\N})$.
Then the definition of the displacement operator reads 
    \begin{equation}
                  \hat D(\vec\chi_x,\vec\chi_p) \Def \e^{\frac{\i}{\hbar}[\vec\chi_p\hat{\vec x}-\vec\chi_x \hat{\vec p}]}.
    \end{equation}
For convenience, we rewrite the displacement operator in terms of the 2N-dimensional displacement vector
  \begin{equation}\label{eq:comp_chi}
      \vec\chi \Def \left(\begin{array}{c} \vec\chi_x \\ \vec\chi_p \end{array}\right) \in \mathbb{R}^{\text{2N}} 
  \end{equation}
and the 2N-dimensional phase-space vector operator
  \begin{equation}\label{eq:comp_xi}
        \hat{\vec\xi} \Def \left(\begin{array}{c} \hat{\vec x} \\ \hat{\vec p} \end{array}\right) \in L^2(\mathbb{R}^{\text{2N}})\, ,  
  \end{equation}
and arrive at the compact expression
 \begin{equation}\label{displacement_operator_conv}
                  \hat D(\vec\chi) = \e^{-\frac{\i}{\hbar}\vec\chi^\T \! \mathcal{J} \hat{\vec \xi}}\, ,
    \end{equation}
where $\mathcal{J}$ is the symplectic form \eqref{symplectic_form}. \\

In the following, we establish some properties and identities 
useful in the context of interferometer sequences in the presence of external quadratic potentials. 
%
%
%
\subsection{Properties and identities}
\label{Properties_and_identities}
%
\paragraph{Displacement} 
As its name already suggests, the displacement operator leads to a displacement 
in phase space
    \begin{align}
	\hat D^\dagger(\vec\chi) \,  \hat{\vec\xi} \, \hat D(\vec\chi) = \hat{\vec\xi} + \vec\chi\, .
    \end{align} 
%
\paragraph{Composition rule} 
Two displacements in sequence are more then just their sum; an additional phase is accumulated
    \begin{align}\label{composition_rule}
	  \hat D(\vec\chi_1) \, \hat D(\vec\chi_2) = \hat D(\vec\chi_1 + \vec\chi_2) ~ \e^{-\frac{\i}{2\hbar}\vec\chi_1^\T  \mathcal{J} \vec\chi_2}\, .
    \end{align}
Indeed, the Baker-Campbell-Hausdorff formula leads to a correction arising from the non-commuting canonical operators $\hat{\vec x}$ 
and  $\hat{\vec p}$ \cite{Wilcox_1967}. 
%
\paragraph{``Sandwich rule''}
Applying the composition rule twice yields
    \begin{align}\label{sandwich_rule}
	\hat D(\vec\chi) ~ \hat D(\vec\chi_0) ~ \hat D(\vec\chi)= \hat D(\vec\chi_0 + 2\vec\chi)
    \end{align} \\
and with $\hat D^\dagger(\vec\chi) = \hat D(-\vec\chi) $ we obtain

    \begin{equation}
	\hat D^\dagger(\vec\chi) ~ \hat D(\vec\chi_0) ~ \hat D(\vec\chi)= \hat D(\vec\chi_0) ~ \e^{+\frac{\i}{\hbar}\vec\chi^\T  \mathcal{J} ~ \vec\chi_0}\, .
    \end{equation}
%
%
\paragraph{Glauber formula}
It can be shown~\cite{Ferraro_2005} that the displacement operator fulfills the identity
    \begin{align}\label{delta}
	\trace{\!\hat D(\vec\chi)} &= (2\pi\hbar)^\N ~ \delta^{(2\N)} (\vec\chi)\, ,
    \end{align}
where the $2\N$-dimensional delta function is given by
 \begin{align}\label{eq:delta_symplectic}
    \delta^{(2\N)}(\vec\chi) & =   \frac{1}{(2\pi\hbar)^{2\N}} \!\!\int\limits_{\mathbb{R}^{2\N}}  \dif^{2\N} \xi ~ \e^{\frac{\i}{\hbar} \vec\chi^\T \mathcal{J} \vec\xi}.
 \end{align}
Moreover, using this expression we can easily prove the completeness of the displacement operator $\hat D (\vec\chi)$ in the sense that 
all operators $\hat O = \hat O(\hat{\vec x},\hat{\vec p})$ acting on states in $L^2(\mathbb{R}^{2\N})$ 
can be decomposed according to the so-called Glauber formula
 \begin{align}\label{glauber_formula}
   \hat O 	&= \frac{1}{(2\pi\hbar)^\N} \!\! \int\limits_{\mathbb{R}^{2\N}}  \dif^{2\N} \chi  \trace \{\hat O \hat D (\vec\chi)\}\, \hat D ^\dagger (\vec\chi)\,.
 \end{align}
The Glauber formula will be useful in the context of the characteristic function in 
Appendix~\ref{app:characteristic_function}.
%
%
\subsection{Displacement operator in the Heisenberg picture}
\label{app:Displacement operator in Heisenberg picture}
The relation $\hat{A}^\dagger \e^{\i \hat{\vec \xi}} \hat{A} = \e^{\i \hat{A}^\dagger\hat{\vec \xi}\hat{A}}$ for an unitary operator $\hat{A}$ allows us to write the displacement operator in the following way
    \begin{align}
	    \hat{D}_{\H,n}  	&\Def \hat{U}_\e^\dagger(t_n,t_0)\,  \hat{D}(\bar{\vec\chi}_n)\,   \hat{U}_\e(t_n,t_0) \nonumber \\
				&= \e^{-\frac{\i}{\hbar}(\bar{\vec\chi}_n)^\T  \mathcal{J} \hat{\vec \xi}_\H}\, ,
    \end{align}
where 
\begin{align}
  \hat{\vec \xi}_\H &= \hat{U}_\e^\dagger(t_n,t_0)\,  \hat{\vec \xi}\,  \hat{U}_\e(t_n,t_0) 
\end{align}
is the phase-space operator in the Heisenberg picture. 
Insertion of the general solution (\ref{general_solution_of_Heisenberg}) of the Heisenberg equations of motion yields 
    \begin{align}\label{displacement_op}
	    \hat{D}_{\H,n}  	&= \e^{-\frac{\i}{\hbar}(\bar{\vec\chi}_n)^\T  \mathcal{J} [\mathcal{T}(t_n,t_0) \hat{\vec\xi} + \!\! \int\limits_{t_0}^{t_n} \dif t' \, \mathcal{T}(t_n,t') \, \mathcal{J} \, \vec{\mathcal{G}}(t')]}.
    \end{align}
Since $\mathcal{T}$ is symplectic, we obtain via Eq.~\eqref{def_sympl} the following relation
    \begin{align}\label{change_relation}
	\mathcal{J}\,  \mathcal{T}(t_n,t_0) &=  (\mathcal{T}^{-1}(t_n,t_0))^\T \mathcal{J} =  \mathcal{T}^\T(t_0,t_n)\, \mathcal{J}.
    \end{align}
Now, we can shift the time-dependence of the phase-space operator  $\hat{\vec \xi}_\H$ to the displacement vector $\bar{\vec\chi}$. 
In this way, the displacement vector $\vec\chi_{n} =  \mathcal{T}(t_0,t_n)\,  \bar{\vec\chi}_n$ 
becomes time-dependent and the displacement operator can be written as
    \begin{align}\label{displacement_op_neu}
	    \hat{D}_{\H,n}  	&= \e^{-\frac{\i}{\hbar} [\vec\chi^\T_{n}(t_0,t_n)  \mathcal{J} \hat{\vec\xi}_0 - \!\! \int\limits_{t_0}^{t_n} \dif t'  \vec\chi^\T_{n}(t',t_n)\, \vec{\mathcal{G}}(t')]}.
    \end{align}
Please keep in mind, the arguments of the displacement vector 
\begin{align}
  \vec\chi_{n} &= \vec\chi_{n}(t_0,t_n) =  \mathcal{T}(t_0,t_n)\,  \bar{\vec\chi}_n
\end{align} 
are exchanged with 
respect to the phase-space operator in the Heisenberg picture
\begin{align}
  \hat{\vec \xi}_\H = \mathcal{T}(t_n,t_0)\, \hat{\vec \xi}.
\end{align}
This allows a description in which the displacement vectors include 
the time evolution and the determination of the beam-splitter matrices is greatly simplified. 
%
%
%
\section{Characteristic function and the Glauber formula}
\label{app:characteristic_function}
The characteristic function of an operator $\hat O$ is defined as \cite{Ferraro_2005}
 \begin{align}\label{eq:def_characteristic_f}
   \eta[\hat O](\vec\chi) & \Def   \trace\!\{\hat O \hat D (\vec\chi)\}\,. 
 \end{align}
Hence, we can rewrite the Glauber formula \eqref{glauber_formula} according to 
 \begin{align}
   \hat O &= \frac{1}{(2\pi\hbar)^\N} \!\! \int\limits_{\mathbb{R}^{2\N}}  \dif^{2\N} \chi ~ \eta[\hat O ](\vec\chi)\, \hat D^\dagger (\vec\chi) \, . 
 \end{align}
For later purposes we briefly discuss the trace rule in phase space based on the characteristic function. We start from the relation
 \begin{align}\label{trace_rule}
   \trace\{\hat O_1 \hat O_2\}  &= \frac{1}{(2\pi\hbar)^{2\N}} \!\! \int\limits_{\mathbb{R}^{2\N}}  \dif^{2\N} \chi_1 \!\! \int\limits_{\mathbb{R}^{2\N}}  \dif^{2\N} \chi_2 ~ \eta[\hat O_1](\vec\chi_1) \nonumber \\
				&\qquad \times\eta[\hat O_2 ](\vec\chi_2)\, \trace\! \{\hat D^\dagger (\vec\chi_1) \hat D^\dagger (\vec\chi_2)\}\, . 
 \end{align}
Using the composition rule (\ref{composition_rule}) and the identity (\ref{delta}) we find
 \begin{align}
   \trace\! \{\hat D^\dagger (\vec\chi_1) \, \hat D^\dagger (\vec\chi_2)\}  &= (2\pi\hbar)^{\N} \delta^{(2\N)}(\vec\chi_1+\vec\chi_2)\, ,  
 \end{align}
which gives rise to the trace rule in phase space based on the characteristic function
 \begin{align}\label{eq:trace_rule_ch_f}
       \trace\{\hat O_1 \hat O_2\}  &= \frac{1}{(2\pi\hbar)^{\N}} \!\! \int\limits_{\mathbb{R}^{2\N}}  \dif^{2\N} \chi ~ \eta[\hat O_1](\vec\chi) ~ \eta[\hat O_2](-\vec\chi)\,. 
 \end{align}
The next appendix is devoted to the Wigner function and its relation to the characteristic function.
%
%
%
%
\section{Wigner function and the corresponding characteristic function}
\label{app:Wigner_func} 
First, we introduce the Wigner function in terms of the characteristic function and show that this expression coincides with the 
standard definition of the Wigner function in the literature~\cite{Quantum_Optics_in_Phase_Space_Schleich_2001}. We then consider 
the Wigner function and the characteristic function for arbitrary Gaussian states.
\subsection{Definition of the Wigner function}
The Wigner function of an operator $\hat O$ is defined as the symplectic Fourier transform of the characteristic function 
$\eta[\hat O](\vec\chi)$ via
 \begin{align}\label{eq:def_Wigner_function}
       \text{W}[\hat O](\vec\xi)  &\Def \frac{1}{(2\pi\hbar)^{2\N}} \!\! \int\limits_{\mathbb{R}^{2\N}}  \dif^{2\N} \chi ~\e^{\frac{i}{\hbar}\vec\chi^\T \mathcal{J}\vec\xi} ~ \eta[\hat O ](\vec\chi)\,.  
 \end{align}
Taking into account the symplectic integral representation of the $2\N$-dimensional delta function~\eqref{eq:delta_symplectic},
the characteristic function is then simply given as inverse symplectic Fourier transform of the Wigner function
 \begin{align}
        \eta[\hat O](\vec\chi) &= \!\!\int\limits_{\mathbb{R}^{2\N}}  \dif^{2\N} \xi ~ \e^{-\frac{\i}{\hbar} \vec\chi^\T \mathcal{J} \vec\xi} ~ \text{W}[\hat O](\vec\xi)\, .
 \end{align}
When we insert this last expression into Eq.~\eqref{eq:trace_rule_ch_f} and apply the relation~\eqref{eq:delta_symplectic}, we obtain 
the trace rule in terms of the Wigner functions corresponding to the operators $\hat{O}_1$ and $\hat{O}_2$
 \begin{align}
        \trace\{\hat O_1 \hat O_2\} &= (2\pi\hbar)^{\N}  \!\!\int\limits_{\mathbb{R}^{2\N}}  \dif^{2\N} \xi ~ \text{W}[\hat O_1](\vec\xi) ~ \text{W}[\hat O_2](\vec\xi)\, .
 \end{align}
Next, we show that the definition of the Wigner function~\eqref{eq:def_Wigner_function} reduces to the well-known 
expression~\cite{Quantum_Optics_in_Phase_Space_Schleich_2001}
 \begin{align}\label{eq:Wigner_f_standard}
       \text{W}[\hat O](\vec x,\vec p)  &= \frac{1}{(2\pi\hbar)^{\N}} \!\! \int\limits_{\mathbb{R}^{\N}}  \dif^{\N} \lambda ~\e^{-\frac{i}{\hbar}\vec p \vec\lambda} \bra{\vec{x}+\frac{\vec\lambda}{2}} \hat O \ket{\vec{x}-\frac{\vec\lambda}{2}}\, ,
  \end{align}
where the $\N$-dimensional integration is taken over $\vec\lambda \in \mathbb{R}^\N$. For this purpose we substitute the definition 
of the characteristic function~\eqref{eq:def_characteristic_f} into~\eqref{eq:def_Wigner_function} and rewrite the $2\N$-dimensional 
phase-space vectors in terms of their corresponding $\N$-dimensional components, see Eq.~\eqref{eq:comp_chi} and~\eqref{eq:comp_xi},
 \begin{align}\label{Wigner_proof}
     \text{W}[\hat O](\vec x,\vec p)	 &=	\frac{1}{(2\pi\hbar)^{2\N}} \!\! \int\limits_{\mathbb{R}^{\N}}  \dif^{\N} \chi_x \!\! \int\limits_{\mathbb{R}^{\N}}  \dif^{\N} \chi_p  ~ \e^{-\frac{i}{\hbar}\left[\vec\chi_p \vec x - \vec\chi_x \vec p\right]} 
				 \trace \{\hat O ~ \e^{\frac{i}{\hbar}\left[\vec\chi_p \hat{\vec x} - \vec\chi_x \hat{\vec p}\right]}\}\, .
 \end{align}
We then evaluate the trace and find
 \begin{align}
        \trace \{\hat O ~ \e^{\frac{i}{\hbar}\left[\vec\chi_p \hat{\vec x} - \vec\chi_x \hat{\vec p}\right]}\}  &=  \frac{\e^{-\frac{i}{2\hbar}  \vec\chi_x \vec\chi_p}}{(2\pi\hbar)^{\N}}  \!\! \iiint\limits_{\mathbb{R}^{3\N}}  \dif^{\N} x' ~~\dif^{\N} x'' ~~\dif^{\N} p' \nonumber  \\  
												    &\hspace{0.5cm}\times \bra{\vec{x}\'} \hat O \ket{\vec{x}\''}\,  \e^{\frac{\i}{\hbar}\vec\chi_p\vec x\''}   \e^{\frac{\i}{\hbar}\left[\vec x\'' -\vec x\'-\vec\chi_x \right]\vec p\'} .
 \end{align}
After substitution of $\vec{p}\' = \vec p\'' -\frac{\vec\chi_p}{2}$ and  $~\dif \vec p\'= ~~\dif \vec p\''$, we obtain the expression
 \begin{align}
        \trace \{\hat O ~ \e^{\frac{i}{\hbar}\left[\vec\chi_p \hat{\vec x} - \vec\chi_x \hat{\vec p}\right]}\}  &=   \frac{1}{(2\pi\hbar)^{\N}}  \!\! \iiint\limits_{\mathbb{R}^{3\N}}  \dif^{\N} x' ~~\dif^{\N} x'' ~~\dif^{\N} p'' \nonumber \\
														&\hspace{0.5cm}\times \bra{\vec{x}\'} \hat O \ket{\vec{x}\''} \e^{\frac{\i}{\hbar}\vec\chi_p \vec x\''} \, \e^{\frac{\i}{\hbar}\left(\vec x\'' -\vec x\'\right)\left(\vec p\''-\frac{\vec\chi_p}{2}\right)} \, \e^{-\frac{\i}{\hbar} \vec\chi_x \vec p\''} .
 \end{align}
Insertion into the Wigner function~\eqref{Wigner_proof} yields after subsequent integration over the variables $\vec\chi_x$, $\vec\chi_p$ 
and $\vec x\''$
 \begin{align}
        \text{W}[\hat O](\vec x,\vec p) &=  \frac{2^\N}{(2\pi\hbar)^{\N}} \!\!\int\limits_{\mathbb{R}^{\N}}  \dif^{\N} x'  \bra{\vec{x}\'} \hat O \ket{2 \vec x-\vec x\'}  \e^{-\frac{\i}{\hbar}\left(2 \vec x\' - 2 \vec x\right) \vec p} .
 \end{align}
When we finally perform the substitution $\vec x\' = \frac{\vec \lambda}{2}+\vec x$, we arrive at the well-known 
expression~\eqref{eq:Wigner_f_standard}.\\ 
 
Next, we discuss the Wigner function and its corresponding characteristic function for Gaussian states, which are of upmost 
experimental importance.  
%
%
%
%
\subsection{Wigner function and characteristic function for Gaussian states}
%
The most general Gaussian state $\hat\rho^{(\G)}$, which includes coherent, squeezed and thermal states, 
is given by the Wigner function
 \begin{align}
    \text{W}[\hat\rho^{(\G)}](\vec\xi)  &=  \frac{1}{\sqrt{(2\pi)^{2\N} \text{det} \Sigma}} ~ \e^{-\frac{1}{2} (\vec\xi-\langle \vec\xi\rangle)^\T \Sigma^{-1} (\vec\xi-\langle \vec\xi\rangle)}
 \end{align}
where
 \begin{align}
    \langle \vec\xi\rangle = \trace \{ \hat\rho^{(\G)} \vec\xi \}  
 \end{align}
denotes the combined expectation value for the position and momentum operators and
 \begin{align}
    \Sigma_{ik} = \frac{1}{2} \langle \xi_i \xi_k + \xi_k \xi_i\rangle -  \langle \xi_i \rangle \langle \xi_k \rangle     
 \end{align}
the corresponding covariance matrix $\Sigma$ with $i,k \in \{1,...,2\N\}$. \\ 

The characteristic function for Gaussian states follows from the inverse symplectic Fourier transform 
 \begin{align}
    \eta[\hat\rho^{(\G)}](\vec\chi) 	&=    \!\!\int\limits_{\mathbb{R}^{2\N}}  \dif^{2\N} \xi ~ \e^{-\frac{\i}{\hbar} \vec\chi^\T \mathcal{J} \vec\xi} ~  \text{W}[\hat\rho^{(\G)}](\vec\xi) \\
					&=    \frac{1}{\sqrt{(2\pi)^{2\N} \text{det} \Sigma}}  \int\limits_{\mathbb{R}^{2\N}}  \dif^{2\N} \xi~ 
					      \e^{-\frac{1}{2} (\vec\xi-\langle \vec\xi\rangle)^\T \Sigma^{-1} (\vec\xi-\langle \vec\xi\rangle)-\frac{\i}{\hbar} \vec\chi^\T \mathcal{J} \vec\xi}\, .
 \end{align}
When we take advantage of the following identity
 \begin{align}\label{integral_identity}
     \!\!\int\limits_{\mathbb{R}^{2\N}}  \dif^{2\N} \xi ~\e^{-\frac{1}{2} \vec \xi^\T A^{-1} \vec \xi +\i \vec s^\T \vec \xi} &= \sqrt{(2\pi)^{2\N} \text{det} A }  ~\e^{-\frac{1}{2} \vec s^\T A \vec s}\, ,
 \end{align}
which holds true for arbitrary vectors $\vec s \in \mathbb{R}^{2\N}$ and positive definite matrices $A \in \mathbb{R}^{2\N\times2\N}$, 
we arrive after the substitution $\vec \xi\' = \vec\xi-\langle \vec\xi\rangle$ and the integration over $\vec \xi\'$ at
 \begin{align}
    \eta[\hat\rho^{(\G)}](\vec\chi)&= \e^{-\frac{1}{2\hbar^2} (\mathcal{J}\vec\chi)^\T \Sigma\, (\mathcal{J}\vec\chi)-\frac{\i}{\hbar}\vec\chi^\T\mathcal{J}\langle \vec\xi\rangle}\, .
 \end{align}
Hence, the symmetric-ordered characteristic function of a Gaussian state $\hat\rho^{(\G)}$ is a complex Gaussian function.
%
%
%
%
%
%
%
%
%
%
%
%
\section{Mach-Zehnder operator}
\label{app:Mach-Zehnder_operator}
By means of the generalized beam-splitter matrix in the Heisenberg picture, Eq.~\eqref{BSM_H_2}, we can calculate the 
Mach-Zehnder operator
   \begin{align}\label{app:U_MZ}
	\hat{U}_{\MZ} 	&= 	\hat{S}^{(\frac{\pi}{2})}_{\H,2} ~ \hat{S}^{(\pi)}_{\H,1} ~ \hat{S}^{(\frac{\pi}{2})}_{\H,0} \nonumber \\[0.25cm]
			&= 	\begin{pmatrix} 
					      \frac{1}{\sqrt{2}} &\!\!\! \frac{-\i}{\sqrt{2}} \, \e^{-\i\Phi_2}  \hat{D}(-\vec\chi_2) \\
					      \frac{-\i}{\sqrt{2}} \, \e^{+\i\Phi_2}  \hat{D}(\vec\chi_2) & \frac{1}{\sqrt{2}} 
				\end{pmatrix} \nonumber 
				\begin{pmatrix} 
					      0 & \!\!\! -\i \, \e^{-\i\Phi_1}  \hat{D}(-\vec\chi_1) \\
					      -\i  \, \e^{+\i\Phi_1}  \hat{D}(\vec\chi_1) & 0
				\end{pmatrix} \nonumber \\ 
			&\hspace{4.75cm} \times	 \begin{pmatrix} 
					      \frac{1}{\sqrt{2}} & \!\!\! \frac{-\i}{\sqrt{2}} \, \e^{-\i\Phi_0}  \hat{D}(-\vec\chi_0) \\
					      \frac{-\i}{\sqrt{2}}  \, \e^{+\i\Phi_0}  \hat{D}(\vec\chi_0) & \frac{1}{\sqrt{2}} 
				\end{pmatrix} \nonumber \\[0.5cm]
			&=\frac{1}{2}		\left(\begin{array}{c} 
						    - \{\e^{+\i[\Phi_0-\Phi_1]} ~ \hat{D}(-\vec\chi_1)\, \hat{D}(\vec\chi_0) + \e^{+\i[\Phi_1-\Phi_2]} ~ \hat{D}(-\vec\chi_2)\, \hat{D}(\vec\chi_1)\}  \\[0.1cm]
						     \i  \{\e^{+\i[\Phi_0-\Phi_1+\Phi_2]} ~ \hat{D}(\vec\chi_2)\, \hat{D}(- \vec\chi_1)\, \hat{D}(\vec\chi_0) - \e^{+\i\Phi_1}\, \hat{D}(\vec\chi_1) \}			
						\end{array} \right. \nonumber \\[0.25cm]
			&\hspace{1.5cm}	\left.\begin{array}{c} 
						     \i \{ -\e^{-\i \Phi_1} ~ \hat{D}(-\vec\chi_1) + \e^{-\i[\Phi_0-\Phi_1+\Phi_2]} ~ \hat{D}(-\vec\chi_2)\, \hat{D}(\vec\chi_1)\, \hat{D}(-\vec\chi_0)\} \\[0.1cm] 
						    - \{\e^{-\i[\Phi_1-\Phi_2]} ~ \hat{D}(\vec\chi_2)\, \hat{D}(-\vec\chi_1) + \e^{-\i[\Phi_0-\Phi_1]} ~ \hat{D}(\vec\chi_1)\, \hat{D}(-\vec\chi_0)\}	
						\end{array} \right) .
    \end{align}
Acting on the initial ground state $\ket{0}= (1,0)^\T$ we get
  \begin{align}\label{app:U_MZ_0_ket_0}
     \hat{U}_{\MZ}  \ket{0}	&=\frac{1}{2}		\left(\begin{array}{c} 
							      - \{\e^{+\i[\Phi_0-\Phi_1]} ~ \hat{D}(-\vec\chi_1)\, \hat{D}(\vec\chi_0) + \e^{+\i[\Phi_1-\Phi_2]} ~ \hat{D}(-\vec\chi_2)\, \hat{D}(\vec\chi_1)\}  \\[0.1cm]
							       \i  \{\e^{+\i[\Phi_0-\Phi_1+\Phi_2]} ~ \hat{D}(\vec\chi_2)\, \hat{D}(- \vec\chi_1)\, \hat{D}(\vec\chi_0) - \e^{+\i\Phi_1}\, \hat{D}(\vec\chi_1) \}			
							\end{array} \right) \nonumber \\[0.25cm]
				&=\frac{1}{2}		\left(\begin{array}{c} 
							      - \e^{+\i[\Phi_1-\Phi_2]} ~ \hat{D}(-\vec\chi_2)\, \hat{D}(\vec\chi_1)   ~ \{1  + \mathfrak{\hat D}_{\MZ} \} \\[0.1cm]
  							      - \i ~ \e^{+\i\Phi_1}\, \hat{D}(\vec\chi_1) ~ \{1 - \mathfrak{\hat D}_{\MZ} \}
							\end{array} \right)\, ,
  \end{align}
where in the last step we have introduced the generalized Mach-Zehnder displacement operator 
  \begin{align}\label{app:gen_displ_1}
     \mathfrak{\hat D}_{\MZ}		&=   \e^{+\i[\Phi_0 - 2 \Phi_1 + \Phi_2]}  ~ \hat{D}(-\vec\chi_1)\,  \hat{D}(\vec\chi_2)\, \hat{D}(-\vec\chi_1)\, \hat{D}(\vec\chi_0)\, .
  \end{align}
The generalized displacement operator in its most compact form is given by Eq.~\eqref{gen_displ_3}.
\section{Total interferometer phase}
\label{app:total_phase}
%
%
\subsection{Mach-Zehnder geometry}
\label{The total Mach-Zehnder phase}
The series expansion of the total Mach-Zehnder phase shift
  \begin{align}\label{Phi_MZ_appendix}
	\Delta\Phi_{\MZ}		&=	\Phi_{\MZ} +  \frac{1}{\hbar} \left[ \frac{\vec\chi_0}{2} + \langle \vec\xi_0\rangle \right]^\T \!\! \mathcal{J}  \vec\chi_{\MZ}
  \end{align}
is determined by the generalized Mach-Zehnder phase \eqref{phi_{MZ}_gamma} and the displacement vector \eqref{chi_MZ_gradient}.
Provided the uniform rotation of the lasers during the pulse sequence is tiny, it suffices to consider terms up to second order in the 
rotation rate in Eq.~\eqref{roation_k}
    \begin{align}\label{k_rotation}
	    \vec k_n 	&= \vec k_0 + (\vec \Omega \times \vec k_0) t_n + \frac{1}{2} \vec\Omega \times (\vec\Omega \times \vec k_0)t_n^2 + \mathcal{O}[\Omega^3]\,; \qquad n\in\{0,1,2\}\, .
    \end{align}
When we assume a time-symmetric pulse sequence ($t_0=0, t_1 = T, t_2 = 2T)$, we finally arrive at the series expansion of the total Mach-Zehnder phase shift
  \begin{align}\label{Delta_Phi_MZ}
	\Delta\Phi_{\MZ}		&= \varphi_{\MZ} 		+ \left\{ \vec k_0 + 3 (\vec \Omega \times \vec k_0) T + \frac{7}{2} [\vec \Omega \times (\vec \Omega \times \vec k_0)] T^2 \right\}^\T \vec g ~ T^2 \nonumber \\
				&\hphantom{\ = \varphi_{\MZ}}  	-  \left\{ 7 \vec k_0 + 15 (\vec \Omega \times \vec k_0) T + \frac{31}{2} [\vec \Omega \times (\vec \Omega \times \vec k_0)] T^2 \right\}^\T \frac{\Gamma}{12} ~ \vec g ~ T^4 \nonumber \\
				&\hphantom{\ = \varphi_{\MZ}}  	+  \left\{ 31 \vec k_0 + 63 (\vec \Omega \times \vec k_0) T + \frac{127}{2} [\vec \Omega \times (\vec \Omega \times \vec k_0)] T^2 \right\}^\T \frac{\Gamma^2}{360} ~ \vec g ~ T^6 \nonumber \\
				&\hphantom{\ = } 		+  \langle \vec x_0 \rangle ^\T  \left[ [\vec \Omega \times (\vec \Omega \times \vec k_0)] T^2 \vphantom{\frac{7}{2}}\right. \nonumber \\  
				&\hphantom{\ = \   +  \langle \vec x_0 \rangle ^\T } \left.  -  ~ \Gamma\left\{ \vec k_0 + 3(\vec \Omega \times \vec k_0)T + \frac{7}{2} [\vec \Omega \times (\vec \Omega \times \vec k_0)] T^2 \right\}  ~ T^2     \right.\nonumber \\
				&\hphantom{\ = \   +  \langle \vec x_0 \rangle ^\T } \left.  +  ~ \frac{\Gamma^2}{12} \left\{ 7 \vec k_0 + 15(\vec \Omega \times \vec k_0)T + \frac{31}{2} [\vec \Omega \times (\vec \Omega \times \vec k_0)] T^2 \right\}   ~ T^4    \right] \nonumber \\
				&\hphantom{\ = }  		+  \left( \frac{\hbar \vec k_0}{2m} + \frac{\langle \vec p_0 \rangle}{m} \right)^\T  	  \left[ 2 \left\{ (\vec \Omega \times  \vec k_0 ) T + \frac{3}{2}  [\vec \Omega \times (\vec \Omega \times \vec k_0)] T^2 \right\} T \right. \nonumber \\ 
				&\hspace{-1.5cm}\hphantom{\ = \ 	+  \left( \frac{\hbar \vec k_0}{2m} + \frac{\langle \vec p_0 \rangle}{m} \right)^\T }                  -  ~ \frac{\Gamma}{3} \left\{ 3 \vec k_0 + 7(\vec \Omega \times \vec k_0)T + \frac{15}{2} [\vec \Omega \times (\vec \Omega \times \vec k_0)] T^2 \right\}  ~ T^3  \nonumber \\
				&\hspace{-1.5cm}\hphantom{\ = \ 	+  \left( \frac{\hbar \vec k_0}{2m} + \frac{\langle \vec p_0 \rangle}{m} \right)^\T } \left.           +  ~ \frac{2\Gamma^2}{5!} \left\{ 15 \vec k_0 + 31(\vec \Omega \times \vec k_0)T + \frac{63}{2} [\vec \Omega \times (\vec \Omega \times \vec k_0)] T^2 \right\}  ~ T^5 \right] \nonumber \\
				&\hphantom{\ = }  		+  \mathcal{O}[\Omega^3] +  \mathcal{O}[\Gamma^3]\,.  
  \end{align}
Here, the first three lines correspond to the generalized Mach-Zehnder phase $\Phi_{\MZ}$ of Eq.~\eqref{phi_{MZ}_gamma}. Therein, the curly brackets take into account the rotating lasers. Moreover, 
each line stands for different orders in the gradient~$\Gamma$. 

The following two blocks are arranged in the same way but correspond to a non-vanishing displacement vector 
$\vec\chi_{\MZ} \neq 0$: They depend on 
the initial position of the atomic wave function $\langle \vec x_0 \rangle $ as well as on the 
initial momentum $\langle \vec p_0 \rangle$. Additionally, the ``Baker-Campbell-Hausdorff'' correction (last three lines with prefactor $\hbar\vec k_0 /2m$) is 
included. 

The total phase shift coincides with the presented terms in \cite{Antoine_2003}\footnote{By comparison with \cite{Antoine_2003} we recognized  that terms based on the formulas (67)-(69) of the mentioned paper 
coincide with our result; however, Eq.~(72) differs by signs. 
Since Eq.~(72) is the series expansion of (69), we expect misprints in Eq.~(72).}. 
Since in \cite{Peters_2001} only gravity gradients are considered, we get their result by setting $\Omega$ to zero.
%
%
\subsection{Butterfly geometry}
\label{app:Butterfly}
The series expansion of the total Butterfly phase shift
  \begin{align}
	\Delta\Phi_{\BU}		&=	\Phi_{\BU} +  \frac{1}{\hbar} \left[ \frac{\vec\chi_0}{2} + \langle \vec\xi_0\rangle \right]^\T \!\! \mathcal{J}  \vec\chi_{\BU}
  \end{align}
includes the generalized Butterfly phase~\eqref{generalized_phi_BU} and the displacement vector~\eqref{chi_BU_gradient}.
In addition, we allow a uniform rotation rate of our lasers (Sagnac effect with time-independent axis of rotation) 
    \begin{align}
	    \vec k_n 	&= \vec k_0 + (\vec \Omega \times \vec k_0) t_n + \frac{1}{2} \vec\Omega \times (\vec\Omega \times \vec k_0) t_n^2 + \mathcal{O}[\Omega^3] 
    \end{align}
and assume the following experimentally typical timings: $t_0=0,~t_1=T,~t_2=3T,~t_3=4T$ (for equal loop areas). As a result, the series expansion of the total 
Butterfly phase shift reads
  \begin{align}
	\Delta\Phi_{\BU}	&= \varphi_{\BU} 		-  \left\{ 6 (\vec \Omega \times \vec k_0) T + 24 [\vec \Omega \times (\vec \Omega \times \vec k_0)] T^2 \right\}^\T \vec g ~ T^2 \nonumber \\
				&\hphantom{\ = \varphi_{\MZ}}  	+  \left\{ 4 \vec k_0 + \frac{45}{2} (\vec \Omega \times \vec k_0) T +  55 [\vec \Omega \times (\vec \Omega \times \vec k_0)] T^2 \right\}^\T \Gamma ~ \vec g ~ T^4 \nonumber \\
				&\hphantom{\ = \varphi_{\MZ}}  	-  \left\{ \frac{11}{3} \vec k_0 + \frac{1001}{60} (\vec \Omega \times \vec k_0) T + \frac{182}{5} [\vec \Omega \times (\vec \Omega \times \vec k_0)] T^2 \right\}^\T \Gamma^2 ~ \vec g ~ T^6 \nonumber \\
				&\hphantom{\ = } 		+  \langle \vec x_0 \rangle ^\T  \left[   ~ \Gamma\left\{  6(\vec \Omega \times \vec k_0)T + 24 [\vec \Omega \times (\vec \Omega \times \vec k_0)] T^2 \right\}  ~ T^2 \phantom{\frac{a}{b}}\right. \nonumber \\  
				&\hphantom{\ = \   +  \langle \vec x_0 \rangle ^\T } \left.  -  ~ \Gamma^2 \left\{   4 \vec k_0 + \frac{45}{2}(\vec \Omega \times \vec k_0)T + 55 [\vec \Omega \times (\vec \Omega \times \vec k_0)] T^2 \right\}   ~ T^4    \right] \nonumber \\
				&\hphantom{\ = }  		-  \left( \frac{\hbar \vec k_0}{2m} + \frac{\langle \vec p_0 \rangle}{m} \right)^\T  	  \left[ 6 [\vec \Omega \times (\vec \Omega \times \vec k_0)] T^3 \phantom{\frac{a}{b}}\right. \nonumber \\ 
				&\hspace{-1.5cm}\hphantom{\ = \ 	+  \left( \frac{\hbar \vec k_0}{2m} + \frac{\langle \vec p_0 \rangle}{m} \right)^\T }                  -  ~ \Gamma \left\{ 2 \vec k_0 + 16 (\vec \Omega \times \vec k_0)T + 45 [\vec \Omega \times (\vec \Omega \times \vec k_0)] T^2 \right\}  ~ T^3  \nonumber \\
				&\hspace{-1.5cm}\hphantom{\ = \ 	+  \left( \frac{\hbar \vec k_0}{2m} + \frac{\langle \vec p_0 \rangle}{m} \right)^\T } \left.           +  ~ \Gamma^2 \left\{ \frac{9}{2} \vec k_0 + 22 (\vec \Omega \times \vec k_0)T + \frac{1001}{20} [\vec \Omega \times (\vec \Omega \times \vec k_0)] T^2 \right\}  ~ T^5 \right] \nonumber \\
				&\hphantom{\ = }  		+  \mathcal{O}[\Omega^3] +  \mathcal{O}[\Gamma^3]\,.  
  \end{align}
The first three lines correspond to the generalized Butterfly phase $\Phi_{\BU}$, Eq.~\eqref{generalized_phi_BU}, where 
the curly brackets take into account the rotating lasers. Moreover, each line corresponds to different orders in the 
gradient~$\Gamma$. The following two blocks are arranged in the same way and depend on 
the initial position of the atomic wave function $\langle \vec x_0 \rangle $ as well as on the 
initial momentum $\langle \vec p_0 \rangle$. They are only present for a non-vanishing displacement vector 
$\vec\chi_{\BU} \neq 0$.
%
%
\subsection{Mach-Zehnder phase shift in non-inertial frames}
\label{app:Phi_non-inertial}
We recall that the total phase shift for the Mach-Zehnder geometry is given by Eq.~\eqref{Phi_MZ_appendix}. Applying 
the expressions Eqs.~\eqref{tau_expansion_co-moving_orbiting}-\eqref{phi_n_different_omega}, valid for an 
observer in a frame in which the lasers are at rest\footnote{In Section~\ref{sssec:Experiments_on_Earth} we 
study this scenario by an alternative approach based on co-moving frames.} 
(constant rotation rate), we can determine the generalized phase 
$\Phi_\MZ$ and the displacement vector $\vec\chi_\MZ$ which yield as total Mach-Zehnder phase shift
\begin{align}\label{Phi_MZ_rotating}
 \Delta\Phi_\MZ &= \varphi_\MZ - \vec k^{\T}_0 \left[ 
				      -   1 
				      + \left(3 \Omega_k - \Omega_g\right) T 
				      +  \frac{7}{12}  \left(\Gamma_0 + 4 \Omega_k  \Omega_g - \Omega_g^2 - 6 \Omega_k^2  \right) T^2  \right.\nonumber\\
		&\hspace{-0.0cm}\qquad\qquad\qquad         +    \frac{1}{4}  \left( 3 \Omega_{\Gamma}  \Gamma_0 - 3 \Gamma_0  \Omega_{\Gamma} + \Gamma_0  \Omega_g + 5 \Omega_k  \Omega_g^2 - 10 \Omega_k^2 \Omega_g - \Omega_g^3 \right. \nonumber\\
		&\hspace{-0.25cm}\qquad\qquad\qquad\qquad\quad   \left.  - 5 \Omega_k   \Gamma_0 + 10 \Omega_k^3 \right)  T^3  \nonumber \\
		&\hspace{-0.75cm}\qquad\qquad\qquad	      -\frac{31}{360} \left(12 \Omega_{\Gamma} \Gamma_0  \Omega_{\Gamma} - 6 \Gamma_0 \Omega_{\Gamma}^2 - 6 \Omega_{\Gamma}^2 \Gamma_0 + \Gamma_0^2 - \Gamma_0  \Omega_g^2 \right. \nonumber\\
		&\hspace{-0.75cm}\qquad\qquad\qquad\qquad\quad	      \left. + 4 \Gamma_0  \Omega_{\Gamma} \Omega_g - 4 \Omega_{\Gamma}  \Gamma_0  \Omega_g + 6 \Omega_k \Gamma_0 \Omega_g - 6 \Omega_k  \Omega_g^3  \right. \nonumber\\
		&\hspace{-0.75cm}\qquad\qquad\qquad\qquad\quad	      \left. + 15 \Omega_k^2 \Omega_g^2 - 20 \Omega_k^3  \Omega_g + \Omega_g^4 - 15 \Omega_k^2  \Gamma_0 - 18 \Omega_k \Gamma_0 \Omega_{\Gamma} \right. \nonumber\\
		&\hspace{-0.75cm}\qquad\qquad\qquad\qquad\quad        \left. + 18 \Omega_k \Omega_{\Gamma} \Gamma_0 + 15 \Omega_k^4 \right) T^4 \nonumber\\
		&\hspace{-0.75cm}\qquad\qquad\qquad		      +\frac{1}{40}  \left( 26 \Gamma_0  \Omega_{\Gamma} \Gamma_0 - 10 \Gamma_0 \Omega_{\Gamma}^3 - 9 \Gamma_0^2 \Omega_{\Gamma} - 17 \Omega_{\Gamma} \Gamma_0^2 + 10 \Omega_{\Gamma}^3 \Gamma_0 \right. \nonumber\\
		&\hspace{-0.75cm}\qquad\qquad\qquad\qquad\quad             	\left. +30 \Omega_{\Gamma}  \Gamma_0 \Omega_{\Gamma}^2 - 30 \Omega_{\Gamma}^2 \Gamma_0 \Omega_{\Gamma} + \Gamma_0 \Omega_g^3 - \Gamma_0^2 \Omega_g - 5 \Gamma_0  \Omega _{\Gamma} \Omega_g^2 \right. \nonumber\\ 
		&\hspace{-0.75cm}\qquad\qquad\qquad\qquad\quad             	\left. + 10 \Gamma_0 \Omega_{\Gamma}^2 \Omega_g+5 \Omega_{\Gamma} \Gamma_0 \Omega_g^2 + 10 \Omega_{\Gamma}^2  \Gamma_0 \Omega_g - 20 \Omega_{\Gamma} \Gamma_0 \Omega_{\Gamma} \Omega_g \right. \nonumber\\ 
		&\hspace{-0.75cm}\qquad\qquad\qquad\qquad\quad             	\left. - 7 \Omega_k \Gamma_0  \Omega_g^2 + 21 \Omega_k^2  \Gamma_0 \Omega_g + 28 \Omega_k \Gamma_0 \Omega_{\Gamma} \Omega_g - 28 \Omega_k \Omega_{\Gamma}  \Gamma_0 \Omega_g \right. \nonumber\\
		&\hspace{-0.75cm}\qquad\qquad\qquad\qquad\quad             	\left. + 7 \Omega_k  \Omega_g^4 - 21 \Omega_k^2 \Omega_g^3 + 35 \Omega_k^3 \Omega_g^2 - 35 \Omega_k^4 \Omega_g - \Omega_g^5 + 7 \Omega_k \Gamma_0^2 \right. \nonumber\\ 
		&\hspace{-0.75cm}\qquad\qquad\qquad\qquad\quad             	\left. - 35 \Omega_k^3 \Gamma_0 - 42 \Omega_k \Gamma_0 \Omega_{\Gamma}^2 - 42 \Omega_k \Omega_{\Gamma}^2 \Gamma_0 - 63 \Omega_k^2 \Gamma_0 \Omega_{\Gamma} \right. \nonumber\\ 
		&\hspace{-0.75cm}\qquad\qquad\qquad\qquad\quad             	\left. + 63 \Omega_k^2 \Omega_{\Gamma} \Gamma_0 + 84 \Omega_k  \Omega_{\Gamma} \Gamma_0 \Omega_{\Gamma} + 21 \Omega_k^5 \right) T^5 
						      + \dots \left.\vphantom{\frac{1}{2}}\right] \vec g'_0\, T^2 \nonumber\\ 
	&\hspace{-0.5cm}\hphantom{=}
		+\vec k^{\T}_0 \left[\vphantom{\frac{1}{2}}
			       \left(\Omega_k^2 - \Gamma_0 \right) T^2 
			      +\left(\Gamma_0 \Omega_{\Gamma} - \Omega_{\Gamma} \Gamma_0 + 3 \Omega_k \Gamma_0 - \Omega_k^3 \right)T^3 \phantom{\frac{a}{b}}\right.  \nonumber \\
		&\hspace{-0.75cm}\qquad\qquad         +  \frac{7}{12} \left( 2 \Omega_{\Gamma} \Gamma_0 \Omega_{\Gamma} - \Gamma_0  \Omega_{\Gamma}^2 - \Omega_{\Gamma}^2  \Gamma_0 + \Gamma_0^2 - 6 \Omega_k^2 \Gamma_0 - 4 \Omega_k  \Gamma_0 \Omega_{\Gamma} \right. \nonumber\\
		&\hspace{-0.75cm}\qquad\qquad\qquad\quad         \left. + 4 \Omega_k \Omega_{\Gamma} \Gamma_0 + \Omega_k^4\right) T^4  \nonumber \\ 
		&\hspace{-0.75cm}\qquad\qquad	  +  \frac{1}{4} \left( \Gamma_0 \Omega_{\Gamma}^3 - \Gamma_0^2 \Omega_{\Gamma} + 3 \Omega_{\Gamma} \Gamma_0^2 - \Omega_{\Gamma}^3 \Gamma_0 - 2 \Gamma_0 \Omega_{\Gamma} \Gamma_0 -3 \Omega_{\Gamma}  \Gamma_0  \Omega_{\Gamma}^2  \right. \nonumber \\
		&\hspace{-0.75cm}\qquad\qquad\qquad	   \left. + 3 \Omega_{\Gamma}^2 \Gamma_0  \Omega_{\Gamma} - 5 \Omega_k  \Gamma_0^2 + 10 \Omega_k^3  \Gamma_0 + 5 \Omega_k \Gamma_0 \Omega_{\Gamma}^2 + 5 \Omega_k \Omega_{\Gamma}^2 \Gamma_0 \right. \nonumber\\ \displaybreak
		&\hspace{-0.75cm}\qquad\qquad\qquad     \left.    + 10 \Omega_k^2 \Gamma_0  \Omega_{\Gamma} - 10 \Omega_k^2 \Omega_{\Gamma} \Gamma_0 - 10 \Omega_k \Omega_{\Gamma} \Gamma_0 \Omega_{\Gamma} - \Omega_k^5 \right)  T^5 
						       + \dots\left.\vphantom{\frac{1}{2}}\right] \langle \vec x_0 \rangle \nonumber\\ 
	&\hspace{-0.25cm}\hphantom{=}
		+\vec k^{\T}_0 \left[ \vphantom{\frac{1}{2}}
			  - 2 \Omega_k  T 
			  +   \left(3 \Omega_k^2 - \Gamma_0 \right) T^2  + \frac{7}{6}  \left( \Gamma_0 \Omega_{\Gamma} - \Omega_{\Gamma} \Gamma_0 + 2 \Omega_k \Gamma_0 - 2 \Omega_k^3 \right) T^3   \right. \nonumber \\
		&\hspace{-0.25cm}\qquad\qquad\quad	  +   \frac{1}{4} \left( 6 \Omega_{\Gamma} \Gamma_0  \Omega_{\Gamma} + \Gamma_0^2 - 3 \Gamma_0  \Omega_{\Gamma}^2 - 3 \Omega_{\Gamma}^2  \Gamma_0 - 10 \Omega_k^2  \Gamma_0 - 10 \Omega_k \Gamma_0  \Omega_{\Gamma} \right.\nonumber\\
		&\hspace{-0.25cm}\qquad\qquad\qquad\qquad        \left.  + 10 \Omega_k  \Omega_{\Gamma} \Gamma_0 + 5 \Omega_k^4 \right) T^4 \nonumber\\
		&\hspace{-0.25cm}\qquad\qquad\quad        -\frac{31}{180} \left( \Gamma_0 \Omega_{\Gamma} \Gamma_0 - 2 \Gamma_0 \Omega_{\Gamma}^3 + \Gamma_0^2 \Omega_{\Gamma} - 2 \Omega_{\Gamma}  \Gamma_0^2 + 2 \Omega_{\Gamma}^3 \Gamma_0  \right. \nonumber\\
		&\hspace{-0.25cm}\qquad\qquad\qquad\qquad	  \left. + 6 \Omega_{\Gamma} \Gamma_0 \Omega_{\Gamma}^2 - 6 \Omega_{\Gamma}^2  \Gamma_0 \Omega_{\Gamma} + 3 \Omega_k \Gamma_0^2 - 10 \Omega_k^3 \Gamma_0 - 9 \Omega_k  \Gamma_0  \Omega_{\Gamma}^2  \right. \nonumber\\
		&\hspace{-0.25cm}\qquad\qquad\qquad\qquad	  \left. - 9 \Omega_k \Omega_{\Gamma}^2 \Gamma_0 - 15 \Omega_k^2 \Gamma_0 \Omega_{\Gamma} + 15 \Omega_k^2 \Omega_{\Gamma} \Gamma_0 + 18 \Omega_k  \Omega_{\Gamma}  \Gamma_0 \Omega_{\Gamma} \right. \nonumber\\
		&\hspace{-0.25cm}\qquad\qquad\qquad\qquad	  \left. + 3 \Omega_k^5  \right) T^5 
			 + \dots   \left.\vphantom{\frac{1}{2}}\right] T 
			 \left( \frac{\hbar \vec k_0}{2m} + \frac{\langle \vec p_0 \rangle}{m} \right)\, .           
\end{align}
In order to compare our general result, Eq.~\eqref{Phi_MZ_rotating}, with well-known results for atomic fountains 
fixed on Earth, we have expanded the phase shift in suitable orders in $\Gamma_0$ and $\Omega$. In this way, 
the case of an atomic fountain which is fixed on the (uniformly) rotating Earth is obtained by choosing the following 
parameters: $\Omega=\Omega_k=\Omega_g=\Omega_\Gamma$. The corresponding expression for the total 
Mach-Zehnder phase shift is presented in Eq.~\eqref{phi_atomic_fountain}.
%
%
%
%

\section{The rotation group SO(3)}
\label{app:Rotation_group_SO(3)}
In this appendix we take a closer look at the rotation group. For a more detailed study we refer to \cite{Relativity_Groups_Particles_Sexl_Urbantke_1992}. 
Rotations are linear, homogeneous transformations of the form
    \begin{align}\label{rotation_transformation}
	\vec x' = \mathcal{R}\, \vec x
    \end{align}
which map the 3-dimensional Euclidean vector space onto itself, thereby preserving the length of and angles between individual 
vectors. The rotation matrix $\mathcal{R}$ is an element of the group of special orthogonal matrices SO(3), which do not change 
the orientation (a right-handed coordinate system remains right-handed).
%
%
\paragraph{Active and passive rotations}
%
The linear, homogeneous transformation \eqref{rotation_transformation} can be interpreted in two different ways: 
(i) As an active transformation (change of the vector itself), in which the matrix $\mathcal{R}$ rotates the components of the vector $\vec x$ 
to the transformed components $\vec x'$ while the coordinate system is fixed, or (ii) as a passive transformation (changing the 
reference frame), 
in which the matrix $\mathcal{R}$ describes the change of the vector components when the same vector is described from the perspective 
of two different coordinate systems, the original and the rotating one. Both point of views are mathematically equivalent. With no loss of generality, 
we choose the active transformation as our preferred viewpoint, so that a rotation is implemented by 
Eq.~\eqref{rotation_transformation} where the vector notation of $\vec x$ and $\vec x'$ is supposed to be a 
short notation for the components with respect to a non-rotating frame.
%
%
\paragraph{SO(3) as a Lie group}
\label{sec:Lie_group_of_SO(3)}
%
%
%
%
The implementation of rotations requires a suitable parametrization. A well-established method is the parametrization by means of the rotation vector $\vec \alpha$. The absolute value
of the rotation vector defines the rotation angle $\alpha=|\vec \alpha|$ and the unit vector $\vec n = \vec \alpha/ |\vec \alpha|$ 
describes the rotation axis. In this way, the rotation around an arbitrary axis $\vec n$ by an amount of $\alpha$ is given by 
the linear transformation
    \begin{align}\label{vector_rotation}
	    \vec x' 	&= \cos(\alpha) \vec x + \sin(\alpha) (\vec n \times \vec x) + (1-\cos(\alpha))(\vec n \cdot \vec x)\vec n\, ,
    \end{align}
which reads in components with the Einstein summation convention and $i,j,k\in\{1,2,3\}$
    \begin{align}\label{active_rotation_components}
	  x'_{i}	&= \cos(\alpha) x_i + \sin(\alpha) \varepsilon_{ijk} n_j x_k + (1-\cos(\alpha))n_k x_k n_i\, .
    \end{align}
By comparison with \eqref{rotation_transformation} we obtain the following expression for the components of the rotation matrix
    \begin{align}\label{rotation_matrix_comp}
	  \mathcal{R}_{ik}	&= \cos(\alpha) \delta_{ik} + \sin(\alpha) \varepsilon_{ijk} n_j + (1-\cos(\alpha))n_i n_k.
    \end{align}
It is necessary to constrain the rotation angle $\alpha$ to the values $0\leq\alpha < \pi$ to guarantee an unambiguous 
characterization of the elements of the rotation group. 
Hence, the homomorphism $\vec\alpha\rightarrow \mathcal{R}(\vec\alpha)$ as a structure preserving map between two algebraic structures (such as groups, vector spaces, etc.) 
becomes bijective and therefore an isomorphism. We can identify the abstract group SO$(3,\mathbb{R})$ as a 3-dimensional compact connected Lie group. \\

We note that besides the parametrization by the rotation vector $\vec \alpha$ there exist other equivalent parametrizations of the 
group manifold SO(3) such as the one given by the Euler angles. 
%
%
%
%
\paragraph{Infinitesimal transformation}
%
The generators $\Lambda_j$ of the Lie algebra of the rotation group follow from the rotation matrix~\eqref{rotation_matrix_comp} via 
\begin{align}
 (\Lambda_j)_{ik}	&:= \frac{\partial \mathcal{R}_{ik}}{\partial \alpha_j}\Big\vert_{\vec\alpha =\vec 0} = -\varepsilon_{jik}
\end{align}
and read in explicit matrix form 
    \begin{align}\label{lambda_i}
	    \Lambda_1	= 	\left( \begin{matrix} 0 & 0 & 0 \\ 0 & 0 & -1 \\ 0 & 1 & 0 \end{matrix} \right), \qquad
	    \Lambda_2	= 	\left( \begin{matrix} 0 & 0 & 1 \\ 0 & 0 & 0 \\ -1 & 0 & 0 \end{matrix} \right), \qquad  
	    \Lambda_3	= 	\left( \begin{matrix} 0 & -1 & 0 \\ 1 & 0 & 0 \\ 0 & 0 & 0 \end{matrix} \right).
    \end{align} 
With $\vec \Lambda=(\Lambda_1,\Lambda_2,\Lambda_3)^\T$ as shorthand notation, infinitesimal rotations can be written to first order 
in $\varepsilon\ll1$ 
    \begin{align}\label{eq:R_inf_rot}
	    \mathcal{R}		&= 	\mathds{1} + \vec \alpha \cdot \vec \Lambda + \mathcal{O}[\varepsilon^2]
    \end{align} 
where $\vec\alpha= \varepsilon \vec n$. 
The matrix scalar product used in the last expression is given by 
    \begin{align}\label{Delta_Omega}
	    \vec \alpha \cdot \vec \Lambda	&=  \alpha_j \Lambda_j = 	\left( \begin{matrix} 0 & -\alpha_3 & \alpha_2 \\ \alpha_3 & 0 & -\alpha_1 \\ -\alpha_2 & \alpha_1 & 0 \end{matrix} \right).
    \end{align} 
Eq.~\eqref{eq:R_inf_rot} can be likewise obtained from the expansion of Eq.~\eqref{active_rotation_components} up to first order in 
$\alpha_j$. The resulting infinitesimal rotation reads
    \begin{align}\label{infinite_rotation_transformation_comp}
	x'_{i} 	&= x_i +  \alpha_j (-\varepsilon_{jik}) x_k + ... = x_i +   \varepsilon_{ijk} \alpha_j x_k + \mathcal{O}[\varepsilon^2]
    \end{align}
and in vector notation corresponds to
    \begin{align}\label{infinite_rotation_transformation_vec}
	\vec x' 	&= \vec x + \vec\alpha \times \vec x + \mathcal{O}[\varepsilon^2] \, .
    \end{align}
Conversely, the finite elements of the Lie group can be obtained by exponentiating
    \begin{align}
	\mathcal{R}_{\vec\alpha}	&= \e^{\vec \alpha \cdot \vec \Lambda}.
    \end{align}
The Taylor expansion of the exponential function in $\vec x' = \e^{\vec \alpha \cdot \vec \Lambda}\, \vec x$ 
yields finally the transformation~\eqref{vector_rotation}.
%
%
%
\paragraph{Second derivative of the rotation matrix}
\label{app:Second_derivative_R}
%
So far, we studied the generators of the rotation group $SO(3)$. Hence, we 
get via 
the rotation matrix (\ref{rotation_matrix_comp}) and the parametrization $\vec \alpha = \vec\Omega t$
  \begin{align}
	\frac{\text{d} \mathcal{R}_{\vec\Omega t}}{\dif t} &= \frac{\text{d}}{\text{d} t} \left(\e^{ (\vec\Omega t) \cdot \vec \Lambda}\right) = \left(\vec\Omega \cdot \vec \Lambda\right) \mathcal{R}_{\vec\Omega t} \, .
 \end{align}
This reads in components
  \begin{align}
	\left( \frac{\text{d}\mathcal{R}_{\vec\Omega t}}{\text{d} t} \right)_{ij} 
				    &=      	- \Omega_l  \, \varepsilon_{lim} \left(\mathcal{R}_{\vec\Omega t}\right)_{mj} 
				     = 		 \varepsilon_{ilm} \Omega_{l}  \left(\mathcal{R}_{\vec\Omega t}\right)_{mj}. 
  \end{align}
The action of the time derivative of the rotation matrix on a vector $\vec x$ is therefore given by
  \begin{align}
	\dot{\mathcal{R}}_{\vec\Omega t} \vec x =  \left( \frac{\text{d}\mathcal{R}_{\vec\Omega t}}{\text{d} t} \right) \vec x 
				    &=      	\vec\Omega \times ( \mathcal{R}_{\vec\Omega t} \vec x )\, . 
  \end{align}
This agrees with the result of the infinitesimal rotation \eqref{infinite_rotation_transformation_vec}.
Consequently, the action of the second derivative of the rotation matrix on $\vec x$ reads
  \begin{align}\label{second_derivative_R}
	\ddot{\mathcal{R}}_{\vec\Omega t} \vec x =  \left( \frac{\text{d}^2\mathcal{R}_{\vec\Omega t}}{\text{d} t^2} \right) \vec x 
				    &=      	\vec\Omega \times \left(\vec\Omega \times ( \mathcal{R}_{\vec\Omega t} \vec x ) \right)=\Omega^2 \mathcal{R}_{\vec\Omega t} \vec x\, , 
  \end{align}
which is the negative centrifugal acceleration $\vec a_{\text{cf}}(t) = -\vec\Omega \times \left(\vec\Omega \times  \vec x(t)  \right) = -\Omega^2\, \vec x(t)$. 
In the last step we took advantage of the shorthand notation~\eqref{eq:def_generator_rotation}.
\newpage 
%
%
%
\bibliography{BibTex/literature-kleinert-ulm.bib}
\end{document}

%% file: macro.inc
\newcommand{\ket}[1]{\mid\!#1\,\rangle}
\newcommand{\bra}[1]{\langle\, #1 \! \mid}
\newcommand{\braket}[2]{~\langle\, #1\! \mid\! #2 \,\rangle}
\newcommand{\ketbra}[2]{\ket{ #1} \bra{#2}}
\renewcommand{\i}{\text{i}}
\newcommand{\e}{\text{e}}
\newcommand{\T}{\text{T}}

\newcommand{\fig}[1]{Fig.~#1}
\newcommand{\figs}[1]{Figs.~#1}

\renewcommand{\'}{\,'}
\renewcommand{\Re}{\mathfrak{Re}}

\renewcommand{\vec}[1]{\boldsymbol{#1}} 
\renewcommand{\'}{'}

\newcommand{\indket}[1]{\mid #1 \rangle}

\newcommand{\trace}{\operatorname{Tr~}}

\newcommand{\dif}{\!\!\!\text{d}}

\newcommand{\MZ}{\text{\tiny MZ}}
\newcommand{\BU}{\text{\tiny BU}}

\newcommand{\g}{\text{g}}
\renewcommand{\a}{\text{a}}
\newcommand{\G}{\text{G}}
\newcommand{\E}{\text{E}}
\newcommand{\N}{\text{N}}
\renewcommand{\H}{\text{H}}

\renewcommand{\S}{\text{S}}
\newcommand{\I}{\text{\tiny I}}

\newcommand{\Def}{\mathrel{\mathop{:}}=}